\documentclass[12pt]{article}
\usepackage{amssymb,graphicx}
\usepackage{amsmath,amsfonts}
\usepackage{epsfig}
\def\const{\mbox{const}}
\def\H{{\cal H}}

\def\d{\partial}
\def\l{\left(}
\def\r{\right)}

\newcommand{\be}{\begin{equation}}
\newcommand{\ee}{\end{equation}}
\newcommand{\bea}{\begin{eqnarray}}
\newcommand{\eea}{\end{eqnarray}}
\newcommand{\bg}{\begin{gather}}
\newcommand{\eg}{\end{gather}}
\newcommand{\bseq}{\begin{subequations}}
\newcommand{\eseq}{\end{subequations}}
\renewcommand{\ln}{\mathop{\rm ln}\nolimits}

\def\half{\frac{1}{2}}

\begin{document}
\begin{flushright}
\end{flushright}
\vspace{10pt}
\begin{center}
  {\LARGE \bf   Infrared-modified gravities\\
and massive gravitons} \\
\vspace{20pt}
V.A.~Rubakov$^a$ and P.G.~Tinyakov$^{b,a}$\\
\vspace{15pt}
  $^a$\textit{Institute for Nuclear Research\\ 
of the Russian Academy of Sciences,\\
60th October Anniversary Prospect, 7a, Moscow, 117312, Russia
  }\\
  $^b$\textit{
Service de Physique Th\'eorique, Universit\'e Libre de Bruxelles (ULB),\\
                     CP225, bld. du Triomphe, B-1050 Bruxelles, Belgium
}\\
    \end{center}
    \vspace{5pt}

\begin{abstract}
We review some theoretical and phenomenological aspects of massive
gravities in 4 dimensions. We start from the Fierz--Pauli theory with
Lorentz-invariant mass terms and then proceed to Lorentz-violating
masses. Unlike the former theory, some models with Lorentz-violation
have no pathologies in the spectrum in flat and nearly flat
backgrounds and lead to interesting phenomenology.

\end{abstract}

\newpage

\tableofcontents


\section{Introduction}
\label{sec:introduction}

Recently, there has been revival of interest in attempts to construct
models of gravity which deviate from General Relativity at ultra-large
distance and time scales, that is, models with infrared-modified
gravity. The general approach is to view these models as possible low
energy limits of an unknown fundamental theory, and at exploratory
stage not worry too much about issues like renormalizability,
embedding into an ultraviolet-complete theory, etc.  Yet there are
other self-consistency problems that limit severely the classes of
acceptable models, while phenomenology of remaining ones turns out to
be rather reach and interesting.

Besides pure curiosity, there were several original motivations for
the recent increase of this activity. One of them has to do with the
cosmological constant problem and the observational evidence for the
accelerated expansion of the Universe at the present epoch (for
reviews from theoretical prospective see, e.g.,
Refs.~\cite{Weinberg:1988cp,Sahni:1999gb,Dolgov:1997za,Chernin:2001sy,Padmanabhan:2002ji,Peebles:2002gy,Copeland:2006wr}).
This accelerated expansion may well be due to the cosmological
constant $=$ vacuum energy density, new weakly interacting field or
some other kind of dark energy, which contributes, according to
Ref.~\cite{Spergel:2006hy}, about 75\% to the total energy density
$\rho_c$ in the present Universe. The problem is that the value
$\epsilon_\Lambda$ of the dark energy density is very small,
\[
  \epsilon_\Lambda \simeq 0.75 \rho_c 
\simeq 4\cdot 10^{-6}~\frac{\mbox{GeV}}{\mbox{cm}^3}.
\]
This is by many orders of magnitude smaller than the values one would
associate, on dimensional grounds, with fundamental interactions ---
strong, electroweak and gravitational,
\[
\epsilon_\Lambda \sim 10^{-46} \epsilon_{QCD}
\sim 10^{-54} \epsilon_{EW} \sim 10^{-123} \epsilon_{grav}.
\]
In other words, the energy scale characteristic of dark
energy\footnote{Hereafter we set $\hbar=c=1$.}, $M_\Lambda =
\epsilon_\Lambda^{1/4} \sim 10^{-3}$~eV, is much smaller than the
energy scales of known interactions, $\Lambda_{QCD} \simeq 200$~MeV,
$M_W \simeq 80$~GeV, $M_{Pl} \simeq 10^{19}$~GeV. The unnatural
smallness of $\epsilon_\Lambda$ (or $M_\Lambda$) is precisely the
cosmological constant problem.

In fact, there are two parts of this problem. One is that the
contributions from strong (QCD), electroweak and gravitational sectors
to vacuum energy density should be of order of $\epsilon_{QCD} \simeq
\Lambda_{QCD}^4$, $\epsilon_{EW} \sim M_W^4$ and $\epsilon_{grav} \sim
M_{Pl}^4$, respectively\footnote{As an example, QCD vacuum has complex
structure --- there are quark and gluon condensates whose values are
determined by highly complicated (and largely unknown) dynamics and
depend on QCD parameters ($\Lambda_{QCD}$ and quark masses) in a
complicated way. The difference between the energy densities of this
vacuum and the naive, perturbative one is certainly of order
$\epsilon_{QCD}$, so there is no reason whatsoever for the energy
density of the physical vacuum to be 46 orders of magnitude smaller
than $\epsilon_{QCD}$.}. Thus, the first part of the cosmological
constant problem is to explain why $\epsilon_\Lambda$ is essentially
zero.  The second part is to understand why $\epsilon_\Lambda$ is in
fact non-zero, and what physics is behind the energy scale
$M_\Lambda$.

It is not inconceivable that the first part of the dark energy problem
may be solved by one or another mechanism that drives the cosmological
constant to zero (for a review see, e.g., Ref.~\cite{Dolgov:2006xi});
such a mechanism most probably would operate at a cosmological epoch
which preceeded any known stage of the cosmological evolution, but at
which the state of the Universe was quite similar to the present
one\footnote{A name suggested by G.~Ross for this scenario is ``deja
vu Universe''.}~\cite{Rubakov:1999aq,Steinhardt:2006bf}.

On the other hand, despite numerous attempts, no compelling idea has
been put forward of how the value of $M_\Lambda$ may be related to
other known fundamental energy scales. One possible viewpoint is that
$\epsilon_\Lambda$ is actually the cosmological constant
(time-independent quantity during the known history of our part of the
Universe), and that its value is determined anthropically (for reviews
see, e.g., Refs.~\cite{Weinberg:1988cp,Linde:2002gj}): much larger
values of $|\epsilon_\Lambda|$ would be inconsistent with our
existence.  This viewpoint implies that the Universe is much larger
than its visible part, and that $\epsilon_\Lambda$ takes different
values in different cosmologically large regions; we happen to have
measured small value of $\epsilon_\Lambda$ merely because there is
nobody in other places to measure (larger values of) the cosmological
``constant''.

Another option is that the accelerated expansion of the Universe is
due to new low energy (infrared) physics. Perhaps the best known
examples are quintessence models (for reviews see, e.g.,
Refs.~\cite{Sahni:1999gb,Dolgov:1997za,Chernin:2001sy,Padmanabhan:2002ji,Peebles:2002gy}),
in which gravity is described by General Relativity while the
accelerated expansion is driven by (dark) energy of a new super-weakly
interacting field (most conventionally, but not necessarily, this
field is Lorentz scalar). The original idea of infrared-modified
gravity is that, instead, the gravitational laws get changed at
cosmological distance and time scales, hopefully leading to the
accelerated expansion without dark energy at all.  This would
certainly be an interesting alternative to dark energy, which might
even be observationally testable.

Another original motivation for infrared-mo\-di\-fied gravities came from
theories with brane-worlds and extra dimensions of large or infinite
size (for a review see, e.g., Ref.~\cite{Rubakov:2001kp}). In these
theories, ordinary matter is trapped to a three-dimensional
hypersurface (brane) embedded in higher-dimensional space. The
idea~\cite{Charmousis:1999rg,Kogan:1999wc,Gregory:2000jc} is that
gravitons may propagate along ``our'' brane for finite (albeit long)
time, after which they escape into extra dimensions. This would modify
the brane-to-brane graviton propagator at large distances and time
intervals, thus changing the gravitational interactions between
particles on ``our'' brane. If successful, models with this property
would provide concrete and calculable examples of the infrared
modification of gravity. Again, this idea is very hard to implement in
a self-consistent way, and the models constructed so far have their
intrinsic problems. In this regard, it is worth mentioning that there
are claims for an exception: it has been
argued~\cite{Gabadadze:2007dv} that the Dvali--Gabadadze--Porrati
``brane-induced gravity'' model~\cite{Dvali:2000hr} (for a review see
Ref.~\cite{Gabadadze:2003ii}) may be fully self-consistent despite the
fact that on the face of it, this model becomes strongly coupled at
unacceptably large distances~\cite{Luty:2003vm,Rubakov:2003zb,Nicolis:2004qq}.
Interestingly, the DGP model has a self-accelerating branch of
cosmological
solutions~\cite{Deffayet:2000uy,Deffayet:2001pu,Dvali:2002pe}, which,
however, has phenomenologically unacceptable ghosts among
perturbations about these solutions~\cite{Luty:2003vm,Gorbunov:2005zk}.

Among other lines of thought we mention theories attempting to
incorporate
MOND~\cite{Bekenstein:1984tv,Bekenstein:1988zy,Bekenstein:2004ne,Bekenstein:2004ca,Bekenstein:2005nv},
which modify gravity for explaining rotation curves of galaxies
without dark matter, and RTG~\cite{Logunov:1998ge,Logunov:2001cx}
motivated by the desire to restore the full generality of energy and
momentum conservation laws. It remains to be seen whether these
theories can be made fully self-consistent and phenomenologically
acceptable.

Recently, massive graviton has been motivated from quite a different
prospective~\cite{'tHooft:2007bf}. Namely, there is a fairly widespread
expectation that Quantum Chromodynamics (QCD) may have a formulation
in terms of a string theory of some sort. Known string theories,
however, often have massless spin-2 state in the spectrum, while QCD
does not.  The argument is that it is desirable to remove this state
from the massless sector of string theory by giving it a mass. In
terms of effective four-dimensional low energy theory, this task
appears very similar to giving a mass to graviton.

One naturally expects that the infrared modification of gravity may be
associated with the modification of the dispersion law $\omega =
\omega ({\bf p})$ of metric perturbations at low spatial momenta ${\bf
p}$, the simplest option being the graviton mass. In this review we
will mostly discuss this type of theories, and stay in 4 dimensions.
We stress, however, that theories of this type by no means exhaust all
possible classes of theories with infrared-modified gravity.  Other
classes include, e.g., scalar-tensor theories, in which the
modification of gravity occurs due to the presence of extra field(s)
(scalars), over and beyond the space-time metric, that are relevant in
the infrared domain. There are examples of models belonging to the
latter class that not only are phenomenologically acceptable but also
lead to interesting cosmological dynamics, including the accelerated
expansion of the
Universe~\cite{Boisseau:2000pr,Gannouji:2006jm,Carloni:2007eu}.
Another class of models involves condensates of vector and/or tensor
fields, see, e.g., Refs.~\cite{Jacobson:2008aj,Gripaios:2004ms} and
references therein.  The discussion of these and similar models is
beyond the scope of this review.

As it often happens, irrespectively of the original motivations,
theoretical developments have lead to new insights. In the case of
infrared-modified gravity and modified graviton dispersion law, these
are insights into self-consistency issues, on the one hand, and
phenomenological implications, on the other. The reason behind
self-consistency problems is the lack of explicit invariance under
general coordinate transformations (or non-trivial realization of
these transformations). Indeed, unless extra fields are added to the
gravitational sector of the theory, straightforward implementation of
the requirement of this gauge symmetry leads in a unique way to
General Relativity (with cosmological constant) plus possible
higher-order terms irrelevant in the infrared domain. Once this gauge
symmetry is broken, explicitly or spontaneously, gravity gets
infrared-modified, but new light degrees of freedom may appear among
metric perturbations, over and beyond spin-2 gravitons. These new
degrees of freedom may be ghosts or tachyons, which is often
unacceptable.  Another dangerous possibility is that they may be
strongly interacting at energy scales above a certain ``ultraviolet''
scale $\Lambda_{UV}$.  This would mean that the theory of gravity gets
out of control at energies above $\Lambda_{UV}$.  If $\Lambda_{UV}$ is
too low, and if the new degrees of freedom do not effectively
decouple, the theory cannot be claimed phenomenologically acceptable.
We will see that the problems of this sort are quite generic to
theories admitting Minkowski background and having explicit
Lorentz-invariance in this background.

As far as four-dimensional models with infrared-modified gravity are
concerned, avoiding the self-consistency problems is relatively easy
if Lorentz-invariance is broken for excitations about flat
background. The main emphasis of this review is on models of this type
\cite{Arkani-Hamed:2003uy,Rubakov:2004eb,Dubovsky:2004sg,Dubovsky:2005dw,Dubovsky:2004ud}.
Breaking of Lorentz-invariance is in fact quite natural in this
context.  Indeed, infrared modification of gravity may be thought of
as an analog of the broken (Higgs) phase in gauge theories, gravity in
a certain sense being the gauge theory of the Lorentz group.  Gravity
in the Higgs phase is thus naturally expected to be
Lorentz-violating. We will discuss various aspects of theories of this
type, including self-consistency, technical naturalness and
phenomenology. The latter is quite interesting in a number of cases,
as intuition gained in Lorentz-invariant field theories is often
misleading when Lorentz-invariance does not hold. In the end of this
review we come back to the issue of the accelerated expansion of the
Universe.

\section{Fierz-Pauli model} 
\label{sec:fierz-pauli-model}

To understand better the problems arising when one attempts to modify
the gravitational interaction at large distances, it is instructive to
consider first the {\it Lorentz-invariant} massive gravity. The
Lorentz-invariant graviton mass term was proposed by Fierz and Pauli
\cite{Fierz:1939ix}; we will refer to the corresponding model as the
Fierz--Pauli model.

One may think of a theory, belonging to the class we discuss in this
Section, in various ways. One may simply add graviton mass terms to
the Einstein--Hilbert action, as we do in what follows.  Equivalently,
one may consider General Relativity interacting with extra massless
fields (see, e.g., Refs.~\cite{Arkani-Hamed:2002sp,Kakushadze:2007hf}
and references therein).  Once these fields obtain background values
which depend on space-time coordinates, general covariance is broken,
and graviton gets a mass in a manner reminiscent of the Higgs
mechanism. In either approach, one arrives at one and the same class
of theories, provided that one imposes the following requirements: (i)
Minkowski space-time is a legitimate background, i.e., flat metric
solves the field equations\footnote{It is worth noting that massive
gravities with legitimate backgrounds other than Minkowski may have
properties quite different from those exposed in this Section. A well
studied example is massive gravity about (anti-)~de~Sitter
space~\cite{Porrati:2001db}.}; (ii) in Minkowski background, there are
no light fields except for metric perturbations; (iii) Lorentz
invariance is unbroken in this background.

The issues which generically arise in modified gravities are: the
ghosts/instabilities; the absence of the zero mass limit (the
van~Dam--Veltman--Zakharov discontinuity); strong coupling at
parametrically low ultraviolet (UV) energy scale; the existence of a
``hidden'' Boulware--Deser mode which is not seen in the analysis of
perturbations about Minkowski background but becomes propagating ---
and dangerous --- once the background is curved, even slightly.  In
this Section we discuss these issues using the Fierz--Pauli model as
an example.

\subsection{Lorentz-invariant massive gravity in Minkowski 
background}
\label{sub:LI-1}

Consider General Relativity plus the most general Lorentz-invariant
graviton mass term added to the action.  We parameterize nearly
Minkowski metric as follows,
\[
   g_{\mu \nu} = \eta_{\mu \nu} + h_{\mu \nu} \; .
\]
We will sometimes need the expressions for $g^{\mu \nu}$ and
$\sqrt{-g}$ up to the second order,
\begin{eqnarray}
    g^{\mu \nu} &=& \eta^{\mu \nu} - h^{\mu \nu}
+ h^{\mu \lambda} h_{\lambda}^\nu \; ,
\nonumber \\
\sqrt{-g} &=& 1 + \half h^\mu_\mu + \frac{1}{8} h^\mu_\mu h_\lambda^\lambda
- \frac{1}{4} h^{\mu \nu} h_{\mu \nu},
\label{eq:h-parameterization}
\end{eqnarray}
where indices are raised and lowered with Minkowski metric $\eta_{\mu
\nu} = \mbox{diag}(1,-1,-1,-1)$.  At the quadratic level about
Minkowski background, the general action for Lorentz-invariant massive
gravity reads
\begin{equation}
S=M_{\rm Pl}^2 \int d^4x \left\{ 
L^{(2)}_{EH}(h_{\mu\nu}) + \frac{\alpha}{4} 
h_{\mu\nu} h^{\mu \nu} + \frac{\beta}{4} (h_\mu^\mu)^2
\right\},
\label{eq:LI-mass-term}
\end{equation}
where $M_{Pl}^2 = 1/(16 \pi G)$, $\alpha$ and $\beta$ are arbitrary
coefficients of  dimension [mass squared], while $L^{(2)}_{EH}$ is
the standard graviton kinetic term coming from the Einstein--Hilbert
action. The latter can be written in the following form,
\begin{equation}
L^{(2)}_{EH} = \frac{1}{4}
\left( \d_\lambda h^{\mu \nu} \, \d^\lambda h_{\mu \nu}
- 2 \d_\mu h^{\mu \nu} \, \d_\lambda h^\lambda_\nu
+ 2 \d_\mu h^{\mu \nu} \, \d_\nu h^\lambda_\lambda -
\d_\mu h^\nu_\nu \, \d^\mu h^\lambda_\lambda \right) \; .
\label{EHquadratic-LI}
\end{equation} 
When discussing metric
perturbations, we will use the convention that the Lagrangian and the
action are related by
\begin{equation}
    S = M_{Pl}^2 \int~d^4 x~ L \; .
\label{dec9-6}
\end{equation}
This will simplify formulas; many of them will not include $M_{Pl}$.

In what follows it will be convenient to use both the above
Lorentz-covariant form of the Lagrangian and the form corresponding to
the (3+1) decomposition.  The metric perturbations in the latter
formalism are traditionally parameterized as
follows~\cite{Mukhanov:1990me},
\begin{eqnarray}
\nonumber
 h_{00} &=& 2\varphi,\\
\nonumber
 h_{0i} &=&  S_i + \d_i B,\\
 h_{ij} &=&  h^{TT}_{ij} - \d_i F_j - \d_j F_i 
- 2 (\psi \delta_{ij} - \d_i\d_j E).
\label{nov14-1}
\end{eqnarray}
Here $h^{TT}_{ij}$ is a transverse-traceless 3-tensor,
\[
    \partial_i h_{ij}^{TT} = 0 \; , \;\;\;\; h^{TT}_{ i i} = 0 \; ,
\]
$S_i$ and $F_i$ are transverse 3-vectors, 
\[
   \partial_i S_i = \partial_i F_i = 0 \; ,
\]
and other variables are 3-scalars; hereafter summation over spatial
indices $i,j = 1,2,3$ is performed with Euclidean metric.
Accordingly, the quadratic part of the Einstein--Hilbert term
decomposes into tensor, vector and scalar parts,
\begin{equation}
    L_{EH}^{(2)} = L_{EH}^{(T)} + L_{EH}^{(V)} + L_{EH}^{(S)},
\label{jan25-11}
\end{equation}
where 
\begin{eqnarray}
    L_{EH}^{(T)} &=& \frac{1}{4} \left(\d_0 h_{ij}^{TT} \,
\d_0 h_{ij}^{TT}
- \d_k h_{ij}^{TT} \, \d_k h_{ij}^{TT} \right),
\label{T-kinetic}\\
   L_{EH}^{(V)}  &=& \frac{1}{2} \d_k (S_i + \d_0 F_i) \,
  \d_k (S_i + \d_0 F_i),
\label{V-kinetic}\\
   L_{EH}^{(S)} &=& 2
[\d_k \psi \d_k \psi - 3 \d_0 \psi \d_0 \psi
 + 2 \d_k (\varphi - \d_0 B + \d_0^2 E) \d_k \psi].
\label{S-kinetic}
\end{eqnarray}
Likewise, the mass terms decompose as 
\begin{eqnarray}
L_{m} = && M_{Pl}^2  \left\{ \frac{\alpha}{4} h_{ij}^{(TT)}  h_{ij}^{(TT)} 
   \right. \\
&& +  \frac{\alpha}{2} (\d_i F_j \d_i F_j - S_i S_i) 
\label{nov15-1} \\
&& +  \left[ (\alpha + \beta) \varphi^2 + 2\beta (3 \psi - \Delta E) 
\varphi
+  (\alpha + \beta) (\Delta E)^2  
\right.
\nonumber \\
&& \left. \left. - 2(\alpha + 3 \beta) \psi \Delta E
  + 3 (\alpha + 3\beta) \psi^2  + \frac{\alpha}{2} B \Delta B
\right] \right\}.
\label{nov15-4}
\end{eqnarray}
Hereafter Lagrangians differing by total derivative are not distinguished.

In General Relativity, most of the fields entering the Lagrangian
$L_{EH}$ do not propagate: the only propagating degrees of freedom are
conveniently parameterized by $h^{TT}_{ij}$, these are transverse
traceless gravitational waves. This feature is of course a consequence
of the gauge invariance of General Relativity, with gauge
transformations, at the linearized level about Minkowski background,
having the form
\begin{equation}
h_{\mu\nu}(x)  \to h_{\mu\nu}(x)  + \partial_\mu \zeta_\nu(x) +
\partial_\nu \zeta_\mu(x),
\label{Mink-gaugetransf}
\end{equation}
where $\zeta_\mu (x)$ are arbitrary functions of coordinates.

Once the mass terms are added, gauge invariance is lost, and there
emerge extra propagating degrees of freedom.  Indeed, one expects that
massless spin-2 graviton will be promoted to massive spin-2
particle. While the former has two polarization states, the latter has
five, with helicities $\pm 2$, $\pm 1$ and $0$. In (3+1)-language this
corresponds to two tensor modes, two vector modes and one scalar mode,
respectively. For general $\alpha$ and $\beta$, however, there is one
more scalar mode, which is necessarily a ghost.  Let us consider this
issue, within the (3+1)-formalism first.

In the tensor sector, $h_{ij}^{TT}$ remain two propagating degrees of
freedom, whose mass is now given by
\begin{equation}
m_G^2 = - \alpha \; .
\label{oct23-add2}
\end{equation}
To avoid tachyons, we consider the case $\alpha <0$ in what
follows\footnote{We will see shortly that $\alpha>0$ leads to even
  more severe problem in the vector sector of the model.}. 
The tensor modes have healthy kinetic term (\ref{T-kinetic}), so this
sector is not problematic.

The vector sector contains a non-dynamical field $S_i$ which enters
the action without time derivatives. 

Let us pause at this point to discuss, in general terms, two kinds of
non-dynamical fields. At the level of quadratic action, a
non-dynamical field may enter the action either linearly or
quadratically.  An example of the latter situation is given by the
vector sector of massive gravity: there is a term $S_i^2$ in the
Lagrangian (\ref{nov15-1}), and also a term $(\d_k S_i)^2$ in
(\ref{V-kinetic}).  In that case the non-dynamical field can be
integrated out: the field equation obtained by varying this field can
be used to express this field through dynamical fields (the latter
enter the action with time derivatives), and then one gets rid of this
field by plugging the resulting expression back into the action. The
number of dynamical fields is not, generally speaking, reduced in this
way (there are important exceptions in gauge-invariant theories, which
we will encounter a number of times in this review).

Another possibility is that the action does not contain a term which
is quadratic in non-dynamical field.  This is the case, e.g., in the
scalar sector of General Relativity, whose action (\ref{S-kinetic}) is
linear in the field $\varphi$ (and $B$, which is also a non-dynamical
field, since, after integration by parts, it enters without time
derivatives).  Unlike in the quadratic case, the corresponding field
equation is a constraint imposed on dynamical fields, and the
non-dynamical field itself is a Lagrange multiplier.  An important
feature here is that the constraint reduces the number of dynamical
fields, i.e., the number of degrees of freedom.

This discussion is straightforwardly generalized to the case of
several non-dynamical fields: if the part of the action, which is
quadratic in these fields, is non-degenerate, all these fields belong
to the first category, otherwise there are Lagrange multipliers whose
number equals the degree of degeneracy.
 
After this general remark, let us come back to the vector sector, and
integrate out the field $S_i$.  Its field equation is
\begin{equation}
 (\Delta - m_G^2) S_i = - \Delta \d_0 F_i \; ,
\nonumber
\end{equation}
where $\Delta$ is the 3-dimensional Laplacian.  Expressing $S_i$
through $F_i$ by making use of this equation, and plugging it back
into the action, one obtains the action for the remaining field
$F_i$. In massless theory, the latter action is identically zero, so
that $F_i$ does not have to obey any equation and thus is
arbitrary. This arbitrariness is of course a consequence of the gauge
freedom (\ref{Mink-gaugetransf}), in this case with transverse
$\zeta_i$.  With the mass terms added, the field $F_i$ is dynamical.
The Lagrangian for $F_i$, in 3-dimensional momentum representation,
is\footnote{Hereafter ${\bf p}$ denotes 3-momentum, while $p$ is
reserved for 4-momentum.}
\begin{equation}
L_F = \frac{m_G^2}{2} \left[ \frac{{\bf p}^2}{{\bf p}^2 + m_G^2}
\d_0 F_i^* ({\bf p})  \d_0 F_i ({\bf p}) - {\bf p}^2
 F_i^* ({\bf p})  F_i ({\bf p}) \right].
\label{LF-feb23}
\end{equation}
To convert it into the standard form, one introduces canonically
normalized field
\begin{equation}
{\cal F}_i ({\bf p})
= M_{Pl} m_G \sqrt{\frac{{\bf p}^2}{{\bf p}^2 + m_G^2}}
F_i ({\bf p})
\label{O5-3*}
\end{equation}
and finds that the linearized action is
\begin{equation}
S_{\cal F} =
\int~d^3p~ \half \left[
\d_0 {\cal F}_i^*  \d_0 {\cal F}_i
 - ({\bf p}^2 + m_G^2)  {\cal F}_i^*  {\cal F}_i \right].
\nonumber
\end{equation}
Hence, the vector sector has two propagating degrees of freedom of
mass $m_G$ (recall that $F_i$ are transverse, so only two components
of $F_i$ are independent).  The number of propagating degrees of
freedom in the tensor and vector sectors corresponds to the number of
helicity $\pm 2$ and $\pm 1$ states of massive graviton, respectively,
in accord with expectations.

Let us note here that for $m_G^2 \equiv - \alpha <0$ the
Lagrangian (\ref{LF-feb23}) has negative overall sign, so that
vector modes are ghosts in that case. This is even worse than
tachyon behaviour of tensor modes.

Let us come back to the theory with $m_G^2 \equiv - \alpha >0$.  From
(\ref{O5-3*}) we see that the limit $m_G \to 0$ is singular.
Fluctuations of the canonically normalized field ${\cal F}_i$ are
finite in this limit, so fluctuations of the vector part of the
metric, the field $F_i$, diverge as $m_G^{-1}$.  At small but finite
$m_G$ this results in the fact that the quantum theory becomes
strongly coupled at an ultraviolet energy scale $\Lambda_{UV}$ which
is much lower than $M_{Pl}$. In the vector sector this scale is not
unacceptably low, however. We will discuss strong coupling later, as
it becomes a problem in the scalar, rather than vector, sector.

Let us now turn to the scalar sector, and consider first the general
case
\begin{equation}
\alpha \neq 0 \; , \;\;\; 
\alpha \neq - \beta \; , \;\;\; 
\alpha \neq - 2 \beta \; .
\nonumber
\end{equation}
Upon integrating by parts, one obtains the following form of the
Lagrangian, including mass terms,
\begin{eqnarray}
L^{(S)} =&& 2 \left[
-2 \varphi \Delta \psi - 2 \dot{\psi} \Delta B
+ 2 \dot{\psi} \Delta \dot{E}
- 3 \dot{\psi}^2 - \psi \Delta \psi
\right.
\nonumber \\
&& +\left.
\frac{\alpha + \beta}{2} \varphi^2 + \beta (3 \psi - \Delta E) \varphi
+  \frac{\alpha + \beta}{2} (\Delta E)^2  
\right.
\nonumber \\
&&~~~~~\left. - (\alpha + 3 \beta) \psi \Delta E
  + 3 \frac{\alpha + 3\beta}{2} \psi^2  + \frac{\alpha}{4} B \Delta B
\right] .
\label{Minkactfull}
\end{eqnarray}
In the case of General Relativity ($\alpha = \beta =0$), the fields
$\varphi$ and $B$ are Lagrange multipliers, giving one and the same
constraint $\psi = 0$. Then the equation of motion obtained by varying
$\psi$ gives $\varphi= \dot{B} - \ddot{E}$; varying $E$ gives nothing
new.  There are no propagating degrees of freedom, while the fields
$B$ and $E$ remain arbitrary. This is again due to the gauge freedom
(\ref{Mink-gaugetransf}), now with $\zeta_0 \neq 0$ and $\zeta_i =
\d_i \zeta_L$.

In the massive case, the fields $\varphi$ and $B$ are no longer
Lagrange multipliers, but they are still non-dynamical and can be
integrated out. Integrating out $B$ one obtains an additional term in
the Lagrangian,
\begin{equation}
L_B = -\frac{8}{\alpha} \dot{\psi} \Delta \dot{\psi} \; ,
\label{LU}
\end{equation}
while integrating out $\varphi$ gives another additional term
\begin{equation}
L_\varphi = - \frac{1}{\alpha + \beta}
[2\Delta \psi - \beta (3\psi - \Delta E)]^2.
\label{oct23-add1}
\end{equation}
Then the Lagrangian for the remaining fields $\psi$ and $E$ takes the
form
\begin{equation}
L^{(S)} =
L_B + L_\varphi + 2 \left[
2 \dot{\psi} \Delta \dot{E}
- 3 \dot{\psi}^2 - \psi \Delta \psi
  + \frac{\alpha + \beta}{2} (\Delta E)^2  
- (\alpha + 3 \beta) \psi \Delta E
  + 3 \frac{\alpha + 3\beta}{2} \psi^2  
\right] .
\label{Minkactfull-bis}
\end{equation}
Both fields $\psi$ and $E$ are dynamical, so there are two propagating
degrees of freedom in the scalar sector. Thus, there is an extra
scalar mode over and beyond the expected helicity-0 state of massive
graviton. This degree of freedom is actually a ghost (negative sign of
the kinetic term).

To see this, let us concentrate on the terms with time derivatives.
These come from the terms explicit in (\ref{Minkactfull-bis}) and from
the term (\ref{LU}). So, for given spatial momentum, the relevant part
of the Lagrangian has the form
\begin{eqnarray}
L_{kin} &=& \frac{A}{2} \dot{\psi}^2 + B \dot{\psi} \dot{E} 
\nonumber \\
&=& \frac{A}{2} \l \dot{\psi} + \frac{B}{A} \dot{E} \r^2
- \frac{B^2}{2A} \dot{E}^2,
\label{kin-general}
\end{eqnarray}
where $A$ and $B$ are numerical coefficients (depending on spatial
momentum).  One observes that irrespectively of the sign of $A$, one
of the two degrees of freedom is a ghost. Of course, the theory is
Lorentz-invariant, so both modes have $\omega^2 = {\bf p}^2$ at high
spatial momenta, and this ghost exists at arbitrarily high spatial
momenta ${\bf p}$.

It is clear from (\ref{oct23-add1}) that the case $\alpha = - \beta$
is special\footnote{The cases $\alpha=0$ and $\alpha=-2\beta$ are
special too. For $\alpha=-2\beta$ the ghost has the same mass as the
graviton, and because of this degeneracy, its wave function grows in
time, i.e., behaves like $t \cdot \mbox{exp} (i\omega t)$.  For
$\alpha=0$ the graviton mass (\ref{oct23-add2}) is zero, and one can
check that the only degrees of freedom of the linearized theory are
transverse traceless massless gravitons.  So, the theory with
$\alpha=0$ cannot be considered as massive gravity, and we will not
discuss this case further.}.  This is precisely the Fierz--Pauli
theory where
\begin{equation}
  \beta = - \alpha = m_G^2.
\nonumber
\end{equation}
In this case there is no quadratic term in $\varphi$, so $\varphi$ is
a Lagrange multiplier. The corresponding constraint is
\begin{equation}
\Delta E = 3\psi - \frac{2}{\beta}  \Delta \psi.
\label{FP-constraint-Mink}
\end{equation}
This constraint kills one degree of freedom out of two, so the only
mode in the scalar sector is the helicity-0 state of massive graviton
with normal (positive) sign of the kinetic term. Indeed, inserting
(\ref{FP-constraint-Mink}) back into the action\footnote{One may
question whether this procedure is legitimate, since one of the field
equations is apparently lost. This is not the case: in the original
formulation, this would-be-lost equation is an equation determining
the Lagrange multiplier $\varphi$ in terms of $\psi$.}
(\ref{Minkactfull}) and adding the term (\ref{LU}) one finds that the
only remaining degree of freedom is $\psi$, and the Lagrangian has the
kinetic term
\[
L_{kin, \, \psi} = 6 \dot{\psi}^2.
\]
In fact, the complete quadratic Lagrangian for $\psi$ is
\begin{equation}
L_{\psi \, FP} = 
6\left(  \partial_\mu \psi \partial^\mu \psi
- m_G^2 \psi^2 \right),
\label{FPMink}
\end{equation}
in full analogy with, e.g., the Lagrangian for tensor modes
$h^{TT}_{ij}$.

To end up this discussion, we make the following comment.  Of course,
in the case of Minkowski background and Lorentz-invariant mass terms,
the analysis is most straightforwardly performed in Lorentz-covariant
way. The (3+1)-formalism utilized here certainly looks as unnecessary
complication. Our analysis, however, does provide useful
insights. First, it suggests that the problems of massive gravity are
most severe in the scalar sector. Second, it shows that the
Fierz--Pauli miracle occurs due to the special property of the field
$\varphi$, which acts as the Lagrange multiplier and kills the
undesirable degree of freedom.  We will see in Section
\ref{sec:boulware-deser-mode} that this property is lost in curved
backgrounds, and the extra degree of freedom --- Boulware--Deser mode
--- reappears in the spectrum. Finally, our discussion suggests a
possibility that Lorentz-violating mass terms may give rise to a
healthier theory, provided they are chosen in such a way that the
unwanted degree of freedom is killed in a consistent way. We will
discuss this possibility in the following Sections.

\subsection{Ghost via St\"uckelberg trick}
\label{sec:ghost-gener-lorentz}

A convenient way to single out and study dangerous degrees of freedom
is to make use of the St\"uckelberg formalism
\cite{Arkani-Hamed:2002sp}. The idea is to enlarge the field content
of massive gravity in such a way that the gauge invariance is
restored, and then make a judicious choice of gauge fixing
condition. We will use this trick in various Sections of this review,
and here we illustrate the St\"uckelberg approach by re-obtaining the
ghost in the spectrum about Minkowski background.

Let us again consider the theory (\ref{eq:LI-mass-term}) with
general mass terms. In linearized theory, one introduces a new,
St\"uckelberg field $\xi_\mu$ and a new field $\bar{h}_{\mu \nu}$ 
by writing
\begin{equation}
h_{\mu\nu}=\bar h_{\mu\nu} + \partial_\mu\xi_\nu(x) +
\partial_\nu\xi_\mu(x).
\label{eq:2.coord-change}
\end{equation}
Then the linearized theory is invariant under gauge transformations
\begin{eqnarray}
 \bar{h}_{\mu\nu}(x)  &\to& \bar{h}_{\mu\nu}(x)  + \partial_\mu \zeta_\nu(x) +
\partial_\nu \zeta_\mu(x)\; ,
\nonumber \\
\xi_\mu &\to& \xi_\mu - \zeta_\mu.
\nonumber
\end{eqnarray}
Importantly, because of the gauge invariance of General Relativity,
the Einstein--Hilbert part of the quadratic action,
Eq.~(\ref{EHquadratic-LI}), is independent of $\xi_\mu$,
\[
L_{EH}^{(2)} = L_{EH}^{(2)}(\bar{h}_{\mu\nu}).
\]
Note that we did not introduce new degrees of freedom: by imposing the
gauge condition $\xi_\mu = 0$ we get back to the original massive
gravity.  The trick is to impose a gauge condition on $\bar{h}_{\mu
\nu}$ instead, and do so in such a way that all independent components
of $\bar{h}_{\mu \nu}$ obtain non-trivial kinetic terms from the
Einstein--Hilbert Lagrangian. This guarantees that the fields
$\bar{h}_{\mu \nu}$ and $\xi_\mu$ will decouple at high energies (for
$\alpha \neq - \beta$), and the properties of the dangerous modes will
be read off from the Lagrangian involving the fields $\xi_\mu$
only. We will further comment on this procedure later on, see
Eqs.~(\ref{12-1**}), (\ref{12-1*}).

We are interested in relatively high energies and spatial momenta,
$\omega^2, \, {\bf p}^2 \gg |\alpha|, |\beta|$.  We will thus keep the
terms in the action which are of the highest order in
derivatives. Because of the structure of (\ref{eq:2.coord-change}),
these terms come not only from the Einstein--Hilbert part of the
action but also from the mass terms. This is a peculiarity of the
St\"uckelberg formalism.

The choice of the gauge condition for $\bar{h}_{\mu \nu}$ is not very
important. One may think of the gauge $\bar{h}_{00}=0$,
$\bar{h}_{0i}=0$ or of covariant gauges\footnote{Unlike non-covariant
gauges, covariant ones do not fix the gauge completely.  There remain
unphysical modes to worry about.}. In either case, the remaining
components of $\bar{h}_{\mu \nu} $ have non-degenerate terms with two
time derivatives, which come from the Einstein--Hilbert action. As an
example, in the gauge $\bar{h}_{00}=0$, $\bar{h}_{0i}=0$, i.e.,
$\bar{\varphi} =0$, $\bar{S}_i =0$, $\bar{B}=0$, the fields
$\bar{F}_i$, $\bar{\psi}$ and $\bar{E}$ have non-degenerate terms with
two time derivatives, see (\ref{V-kinetic}) and (\ref{S-kinetic}).  We
will sometimes write, schematically,
\[
  L_{EH}^{(2)} = (\d \bar{h})^2.
\]
Kinetic terms for the field $\xi_\mu$ come from the mass terms in the
action (\ref{eq:LI-mass-term}).  They have the following form,
\begin{equation}
\frac{\alpha}{2} (\partial_\mu\xi_\nu)^2 
+\l\frac{\alpha}{2}+ \beta\r(\partial_\mu\xi^\mu)^2.
\label{LXI-LI}
\end{equation}
The mass terms induce also mixing between $\xi_\mu$ and $\bar{h}_{\mu
\nu}$, but as we will discuss shortly, this mixing is unimportant for
$\alpha \neq - \beta$ at high momenta and frequencies, ${\bf p}^2,
\omega^2 \gg |\alpha|, |\beta|$.  Once this mixing is neglected, the
fields $\bar{h}_{\mu \nu}$ and $\xi_\mu$ decouple, as promised, so we
can study the metric and St\"uckelberg sectors, $\bar{h}_{\mu \nu}$
and $\xi^\mu$, independently.

In the metric sector, the kinetic part of the Lagrangian is just the
gauge fixed Einstein--Hilbert Lagrangian. Thus, the only propagating
modes in this sector are $\bar{h}_{ij}^{TT}$.  Other propagating modes
belong to the St\"uckelberg sector.  Once mixing between $\bar{h}_{\mu
\nu}$ and $\xi^\mu$ is taken into account, the latter modes have
non-vanishing $\bar{h}_{\mu \nu}$, but this effect is small and can be
neglected. Let us see explicitly how this works in the gauge
$\bar{h}_{00} = 0$, $\bar{h}_{0i} = 0$. Consider, as an example,
vector sector in (3+1)-decomposition language. The full Lagrangian in
this sector is
\begin{equation}
L^{(V)} = \half (\d_i \d_0 \bar{F}_i)^2 + \frac{\alpha}{2} (\d_i \bar{F}_j)^2
- \alpha \d_i \bar{F}_j \cdot \d_i \xi_j^T -
\frac{\alpha}{2} (\d_\mu \xi_i^T)^2,
\label{nov20-1}
\end{equation}
where we set $S_i=0 $ according to our gauge choice and $\xi_i^T$ is
3-dimensionally transverse, $\d_i \xi_i^T = 0$.  The first term in
(\ref{nov20-1}) is the Einstein--Hilbert term (\ref{V-kinetic}), the
last term comes from (\ref{LXI-LI}), while the third term is precisely
the term that mixes metric and the St\"uckelberg field. The field
equations are
\begin{eqnarray}
  \ddot{\bar{F}}_i + \alpha \xi_i^T - \alpha \bar{F}_i = 0,
\label{nov20-2}
\\
\Box \xi_i^T +  \Delta \bar{F}_i = 0.
\label{nov20-3}
\end{eqnarray}
At $\omega^2, {\bf p}^2 \gg m_G^2\equiv -\alpha$ these equations may
be solved perturbatively in $\alpha$. To the zeroth order, the first
of these equations has no oscillating solutions, so $F_i =0$ and the
only propagating modes are $\xi_i^T$, as expected. These are
helicity-1 states of massive graviton in the St\"uckelberg picture. To
the first order, one has from (\ref{nov20-2})
\[
  \bar{F}_i = \frac{\alpha}{2 \omega^2} \xi_i^T,
\]
so eq.~(\ref{nov20-3}) becomes
\[
   (\omega^2 - {\bf p}^2) \xi_i^T + \frac{\alpha {\bf p}^2}{\omega^2} 
 \xi_i^T = 0.
\]
As promised, the second term here, which appears due to mixing between
$\bar{F}_i$ and $\xi_i^T$, is negligible for $\omega^2, {\bf p}^2 \gg
m_G^2$.

The lesson from this exercise is twofold. First, it shows that
neglecting the metric sector $\bar{h}_{\mu \nu}$ is indeed legitimate
(except for helicity-2 states) as long as modes with $\omega^2, {\bf
p}^2 \gg m_G^2$ are considered. Second, we see that the St\"uckelberg
formalism is useless for studying modes with $\omega^2 \lesssim
m_G^2$; off hand, one might even loose some modes with $\omega^2
\lesssim m_G^2$ by considering the St\"uckelberg sector only. Indeed,
there is no guarantee that the system of equations like
(\ref{nov20-2}), (\ref{nov20-3}) does not have more slowly oscillating
solutions than equation $\Box \xi_i^T =0$; after all, equations
(\ref{nov20-2}), (\ref{nov20-3}) are both second order in time.  In
the theory considered in this Section, the number of modes with high
and low $\omega$ is the same by Lorentz-invariance, but the last
remark should be kept in mind when studying Lorentz-violating massive
gravities.

Let us come back to modes with $\omega^2, {\bf p}^2 \gg m_G^2$ and
consider the St\"uckelberg sector.  One may view the expression
(\ref{LXI-LI}) as the generic Lorentz-invariant Lagrangian for the
vector field $\xi_\mu$.  It is well known that this Lagrangian has a
ghost in the spectrum unless the two terms combine into field strength
tensor $F_{\mu\nu}^2= (\partial_\mu\xi_\nu - \partial_\nu\xi_\mu)^2$.
This occurs when $\alpha = - \beta$.  Thus, we again see that the
no-ghost situation is possible in the Fierz--Pauli case only.

Ghosts are unacceptable in Lorentz-invariant theory. Therefore, we
concentrate on the Fierz--Pauli theory in the rest of this Section.

Let us see how the St\"uckelberg analysis applies to the Fierz--Pauli
case, $\beta =-\alpha = m_G^2$. In that case, the relevant part of the
mass term reads
\begin{eqnarray}
L_{m} = && - \frac{m_G^2}{2}
(\d_\mu \xi_\nu - \d_\nu \xi_\mu) (\d^\mu \xi^\nu - \d^\nu \xi^\mu)
\nonumber \\
&& 
- m_G^2 (\d_\nu \xi^\mu \bar{h}^\nu_\mu - \d_\mu \xi^\mu \bar{h}^\nu_\nu),
\label{13-1*}
\end{eqnarray}
where we have omitted terms without derivatives but kept the kinetic
mixing between $\bar{h}_{\mu \nu}$ and $\xi^\mu$. The 4-dimensionally
transverse part of $\xi^\mu$, that obeys $\d_\mu \xi_{tr}^\mu=0$, has
healthy kinetic term given by the first line in (\ref{13-1*}). On the
other hand, the longitudinal part,
\[
\xi_\mu = \frac{1}{2} \d_\mu \phi ,
\] 
has the kinetic term only due to mixing with the field $\bar{h}_{\mu
\nu}$; this is why the mixing term, which is subdominant at $\alpha
\neq - \beta$, plays a key role now. Let us not impose the gauge
condition on $\bar{h}_{\mu \nu}$ for the time being. Then the kinetic
term for $\bar{h}_{\mu \nu}$ and $\phi$ is given by
\begin{equation}
L_{EH}^{(2)}(\bar{h}_{\mu \nu} ) - \frac{m_G^2}{2}
(\d_\mu \d_\nu \bar{h}^{\mu \nu} - \d_\mu \d^\mu \bar{h}^\nu_\nu) \phi.
\label{13-1**}
\end{equation}
It can be diagonalized~\cite{Arkani-Hamed:2002sp} by noticing that the
combination $(\d_\mu \d_\nu \bar{h}^{\mu \nu} - \d_\mu \d^\mu
\bar{h}^\nu_\nu)$ is proportional to the linearized Riemann curvature,
so the second term in (\ref{13-1**}) has the structure $m_G^2
R(\bar{h}_{\mu\nu}) \cdot \phi$.  Hence, the kinetic term is
diagonalized by a conformal transformation, which at the linearized
level reads
\begin{equation}
   \bar{h}_{\mu \nu} = \hat{h}_{\mu \nu} - \frac{m_G^2}{2}
\eta_{\mu \nu} \phi.
\label{hat-h-def}
\end{equation}
Then the kinetic term becomes
\begin{equation}
 L_{kin}  = L_{EH}^{(2)} (\hat{h}_{\mu \nu}) +
\frac{3}{8} m_G^4 \d_\mu \phi \d^\mu \phi.
\label{13-2*}
\end{equation}
Upon gauge fixing of $\hat{h}_{\mu \nu}$, the longitudinal sector of
the theory contains one degree of freedom $\phi$ with healthy kinetic
term. In this way one recovers the absence of ghosts in the
Fierz--Pauli theory.

Note that for general $\alpha\neq -\beta$, the term (\ref{13-2*}) is
subdominant as compared to the term $(\alpha + \beta) (\Box \phi)^2$
that arises from (\ref{LXI-LI}).  Thus, mixing between scalar parts of
$\bar{h}_{\mu \nu}$ and $\xi^\mu$ is unimportant in that case, just
like in the vector sector.

A general comment is in order. It has to do with the fact that the
St\"uckelberg procedure, with gauge conditions imposed on $\bar
h_{\mu\nu}$, may introduce spurious solutions to the field
equations. Consider, as an example, the gauge $\bar{h}_{00} = 0$,
$\bar{h}_{0i}=0$. With this gauge choice, one has $h_{00} = 2 \d_0
\xi_0$, $h_{0i} = \d_0 \xi_i + \d_i \xi_0$.  Varying the action of the
original theory with respect to $h_{\mu \nu}$, one obtains the field
equations
\begin{equation}
    \frac{\delta S}{\delta h_{00}} = 0 \; , \;\;\;\; 
\frac{\delta S}{\delta h_{0i}} = 0  \; , \;\;\;\; 
\frac{\delta S}{\delta h_{ij}} = 0.
\label{12-1**}
\end{equation}
The first two of these equations do not contain second time
derivatives; these are constraints. On the other hand, substituting
$h_{00} = 2 \d_0 \xi_0$, $h_{0i} = \d_0 \xi_i + \d_i \xi_0$ into the
action and varying then with respect to $\xi_\mu$ and $h_{ij}$ one
finds
\begin{equation}
    \d_0 \l\frac{\delta S}{\delta h_{00}} \r 
+ \d_i \l \frac{\delta S}{\delta h_{0i}} \r = 0  \; , \;\;\;\; 
\d_0 \l \frac{\delta S}{\delta h_{0i}} \r = 0  \; , \;\;\;\; 
\frac{\delta S}{\delta h_{ij}} = 0.
\label{12-1*}
\end{equation}
There are no constraints any longer; instead, all these equations are
second order in time. Hence, the system (\ref{12-1*}) has more
solutions than (\ref{12-1**}). However, we are interested in
propagating modes, i.e., solutions to the linearized field equations
that have the form $\mbox{exp} (i\omega t - i {\bf px})$. In this case
the left hand sides of (\ref{12-1**}) oscillate, unless they are
identically zero, so the left hand sides of (\ref{12-1*}) cannot
vanish unless (\ref{12-1**}) are satisfied. The system (\ref{12-1*})
has the same number of propagating modes as the original system
(\ref{12-1**}).  Also, energies and momenta of the solutions are the
same in the original and St\"uckelberg formalisms: if a propagating
mode is a ghost in the St\"uckelberg formalism, it is a ghost in the
original theory.  Indeed, in general there is unique (modulo terms
which do not contribute to total energy and momentum) energy-momentum
tensor that is conserved on solutions of the field equations. The
latter observation is valid also at non-linear level, and for
adiabatically varying backgrounds.

\subsection{Van~Dam--Veltman--Zakharov discontinuity}
\label{sec:vdvz-discontinuity}

The Fierz-Pauli mass term changes the gravitational interaction both
between two massive bodies and between a massive body and 
light. This interaction is straightforward to calculate in the
weak-field approximation \cite{vanDam:1970vg,Zakharov:1970cc}. The
result is surprising: the prediction for the light bending in the
massive case is different from General Relativity even in the limit of
zero graviton mass. This is known as the van~Dam--Veltman--Zakharov
(vDVZ) discontinuity: the linearized Fierz--Pauli theory does not
approach  linearized General Relativity as $m_G \to 0$. Taken at
face value, this result would mean that the Fierz--Pauli gravity is
ruled out, as the experimental measurement of the light bending agrees
with the General Relativity (see, e.g., \cite{Will:2005va} and
references therein).

Let us consider this phenomenon in more detail. At the linearized
level, the interaction between two sources of the gravitational field
is given by
\[
G T^{\mu\nu}
P_{\mu\nu\lambda\rho}T'^{\lambda\rho},
\]
where $G$ is the gravitational coupling constant,
$P_{\mu\nu\lambda\rho}$ is the propagator of the gravitational field
and $T^{\mu\nu}$ and $T'^{\lambda\rho}$ are the energy-momentum
tensors of the two sources. The point is that the propagators are
different in the massive and massless cases.  Their structure in both
cases is
\[
P_{\mu\nu\lambda\rho} \propto 
\frac{\sum_i e^i_{\mu\nu}e^i_{\lambda\rho}}{p^2-m_G^2},
\]
where $e^i_{\mu\nu}$ are the graviton polarization tensors; in the
massless case the mass in denominator is zero. Since the denominator
is continuous in the limit of zero mass, it is the sum over the
polarizations which is responsible for the discontinuity.

In the massive case there are 5 polarization tensors. The summation
over these tensors gives
\begin{eqnarray}
\label{eq:massive-propagator}
\mbox{FP:}\;\;\; P_{\mu\nu\lambda\rho}&=& \frac{1}{p^2-m_G^2}
\biggl\{
\frac{1}{2} \eta_{\mu\lambda}\eta_{\nu\rho} + 
\frac{1}{ 2} \eta_{\mu\rho}\eta_{\nu\lambda} - 
\frac{1}{3} \eta_{\mu\nu}\eta_{\lambda\rho}
+ (\mbox{$p$-dependent terms})\biggr\},
\end{eqnarray}
where the terms containing $p_{\mu}$ are irrelevant because they do
not contribute when contracted with the conserved energy-momentum
tensors. In the massless case there are only two polarizations.  The
propagator in this case takes the form
\begin{eqnarray}
\label{eq:massless-propagator}
\mbox{GR:}\;\;\;
P_{\mu\nu\lambda\rho} &=& \frac{1}{p^2}
\biggl\{
\frac{1}{2} \eta_{\mu\lambda}\eta_{\nu\rho} + 
\frac{1}{2} \eta_{\mu\rho}\eta_{\nu\lambda} - 
\frac{1}{2} \eta_{\mu\nu}\eta_{\lambda\rho}
+ (\mbox{$p$-dependent terms})\biggr\}.
\end{eqnarray}
The difference between these two expressions is in the coefficient of
the third term.  This difference persists in the limit of zero mass;
this is precisely the vDVZ discontinuity.  Note also that the
difference is in the part of the propagator which couples to the trace
of the energy-momentum tensor\footnote{One may ask whether one can get
around the vDVZ discontinuity by abandoning the weak equivalence
principle, i.e., by modifying the way in which gravity couples to
matter. Indeed, in massive gravity, the consistency of the field
equations does not require the covariant conservation of the source
tensor (unlike in General Relativity, where the gravitational part of
the field equations --- the Einstein tensor --- obeys the Bianchi
identity, so the matter part --- energy-momentum tensor --- must obey
the covariant conservation law).  If, instead of the coupling to the
conserved energy momentum tensor, the field $h_{\mu \nu}$ couples to
some tensor $S^{\mu \nu}$ whose divergence is non-zero at finite $m_G$
and vanishes in the massless limit only, the above analysis does not
go through.  This question has been studied in
Ref.~\cite{Ford:1980up}; the result is that one cannot get rid of the
vDVZ discontinuity in this way.}.

It is worth noting that the vDVZ discontinuity is specific to spin-2
field. In the case of vector field, the zero mass limit of the
propagator coincides, modulo longitudinal piece, with the propagator
of massless field, so the vDVZ discontinuity is absent.

In general, the interaction constants $G_{GR}$ and $G_{FP}$ in the
massless and massive cases are different. The relation between them
can be found by requiring that two non-relativistic bodies interact
with the same strength in the massive and massless theories. In the
non-relativistic limit only the $00$-component of the energy-momentum
tensor contributes, so that one has
\begin{eqnarray}
\nonumber
\mbox{GR:} \quad G_{GR} \,T_{\mu\nu} 
\,P_{\mu\nu\lambda\rho}\, T'_{\lambda\rho}
&=& \frac{1}{2} G_{GR}\, T_{00} T'_{00} \frac{1}{p^2}  ,
\\ \nonumber
\mbox{FP:} \quad  G_{FP} \,T_{\mu\nu} 
\,\tilde P_{\mu\nu\lambda\rho}\, T'_{\lambda\rho}
&=& \frac{2}{3}  G_{FP}\, T_{00} T'_{00} \frac{1}{ p^2-m_G^2}  .
\end{eqnarray}
This implies for the zero-mass limit 
\begin{equation}
G_{FP} = \frac{3}{ 4} G_{GR} \equiv  \frac{3}{ 4} G_{Newton} \; . 
\label{eq:G=tildeG}
\end{equation}
Consider now the prediction for the light bending in both cases. The
energy-momentum tensor of the electromagnetic wave is traceless. Thus,
the third term in the propagator does not contribute, and one finds
the following expressions for the interaction strength:
\begin{eqnarray}
\nonumber
\mbox{GR:} &\quad& G_{GR} T_{00} T_{00}' \frac{1}{ p^2}, 
\\ \nonumber
\mbox{FP:} &\quad& G_{FP} T_{00} T_{00}' \frac{1}{p^2-m_G^2}.
\end{eqnarray}
In view of eq.~(\ref{eq:G=tildeG}), the light bending predicted in the
massive theory in the limit of the vanishing graviton mass is $3/4$ of
that in the massless theory, General Relativity.

Clearly, the discontinuity is related to the longitudinal
polarizations of the graviton, i.e., to the St\"uckelberg field
$\xi_\mu$ discussed in the preceding Section. In what follows we will
shed more light on the mechanism responsible for this phenomenon.

\subsection{Vainshtein radius}
\label{sec:vainshtein-radius}

As we already noticed, if the arguments of the previous Section were
strictly correct, they would imply that the mass of the graviton is
exactly zero in the Lorentz-invariant theory.  These arguments,
however, have a loophole \cite{Vainshtein:1972sx} as they rely on the
linear approximation. In General Relativity this approximation is
valid for distances much larger than the Schwarzschild radius of the
source. Therefore, the gravitational bending of light that passes next
to the surface of the Sun is well described in the linear regime.

The situation is different in the Fierz--Pauli gravity.  It has been
argued in Ref.~\cite{Vainshtein:1972sx} by studying spherically
symmetric classical solutions, that at non-zero graviton mass the
linear approximation actually breaks down already at distance much
larger than the Schwarzschild radius, namely at the distance called
the Vainshtein radius,
\begin{equation}
r_V=\l\frac{M}{M_{\rm Pl}^2 m^4_G}\r^{1/5},
\label{eq:2.r_V}
\end{equation}
where $M$ is the mass of the source.  Note that the smaller the
graviton mass, the larger the distance where the non-linear regime
sets in. Taking the graviton mass to be of order of the present Hubble
parameter one finds $r_V\simeq 100~{\rm kpc}$ for the Sun, so that
bodies orbiting the Sun, as well as light passing not far from the
solar surface feel the non-linear gravitational interaction. The above
argument for the wrong bending of light in massive theory is,
therefore, not directly applicable. On the other hand, non-linearity
of the Fierz-Pauli gravity in the entire solar system is a problem by
itself. 

The origin of the scale $r_V$ is easy to understand by simple power
counting \cite{Arkani-Hamed:2002sp}. Let us recall first how the
Schwarzschild radius $ r_S=2M/M_{\rm Pl}^2$ appears as the expansion
parameter in General Relativity. Schematically, the quadratic
Einstein--Hilbert action (\ref{EHquadratic-LI}) with source term reads
\begin{equation}
\int~d^4x~ \left[ M_{\rm Pl}^2 (\partial h)^2 + T h \right],
\label{eq:2.quadr-action}
\end{equation}
where $h$ stands for the metric perturbation and $T$ for the
energy-momentum tensor. The corresponding equations have the
solution\footnote{Throughout this review, when presenting
power-counting arguments we ignore numerical factors and signs.}
\[
h=\frac{1}{\partial^2} \frac{T}{ M_{\rm Pl}^2}
\]
or, equivalently, 
\begin{equation}
h = \frac{M}{ M_{\rm Pl}^2 r},
\label{eq:2.Newtonslaw}
\end{equation}
where $M$ is the total mass of the source. This is the standard
Newton's law. The non-linear corrections to the action begin with the
terms of the type $M_{Pl}^2 \int~d^4x~ h(\partial h)^2$. Requiring
that these terms are small compared to the quadratic contributions
(\ref{eq:2.quadr-action}) one obtains the condition $h \ll 1$, i.e.
for the perturbation (\ref{eq:2.Newtonslaw}),
\[
\frac{M}{ M_{\rm Pl}^2 r}\ll 1 \; .
\]
This is precisely the condition $r\gg r_S$ ensuring the validity
of the linear approximation in General Relativity.

In the case of the Fierz--Pauli massive gravity this condition has to
be satisfied as well. However, there is a stronger constraint.  In the
St\"uckelberg language of Section \ref{sec:ghost-gener-lorentz}, this
constraint comes from the analysis of the field $\xi_\mu$. Of
particular importance is its scalar part $\xi_\mu = \partial_\mu
\phi$. It follows from the discussion in the end of Section
\ref{sec:ghost-gener-lorentz} that the action for the fields
$\hat{h}_{\mu \nu}$ and $\phi$ in the presence of conserved source
$T_{\mu \nu}$ has schematically the following form
\begin{equation}
\int~d^4x~ \left[  M_{\rm Pl}^2 (\partial \hat h)^2 +  M_{\rm Pl}^2 m_G^4
(\partial \phi)^2  + T \hat h + m^2_G T \phi + \ldots\right],
\label{DiagAction-LI}
\end{equation}
where we again omitted numerical coefficients and did not write
explicitly the mass terms for $\hat{h}_{\mu \nu}$ and $\phi$.  The
kinetic term here is the same as in (\ref{13-2*}), while the source
term is obtained from the standard expression $h^{\mu \nu}T_{\mu \nu}$
by making use of (\ref{eq:2.coord-change}) and (\ref{hat-h-def}),
together with the linearized conservation law $\d_\mu T^{\mu \nu} =
0$.  Note that after the diagonalization of the kinetic term via
(\ref{hat-h-def}), there appears direct interaction of matter with the
field $\phi$. By solving the equations of motion we find that at
distances much smaller than $m_G^{-1}$ the gravitational potential
$\hat h$ is given by eq.~(\ref{eq:2.Newtonslaw}) and that $m^2_G \phi$
is of the same order,
\[
m^2_G \phi = \frac{M}{ M_{\rm Pl}^2 r}.
\] 
The latter formula implies that $\phi$ itself is large and singular in
the limit $m_G\to 0$, cf. Section \ref{sub:LI-1}, Eq.~(\ref{O5-3*}).
This is the origin of the non-linearity at large distances from the
source.

Indeed, a non-linear generalization of the Fierz--Pauli mass term
would contain higher powers of the perturbation $h_{\mu \nu}$.  The
lowest term of this type is just $h^3$. This term gives rise to the
non-linear contribution to the action of the form
\begin{equation}
\int~d^4x~M_{\rm Pl}^2 m_G^2 (\partial^2\phi)^3. 
\label{oct11-*}
\end{equation}
Another source of terms of the same order is the non-linearity of
gauge transformations in General Relativity. In general, the
coordinate transformation $x^\mu \to x^\mu + \zeta^\mu$ corresponds to
the following gauge transformation of metric,
\begin{equation}
  g_{\mu \nu}(x) \to g^\prime_{\mu \nu} (x)= g_{\mu \nu}(x+\zeta) +
\d_\mu \zeta^\lambda \; g_{\nu \lambda}(x+\zeta) + \d_\nu
\zeta^\lambda \; g_{\mu \lambda}(x+\zeta) + \d_\mu \zeta^\lambda \;
\d_\nu \zeta^\rho \; g_{\lambda \rho}(x+\zeta)\; .
\label{nov16-19*}
\end{equation}
Writing $g_{\mu \nu} = \eta_{\mu \nu} + h_{\mu \nu}$, one obtains at
quadratic order (in both $h_{\mu \nu}$ and $\zeta^\mu$)
\begin{eqnarray}
h^{\prime}_{\mu \nu} = h_{\mu \nu} &+&
 \d_\mu \zeta_\nu +  \d_\nu \zeta_\mu
\nonumber \\
&+&  \d_\mu \zeta^\lambda  \; \d_\nu \zeta_\lambda +
 \d_\mu \zeta^\lambda \; h_{\nu \lambda} + \d_\nu \zeta^\lambda \; 
h_{\mu \lambda}
\nonumber
\end{eqnarray}
where indices are still raised and lowered with Minkowski metric.
Accordingly, the change of variables (\ref{eq:2.coord-change}) at this
order has the form
\begin{eqnarray}
h_{\mu \nu} = \bar{h}_{\mu \nu} &+&
 \d_\mu \xi_\nu +  \d_\nu \xi_\mu
\nonumber \\
&+&  \d_\mu \xi^\lambda \; \d_\nu \xi_\lambda +
 \d_\mu \xi^\lambda \; \bar{h}_{\nu \lambda} 
+ \d_\nu \xi^\lambda \; \bar{h}_{\mu \lambda}
\label{oct-11-**}
\end{eqnarray}
The field $\xi^\mu$ still does not enter the Einstein--Hilbert part of
the action, while the mass term receives the contribution, whose
schematic form is
\begin{equation}
\int~d^4x~M_{Pl}^2 m_G^2 (\d \xi)^3
\label{oct11-888}
\end{equation}
that is, the contribution of the form (\ref{oct11-*}) for $\xi_\mu =
\d_\mu \phi$.

Linearized theory is valid when the contribution (\ref{oct11-*}) is
smaller than the quadratic term. This requirement leads to the
condition
\[
\frac{M}{M_{\rm Pl}^2 m_G^4 r^5}\ll 1,
\]
which is equivalent to $r\ll r_V$ with $r_V$ given by
eq.~(\ref{eq:2.r_V}).

The situation may be improved by tuning the explicit $h^3$ terms in
the non-linearly generalized Fierz--Pauli action in such a way that
the leading correction $(\partial^2\phi)^3$ is absent. In this way the
onset of the non-linear regime may be pushed to smaller scales,
namely, to the distance
\begin{equation}
r_*=\left(\frac{M}{ M_{\rm Pl} ^2m_G^2}\right)^{1/3}.
\label{oct24-add1}
\end{equation}
One can show that the situation cannot be improved further
\cite{Arkani-Hamed:2002sp}. For the graviton mass of the order of the
present Hubble parameter\footnote{Hereafter we assume that greater
values of $m_G$ would be inconsistent with cosmology. Even without
this assumption, the requirement that Newton's law remains valid at
the scale of galaxy clusters would give $m_G \lesssim
10~\mbox{Mpc}^{-1}$. The estimates here and in the rest of this
Section would change by about two orders of magnitude, with no change
of conclusions.}  the non-linear regime occurs at distances below
$r_*\sim 10~{\rm pc}$ from the Sun, which still covers the whole solar
system.

Let us make contact of the analysis presented here with the study of
the vDVZ discontinuity. It is clear from (\ref{DiagAction-LI}) that
the gravitational field $\bar h_{\mu \nu}$ that interacts with matter,
is an admixture of the two fields $\hat{h}_{\mu \nu}$ and $m_G^2
\eta_{\mu \nu}\phi$.  The field $\hat{h}_{\mu \nu}$ has the same
kinetic term $L_{EH}(\hat{h}_{\mu \nu})$ as the linearized
gravitational field in General Relativity, while $m_G^2 \phi$ has the
kinetic term of a gravi-scalar.  Both fields interact with matter at
roughly the same strength.  Keeping the part $\hat{h}_{\mu \nu}$ only,
one would obtain, in the massless limit, the propagator of $\bar
h_{\mu \nu}$ (and hence the propagator of the full metric perturbation
$h_{\mu \nu}$ modulo longitudinal terms omitted in
(\ref{eq:massive-propagator}) and (\ref{eq:massless-propagator}))
which has precisely the form of the propagator in the linearized
General Relativity. The field $m_G^2 \phi$ adds extra trace piece into
the propagator, which does not vanish in the massless limit and sums
up with the contribution of $\hat{h}_{\mu \nu}$ to the propagator
(\ref{eq:massive-propagator}).

To summarize, distances below which massive gravity is in non-linear
regime are not less than given by (\ref{oct24-add1}) and hence are
very large.  One may hope that the non-linear interactions would
modify the theory in such a way as to make the massless limit
smooth~\cite{Vainshtein:1972sx}. This indeed happens in some cases, an
example being the DGP model~\cite{Nicolis:2004qq} where the non-linear
interactions affect mainly the gravi-scalar sector and essentially
decouple it from other modes in the small mass limit, eliminating the
extra contribution to the propagator responsible for the vDVZ
discontinuity. This mechanism, however, does not work in the case of
the Fierz--Pauli theory~\cite{Creminelli:2005qk}.  Thus, the
Fierz--Pauli gravity about Minkowski background is problematic already
at the classical level: it most likely contradicts precision tests of
General Relativity. It becomes even more problematic at the quantum
level, as we will discuss below.

\subsection{Strong coupling}
\label{sec:strong-coupling}

At the quantum level the above non-linearity problem manifests itself
as strong coupling at the energy scale which is much lower than the
naive expectation.

Both massless and massive gravity should be treated as low energy
effective theories valid at energies (more precisely, momentum
transfers) below some ultra-violet (UV) scale $\Lambda_{UV}$.  Above
this scale, these theories are meant to be extended into some
``fundamental'' theories (UV-completions) with better UV behavior.
The situation here is analogous to the theory of self-interacting
massive vector field, whose possible UV-completion is non-Abelian
gauge theory with the Higgs mechanism.  In the case of General
Relativity, the UV-completion is most likely the string/M-theory;
whether there exists a UV-completion of massive gravity is not known
(in fact, the discussion in this Section suggests that the
Lorentz-invariant massive gravity does not have UV-completion at all).

One can trust the effective theories only at distances $r \gg
\Lambda_{UV}^{-1}$; at shorter distances one has to deal with putative
UV-complete theory which most likely has quite different
properties. Experimentally, Newtonian gravity has been tested down to
sub-millimeter
distances~\cite{Long:2003dx,Long:2003ta,Smullin:2005iv,Hoyle:2004cw},
so an effective low energy theory should be valid down to those
distances. This implies
\[
\Lambda_{UV}^{-1} \lesssim 10^{-2}~\mbox{cm} \; ,
\;\;\;\; \mbox{or} \;\;\;\;
\Lambda_{UV} \gtrsim 10^{-3}~\mbox{eV}.
\]
On the other hand, an upper limit on the UV energy scale
$\Lambda_{UV}$ can be obtained within the low energy effective theory
itself.  The tree level scattering amplitudes, calculated in the low
energy theory, grow with the center-of-mass energy $E$ and eventually,
at some energy $E\sim \Lambda$, become large and hit the unitarity
limit. Above this energy $\Lambda$ the effective theory becomes
strongly coupled, and hence cannot be trusted. The entire framework
makes sense provided that the UV-completion scale is below the strong
coupling scale,
\begin{equation}
\Lambda_{UV} \lesssim \Lambda.
\label{oct12-1-4*}
\end{equation}
A well known illustration is again non-Abelian gauge theory with the
Higgs mechanism. The strong coupling scale of its effective low energy
theory --- the theory of massive self-interacting vector field --- is
equal to
\begin{equation}
\Lambda = m_V/g \; , 
\label{oct12-1-extra}
\end{equation}
where $m_V$ is the vector boson mass and $g$ is the gauge
coupling. This low energy theory is extended into its UV-completion
--- the gauge theory in the Higgs phase --- at the scale $\Lambda_{UV}
= m_H$, where $m_H$ is the Higgs boson mass: at the latter scale new
degrees of freedom, the Higgs bosons, show up. Since $\Lambda = m_V/g
= v$, where $v$ is the Higgs vacuum expectation value, and $m_H=
\sqrt{\lambda} v$, where $\lambda < 1$ is the Higgs self-coupling,
the inequality (\ref{oct12-1-4*}) indeed holds in this
example.

In General Relativity, the strong coupling scale is $\Lambda =
M_{Pl}$, so its UV-completion may occur well above accessible
energies\footnote{This is not a necessity: the inequality in
(\ref{oct12-1-4*}) may be strong. In this regard, an interesting
possibility is TeV-scale gravities, where $\Lambda_{UV}$ is of order
of a few TeV, and new phenomena occur at collider energies. This
possibility is reviewed, e.g., in 
Refs.~\cite{Rubakov:2001kp,Gherghetta:2006ha,Hewett:2005uc,Rubakov:2005ub}.}.
In massive gravity, the strong coupling scale $\Lambda$, and hence the
UV-completion scale $\Lambda_{UV}$, are certainly much below
$M_{Pl}$. Naively, one would estimate the strong coupling scale as
follows,
\begin{equation}
\Lambda \sim (M_{Pl}m_G)^{1/2}.
\label{oct12-1-3++}
\end{equation}
This is a direct analog of (\ref{oct12-1-extra}). Indeed, consider the
transverse component of the St\"uckelberg field $\xi^\mu$ which obeys
$\d_\mu \xi^\mu_{tr} = 0$. The kinetic term in the Lagrangian for this
component comes from the Fierz--Pauli mass term and schematically has
the form
\begin{equation}
L_{tr}^{(2)} = M_{Pl}^2 m_G^2 (\d \xi_{tr})^2.
\label{oct12-1-3+}
\end{equation}
The terms, which are cubic in $h_{\mu \nu}$ and come from a non-linear
generalization of the Fierz--Pauli term, as well as from the
non-linear change of variables (\ref{oct-11-**}), give rise to the
interaction terms schematically written in (\ref{oct11-888}). Had the
form (\ref{oct12-1-3+}) been common to the kinetic terms for both
transverse and longitudinal components of $\xi^\mu$, one would
introduce canonically normalized field $\chi^\mu = M_{Pl} m_G \xi^\mu$
and find that it enters the kinetic term with unit coefficient, while
the interaction Lagrangian is
\begin{equation}
   L_{int} = \frac{1}{M_{Pl}m_G} (\d \chi)^3.
\label{oct12-1-4**}
\end{equation}
This theory would indeed have the strong coupling scale
(\ref{oct12-1-3++}), since it is this parameter that suppresses the
higher order operator (\ref{oct12-1-4**}). The analysis of other
higher order operators, $(\d \xi)^4$, etc., would lead to the same
conclusion.  The same argument applied to theories of self-interacting
vector fields leads to the estimate (\ref{oct12-1-extra}), hence the
analogy between (\ref{oct12-1-3++}) and (\ref{oct12-1-extra}).

The scale (\ref{oct12-1-3++}) is actually quite interesting
phenomenologically.  For the graviton mass of order of the present
Hubble parameter, the corresponding distance is
\begin{equation}
(M_{Pl} m_G)^{-1/2} \simeq 0.05~\mbox{mm}.
\label{oct12-1-4*+}
\end{equation}
Were this the true scale of the UV-completion, one would expect novel
phenomena in the gravitational sector at sub-millimeter distances. In
the Fierz--Pauli theory, however, the strong coupling scale is
actually much lower than the estimate (\ref{oct12-1-3++}), and the
corresponding distance is much greater than (\ref{oct12-1-4*+}).

The problem occurs in the longitudinal sector, where $\xi_\mu = \d_\mu
\phi$. Let us recall Eqs.~(\ref{DiagAction-LI}) and
(\ref{oct11-*}). According to (\ref{DiagAction-LI}), the canonically
normalized field is $\chi = M_{Pl} m_G^2 \phi$, and from
(\ref{oct11-*}) we see that its self-interaction has the form
\[
   L_{int} = \frac{1}{M_{Pl}m_G^4} (\d^2 \chi)^3.
\]
The scale that suppresses this higher order operator is now
\begin{equation}
\Lambda = (M_{\rm Pl}m^4)^{1/5}.
\label{eq:Lambda_5}
\end{equation}
This is the actual strong coupling scale in generic Fierz--Pauli
theory.  For the graviton mass of order the present Hubble parameter
this scale is of order $10^{-21}~{\rm eV}\sim (10^{16}~{\rm
cm})^{-1}$, which is clearly too low.

\begin{figure}
\begin{picture}(300,110)(0,0)
\put(70,0){
\epsfig{file=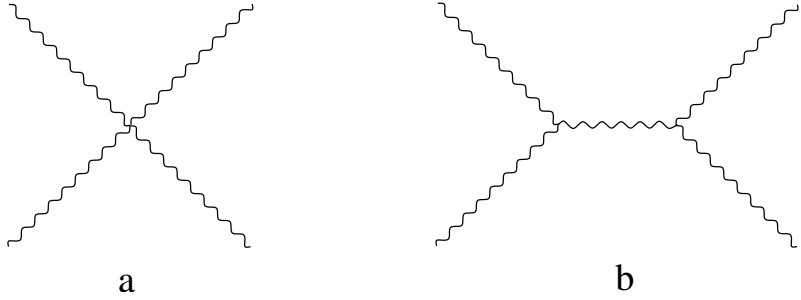,%
width=300pt,height=110pt,%
}}
\end{picture}
\caption{Four-graviton scattering in the first two orders of the
 perturbation theory (panels $a$ and $b$, respectively). Only
 $s$-channel diagram is shown in the second order, panel $b$. }
\label{fig:4graviton-scattering}
\end{figure}
In the framework of perturbation theory, the origin of this strong
coupling is the growth of the propagator and the wave functions of the
longitudinal components of massive graviton with energy, much in
common with the case of massive non-Abelian vector field.  Consider
the amplitude of the four-graviton scattering represented by the
diagrams of Fig.~\ref{fig:4graviton-scattering}. The external lines of
the diagrams for longitudinal gravitons behave like $E^2/m^2_G$. The
4-vertex gives the factor $E^2/M_{\rm Pl}^2$. Therefore, the first of
the diagrams gives the contribution of order
\[
\frac{E^{10}}{ M_{\rm Pl}^2 m_G^8}.
\]
The second diagram is of the same order, since two leading
contributions in the propagator cancel out in the on-shell
amplitude~\cite{Aubert:2003je}. Thus, the scattering amplitude indeed
becomes large at energies of order (\ref{eq:Lambda_5}).  This has been
checked by the explicit calculation of the
amplitude~\cite{Aubert:2003je}.

The scale of strong coupling can be pushed to higher energies by a
judicious choice of the interaction terms.  Indeed, one is able to
choose a non-linear extension of the Fierz--Pauli theory in such a way
that the cubic terms $(\d^2 \phi)^3$ vanish. Then the fourth-order
terms are
\[
L_{int} = M_{Pl}^2 m_G^2 (\d^2 \phi)^4 = \frac{1}{M_{Pl}^2 m_G^6}
(\d^2 \chi)^4
\]
and the strong coupling scale is given by
\be
\Lambda = (M_{\rm Pl}m^2_G)^{1/3}.
\label{finish1}
\ee
This is the best that can achieved~\cite{Arkani-Hamed:2002sp}, since
there are not only self-interactions of the field $\phi$ (these can be
canceled out by an appropriate choice of higher order terms in $h_{\mu
\nu}$) but also interactions between the longitudinal component of the
St\"uckelberg field, $\xi_\mu = \d_\mu \phi$, and the transverse
component $\xi^\mu_{tr}$ .

For the mass of the graviton $m_G$ of the order of the present Hubble
parameter one finds from (\ref{finish1}) that $\Lambda \sim 3\times
10^{-13}{\rm eV} \sim (10^8~{\rm cm})^{-1}$. This is also unacceptably
low\footnote{This discussion refers to flat background. It may, in
principle, happen that effects due to curvature push the strong
coupling scale to higher values, as occurs, for instance, in the DGP
model \cite{Nicolis:2004qq}. This does not happen in the Fierz-Pauli
case \cite{Creminelli:2005qk}.}.  We conclude that the Fierz--Pauli
theory suffers from severe strong coupling problem.

\subsection{Fierz--Pauli theory in curved backgrounds:
Boulware--Deser mode}
\label{sec:boulware-deser-mode}

\subsubsection{Cosmological background}
\label{subsub-cosmo}

If the background is not exactly Minkowskian, the Fierz--Pauli
cancellation no longer works, and the ghost or tachyon mode reappears
in the spectrum.  This mode exists for arbitrarily high spatial
momenta, and hence it is unacceptable phenomenologically.  This
phenomenon is known as the Boulware--Deser
instability~\cite{Boulware:1973my}. Importantly, it occurs
irrespectively of the way the Fierz--Pauli theory is generalized to
curved space-time~\cite{Creminelli:2005qk}.

Let us see the appearance of the Boulware--Deser mode explicitly, by
working out an example of the cosmological background. As we have
discussed above, the scalar sector is the most problematic; indeed, as
we will find shortly, the Boulware--Deser mode emerges precisely in
this sector. So, we concentrate on the scalar sector in what follows.

To warm up, we begin with General Relativity.  Let the metric with
perturbations have the following form
\begin{equation}
ds^2 = a^2(\eta) (\eta_{\mu \nu} + h_{\mu \nu}) dx^\mu dx^\nu,
\label{nov15-3}
\end{equation}
where $\eta$ is conformal time.  Note that we have changed the
definition of $h_{\mu \nu}$ here; in previous Sections $h_{\mu \nu}$
denoted the deviation of $g_{\mu \nu}$ from background metric, while
here this deviation is equal to $a^2 h_{\mu \nu}$.  In what follows we
will raise and lower indices by Minkowski metric, so that by
definition $ h_{\mu}^{\nu} = \eta^{\nu \lambda} h_{\mu \lambda}$.
Another convention is that summation over spatial indices will be
performed using $\delta_{ij}$, and we will never use covariant
derivatives in explicit formulas. Hence, the dependence on the scale
factor will be always explicit.

The linearized gauge transformations (\ref{Mink-gaugetransf}) are
generalized by making use of (\ref{nov16-19*}).  According to our
conventions, we define $\zeta_\mu = \eta_{\mu \nu} \zeta^\nu$ and
write the gauge transformations in this background as follows,
\begin{itemize}
\item[--] 
spatial, $\zeta_i = - \partial_i \zeta_L$:
\begin{equation}
  \delta B = \zeta_L^\prime \; , \;\;\;\;
  \delta E = \zeta_L
\label{gaugetr-L}
\end{equation}

\item[--]
temporal, $\zeta_0$:
\begin{equation}
 \delta B = \zeta_0 \; , \;\;\;
  \delta \varphi = \zeta_0^\prime + \H \zeta_0
\; , \;\;\;
  \delta \psi =  \H \zeta_0,
\label{gaugetr-0}
\end{equation}
\end{itemize}
where we use the notations (\ref{nov14-1}) and specify to the scalar
sector.  Hereafter prime denotes $\d/ \d\eta$ and
\[
\H = \frac{a^\prime}{a} \; .
\]

To consider expanding Universe, we introduce positive cosmological
constant, the corresponding term in the action being
\[
S_\Lambda = - 6 H_0^2 M_{Pl}^2 \int~dx~\sqrt{-g}.
\]
Here the constant $H_0^2$ is, by the Einstein equations, the Hubble
parameter of the de~Sitter space in the theory without graviton mass
and without matter. The background equations are
\begin{eqnarray}
     \H^2 &=& H_0^2 a^2,
\nonumber \\
   2\H^\prime + \H^2  
&=& 3 H_0^2 a^2 \; ,
\label{nov15-2}
\end{eqnarray}
and their solution, the scale factor of the de~Sitter space-time, is
\[
a = - \frac{1}{H_0 \eta} \; .
\]
The quadratic part of the cosmological constant term is
\begin{equation}
S_\Lambda^{(2)} =
 2H_0^2 M_{Pl}^2 \int~d^3x d\eta ~ a^4
\left[
\frac{3}{2} \varphi^2 - 9 \varphi \psi + 3\varphi \Delta E
- \frac{9}{2} \psi^2 + 3\psi \Delta E + \frac{3}{2} (\Delta E)^2
- \frac{3}{2} (\d_i B)^2 \right] \; ,
\label{dec16-1}
\end{equation}
where $\Delta = \d_i \d_i$.

By appearance, the part (\ref{dec16-1}) in the action resembles the
graviton mass term, as it contains no derivatives of the fields.  This
term does not have the Fierz--Pauli structure, and both
$\varphi^2$-term and $(\d_i B)^2$-term are present.  Hence, unlike in
Minkowski background, none of these fields appears to be a Lagrange
multiplier, and one might suspect that the theory has two dynamical
scalars $\psi$ and $E$.  This does not happen in General Relativity:
the two modes remaining after integrating out $\varphi$ and $B$ are
pure gauge modes.

Indeed, let us consider the Einstein--Hilbert and cosmological terms
together. Off shell (that is, for arbitrary background) the quadratic
part is
\begin{eqnarray}
S^{(2)}_{EH + \Lambda} = 2 M_{Pl}^2 \int~d^3x d\eta~a^2&&\left[
\phantom{\frac{}{}} 
-2 \varphi \Delta \psi - 2 \psi^\prime \Delta B
+ 2 \psi^\prime \Delta E^{\prime }
+ 3 \psi \psi^{\prime \prime} - \psi \Delta \psi
\right.
\nonumber \\
&& \left.
 + \H (
 2 \varphi \Delta B - 2\varphi \Delta E^\prime + 6 \phi \psi^\prime)
\right.
\nonumber\\
&& + \l -\frac{9}{2} \H^2
+ \frac{3}{2} H_0^2 a^2 \r \varphi^2
+\l -\frac{9}{2} \H^2
- \frac{9}{2} H_0^2 a^2 \r \psi^2
\nonumber \\
&& \left. + \l \H^2 - H_0^2 a^2 \r
\l 9\varphi \psi - 3\varphi \Delta E + \frac{3}{2} (\d_i B)^2
\r \right.
\nonumber \\
&& \left.
 + \l 2\H^\prime + \H^2 
- 3 H_0^2 a^2 \r
\l - \psi \Delta E - \half (\Delta E)^2 \r \right].
\label{dSaction-nomass}
\end{eqnarray}
Now, because of the background equations of motion, the two last lines
in this expression vanish, and the action simplifies to
\begin{eqnarray}
S^{(2)}_{EH + \Lambda} = 2 M_{Pl}^2 \int~d^4x~a^2&&\left[
-2 \varphi \Delta \psi - 2 \psi^\prime \Delta B
+ 2 \psi^\prime \Delta E^{\prime }
+ 3 \psi \psi^{\prime \prime} - \psi \Delta \psi
\right.
\nonumber \\
&& \left.
 + \H (
 2 \varphi \Delta B - 2\varphi \Delta E^\prime + 6 \varphi \psi^\prime)
\right.
\nonumber\\
&& \left. +  \H^2 (-3 \varphi^2 - 9\psi^2)\right].
\label{dSaction-nomass-end}
\end{eqnarray}
As expected, $B$ and $\varphi$ are non-dynamical fields with
non-degenerate quadratic term. Their equations of motion give
\begin{equation}
   \varphi = \frac{1}{\H} \psi^\prime \; , \;\;\;\;\;
    B = \frac{1}{\H} \psi + E^\prime .
\label{dS-PhiU-PsiE}
\end{equation}
The miracle is that after these expressions are substituted back into
the action (\ref{dSaction-nomass}), and integration by parts is
performed, one arrives at vanishing quadratic Lagrangian,
$L_{EH+\Lambda}^{(2)} (\psi, E) = 0$, where, again, the equations for
background were used.  Thus, $\psi$ and $E$ are arbitrary functions of
$x^\mu$, while $\varphi$ and $B$ are related to them via
(\ref{dS-PhiU-PsiE}).  These configurations are pure gauges of the
form (\ref{gaugetr-L}) and (\ref{gaugetr-0}). To check that, one again
uses equations for the background (in particular, $\H^\prime = \H^2$).

This miracle of course happens because of gauge invariance.  Once
gauge-invariance is broken explicitly by the graviton mass terms,
miracles do not happen, and the Boulware--Deser mode appears.

Let us  introduce the mass term generalized to curved space-time,
\[
   S_m = S_m (g_{\mu \nu} , \, \eta_{\mu \nu}).
\]
There is a lot of arbitrariness at this stage: general covariance is
explicitly broken, and $S_m$ may contain various combinations of
$g_{\mu \nu } \eta^{\mu \nu}$, $g^{\mu \nu } \eta_{\mu \nu}$,
$\sqrt{g}$, etc.  The discussion that follows is not sensitive to the
particular form of the mass term; it is assumed only that it does not
depend on the derivatives of the metric, becomes the Fierz--Pauli term
in Minkowski limit, and, for simplicity, that it is proportional to a
single mass parameter $m_G^2$.  To illustrate the general analysis, we
will use the simplest generalization of the Fierz--Pauli mass term,
\begin{equation}
S_{m} = \half M_{Pl}^2 \int~d^4 x~\left\{
- \frac{m_G^2}{2} \eta^{\mu \lambda} \eta^{\nu \rho}
(g_{\mu \nu} - \eta_{\mu \nu})  (g_{\lambda \rho} - \eta_{\lambda \rho})
+ \frac{m_G^2}{2} [\eta^{\mu \nu} (g_{\mu \nu} - \eta_{\mu \nu})]^2
\right\}.
\label{FPmasscurved}
\end{equation}
We stress that this form will be used for illustration purposes only.

There is a coordinate frame where $\eta_{\mu \nu} =\mbox{diag}
(1,-1,-1,-1)$, and we assume that the background space-time is
homogeneous and isotropic in this frame. Then the metric has the
general Friedman--Robertson--Walker form (plus perturbations) in this
frame,
\begin{equation}
ds^2 = a^2 (t) [n^2 (t) (1+h_{00})dt^2 + 2n(t) h_{0i} dt dx^i
+ (-\delta_{ij} + h_{ij}) dx^i dx^j].
\nonumber
\end{equation}
In this frame, the background is characterized by two metric
functions, $a(t)$ and $n(t)$.  It is still convenient to work with 
conformal time $\eta$, that is, to perform the change of variables
$d\eta = n(t) dt$.  In other words, we will work in the conformal
frame where the background
metric has the form (\ref{nov15-3}). Consistency of the
filed equations implies an equation relating $n (\eta)$ and $a(\eta)$,
which generically has the form $ n^\prime = f(n,a) a^\prime $, see
Appendix~A.  
Note, however, that its solution is not unique: at a given
moment of time one can take $n$ and $a$ arbitrary.

Once the mass term is added, the quadratic part of the action in
cosmological background has a very general structure,
\begin{eqnarray}
S^{(2)}_{EH + \Lambda + m_G} = 2 M_{Pl}^2 \int~d^3x d\eta~a^2&&\left[
-2 \varphi \Delta \psi - 2 \psi^\prime \Delta B
+ 2 \psi^\prime \Delta E^{\prime }
+ 3 \psi \psi^{\prime \prime} - \psi \Delta \psi
\right.
\nonumber \\
&& \left.
 + \H (
2 \varphi \Delta B - 2\varphi \Delta E^\prime + 6 \varphi \psi^\prime)
\right.
\nonumber\\
&& \left. +  \frac{m_\varphi^2}{2} \varphi^2 - \frac{m_B^2}{2} (\d_i B)^2
+ \frac{m_E^2}{2} (\Delta E)^2  + \frac{m_\psi^2}{2} \psi^2 \right. 
\nonumber \\
&& \left. + \mu_1 \varphi \psi  + \mu_2 \varphi \Delta E + \mu_3 \psi \Delta E
\right].
\label{dSaction-mass}
\end{eqnarray}
The terms in the last two lines have three-fold origin.  First, there
are terms that do not vanish in the limit $m_G^2 \to 0$; these are the
terms in the last line of eq.~(\ref{dSaction-nomass-end}).  Second,
the background equations no longer coincide with eq.~(\ref{nov15-2}),
so the last two lines in (\ref{dSaction-nomass}) do not
vanish. Finally, there are contributions due to the mass term $S_m$
itself. We give more detailed treatment of the latter two
contributions in Appendix A.
 
There are generically no specific relations between the terms in the
last two lines of (\ref{dSaction-mass}).  The fields $\varphi$ and $B$
are still non-dynamical, and after integrating them out one no longer
obtains zero action for $\psi$ and $E$.  Instead, the action contains
terms with two time derivatives of $\psi$ and $E$, some of which are
explicit in (\ref{dSaction-mass}) and some emerge after $\varphi$ and
$B$ are integrated out (note that the terms in (\ref{dSaction-mass})
proportional to $\varphi$ and $B$ do not contain second time
derivatives of $\psi$ and $E$, so higher time derivatives of the
latter fields do not appear).  Both $\psi$ and $E$ are dynamical
fields, hence the scalar sector has two propagating modes.  This is in
contrast to the theory in Minkowski background, with a single
propagating mode in the scalar sector. The extra mode is precisely the
Boulware--Deser degree of freedom.

Let us now specify to near-Minkowski background,
\[
\H^2 \ll m_G^2 \;, \;\;\;\; |a-1| \ll 1 \; , \;\;\;\;
|n-1| \ll 1.
\]
Note that in this limit $\H$ coincides with the standard Hubble
parameter. We will be interested in relatively high momenta, ${\bf
p}^2 \gg m_G^2$.  A detailed analysis reveals the following features.
First, the properties of the scalar perturbations are different in the
two ranges of momenta
\begin{eqnarray}
  (i)\; : && {\bf p}^2 \ll \frac{m_G^4}{\H^2} ,
\nonumber \\
    (ii)\; : && {\bf p}^2 \gg \frac{m_G^4}{\H^2} .
\label{nov15-7}
\end{eqnarray}
Hence, the high momentum limit and Minkowski limit do not commute.  We
discuss the range ({\it i}) in Appendix~\ref{appA} and here we briefly
summarize the results. There are indeed two propagating modes. One of
them is the Fierz--Pauli mode whose dispersion relation remains
$\omega^2 = {\bf p}^2$, up to small corrections.  The second mode is a
ghost or tachyon-ghost (tachyon and ghost at the same time). Being
ghost means that the energy is unbounded from below; if the mode is
simultaneously tachyon, it exponentially increases in time.

Let us now consider the range ({\it ii}), i.e., the high momentum
limit.  To integrate out non-dynamical fields, we solve equations
obtained by varying the action with respect to $\varphi$ and $B$.
These equations are written explicitly in Appendix~A, see
eqs.~(\ref{Phieq}) and (\ref{Ueq}).  We need the expression for
$\varphi$ to the leading order in derivatives and the expression for
$B$ to both leading and sub-leading order; the reason is that there
are cancellations.  The corresponding expressions are
\begin{eqnarray}
\varphi &=& \frac{1}{\H} \psi^\prime,
\nonumber \\
B &=& \frac{1}{\H} \psi + E^\prime
- \frac{m_\varphi^2}{2 \H^2} \frac{1}{\Delta} \psi^\prime
- \frac{3}{\Delta} \psi^\prime - \frac{\mu_2}{2\H} E.
\nonumber
\end{eqnarray}
Plugging these back into the action (\ref{dSaction-mass}) and
integrating by parts, we arrive at the action for dynamical fields,
\begin{eqnarray}
S^{(2)}_{EH + \Lambda + m_G} &=& 2 M_{Pl}^2 \int~d^3x d\eta~a^2\left\{
-\left[ 1- \frac{\H^\prime}{\H^2}
+ \frac{m_B^2}{2\H^2} \right] \d_i\psi \d_i \psi
+ \l 3 + \frac{m_\varphi^2}{2\H^2} \r (\psi^\prime)^2 \right.
\nonumber \\
&&~~~~~~~~~~~~~~~~~~~~~~ \left. +  \frac{\mu_2 - m_B^2}{\H} \psi^\prime \Delta E
- \frac{m_B^2}{2} \d_i E^\prime \d_i E^\prime 
+ \frac{m_E^2}{2} (\Delta E)^2 \right\}.
\label{finalaction-mass}
\end{eqnarray} 
Note that the terms with the highest derivatives, $\psi^\prime \Delta
E^\prime$, have canceled out.

We now see explicitly that there are two propagating modes. We also
see that their action (\ref{finalaction-mass}) is singular in
Minkowski limit\footnote{According to the above discussion, this is
the limit in which one first sends ${\bf p}^2$ to infinity, and only
then approaches Minkowski space-time.}. Indeed, by comparing
(\ref{dSaction-mass}) with the Lagrangian in Minkowski space-time
(eq.~(\ref{nov15-4}) with $\alpha = -\beta = -m_G^2$), one finds that
in Minkowski limit
\begin{equation}
m_B^2 \to - \frac{m_G^2}{2} \; , \;\;\;
\mu_2 \to - m_G^2,
\label{vr-feb5}
\end{equation}
while $m_E^2$ and $m_\varphi^2$ tend to zero. Thus, the first and the
third terms in (\ref{finalaction-mass}) have the coefficients that
diverge in the Minkowski limit, in which $\H \to 0$. Furthermore, at
$\H^2 \ll m_G^2$, the first term in (\ref{finalaction-mass}) has the
overall positive sign (because of the first relation in
(\ref{vr-feb5})), which corresponds to negative energy. This energy is
unbounded from below, so there is a ghost or tachyon in the
spectrum. We show in Appendix~A within the model (\ref{FPmasscurved}),
that one of the modes still has the dispersion relation
\[
  \omega^2 = {\bf p}^2,
\]
while the other mode is tachyonic or non-tachyonic, depending on the
relationship between $(a-1)$ and $(n-1)$.

\subsubsection{St\"uckelberg treatment}
\label{susbsub:stuck-BD}

A lesson from the above analysis is that once the gauge invariance is
given up, there are two scalar propagating modes in curved backgrounds
no matter how close these backgrounds are to Minkowski
space-time. Minkowski limit is singular, and for nearly Minkowski
space-time, one of these modes is necessarily pathological. Another
lesson is that the explicit analysis of this Boulware--Deser mode is
rather messy.  On the contrary, the Boulware--Deser phenomenon is
relatively straightforward to see within the St\"uckelberg
formalism~\cite{Arkani-Hamed:2002sp,Dubovsky:2005dw,Creminelli:2005qk,Deffayet:2005ys}.

Let us consider backgrounds which are only slightly different from
Minkowski space-time. In that case the perturbation theory in $h_{\mu
\nu}$ is adequate. The quadratic Lagrangian has been discussed in
previous Sections.  Generically, at the cubic order one has the
following contributions to the Lagrangian
\begin{equation}
m_G^2 \left[ \lambda_1 (h^\mu_\mu)^3
+ \lambda_2 h^\mu_\mu \, h_{\nu \lambda} h^{\nu \lambda}
+ \lambda_3 h_{\mu \nu}h^\nu_\lambda h^{\lambda \mu} \right]
\label{nov16-1+}
\end{equation}
with $\lambda_{1,2,3}$ of order one. For non-trivial background the
field $h_{\mu \nu} = g_{\mu \nu}- \eta_{\mu \nu}$ has non-zero
background part $h_{\mu \nu}^{(c)}$.  To perform the St\"uckelberg
analysis, one changes variables in the way dictated by
(\ref{nov16-19*}),
\begin{equation}
  g_{\mu \nu}(x) =
\bar{g}_{\mu \nu}(x+\xi) + \d_\mu \xi^\lambda \; \bar{g}_{\nu \lambda}(x+\xi)
 + \d_\nu \xi^\lambda \; \bar{g}_{\mu \lambda}(x+\xi) +
   \d_\mu \xi^\lambda \; \d_\nu \xi^\rho \; \bar{g}_{\lambda \rho}(x+\xi),
\label{nov16-1*}
\end{equation}
where
\begin{eqnarray}
\bar{g}_{\mu \nu} (x+\xi) &=&
\eta_{\mu \nu} + h_{\mu \nu}^{(c)} (x+\xi)
+ \bar{h}_{\mu \nu} (x+\xi) 
\nonumber \\
&=&\eta_{\mu \nu} + h_{\mu \nu}^{(c)} (x)
+ \bar{h}_{\mu \nu} (x) +   \d_\lambda h_{\mu \nu}^{(c)} (x) \xi^\lambda
+ \d_\lambda \bar{h}_{\mu \nu} (x) \xi^\lambda + \dots
\nonumber
\end{eqnarray}
Here $\bar{h}_{\mu \nu}$ and $\xi^{\mu}$ are perturbations, and
$\bar{h}_{\mu \nu} (x)$ is meant to be gauge fixed.  As before, the
Einstein--Hilbert action does not contain the field
$\xi^\mu$. Concentrating on the longitudinal St\"uckelberg field
$\xi_\mu = \d_\mu \phi$, and inserting the decomposition
(\ref{nov16-1*}) into both quadratic Fierz--Pauli term and cubic term
(\ref{nov16-1+}), one obtains the quadratic action for $\phi$,
\begin{equation}
m_G^2 \left[ \frac{3}{8} m_G^2 \d_\mu \phi \d^\mu \phi
+ \tilde{\lambda}_1 h_{\mu\nu}^{(c)} \cdot \d^\mu \d^\nu \phi \cdot
\Box \phi + \tilde{\lambda}_2 h_\mu^{(c) \, \mu} (\Box \phi)^2 \right],
\label{24-3*}
\end{equation}
where we keep the Fierz--Pauli contribution (\ref{13-2*}) that is
independent of $h_{\mu \nu}^{(c)}$, as well as the part which is
proportional to the background $h_{\mu \nu}^{(c)}$ and has the largest
number of derivatives.  Omitting other terms is legitimate for
studying slowly varying backgrounds and perturbations whose momenta
obey $\omega^2, {\bf p}^2 \gg m_G^2$.

One point to note is that any configuration obeying $\Box \phi = 0$
solves the field equation following from (\ref{24-3*}). This explains
why we have always found a mode with the dispersion relation $\omega^2
= {\bf p}^2$ when studying the theory in cosmological background. More
important is the fact that the Lagrangian (\ref{24-3*}) is of the
fourth order in derivatives of $\phi$, so there is a ghost in the
spectrum.  To see this explicitly and estimate the mass of the ghost,
consider a simplified version\footnote{The argument below is
straightforwardly generalized to the case (\ref{24-3*}).  One makes
use of the spatial Fourier representation, and writes the Lagrangian
in the form $A \ddot{\phi}^2 + B \dot{\phi}^2 + C \phi^2$, where
coefficients depend on ${\bf p}$.  One gets rid of the second time
derivatives by introducing a new field $\chi$ and finds that the
resulting structure of the kinetic terms is $B \dot{\phi}^2 + 2 A
\dot{\chi} \dot{\phi}$.  This structure implies that there is a ghost,
cf. (\ref{kin-general}).}  of (\ref{24-3*}),
\[
m_G^2 \left[ m_G^2  \d_\mu \phi \d^\mu \phi
 + \tilde{\lambda} \, h^{(c)} \cdot (\Box \phi)^2 \right].
\]
This Lagrangian is equivalent to
\begin{eqnarray}
&& m_G^2\left[ m_G^2  \d_\mu \phi \d^\mu \phi
 + 2\tilde{\lambda} \, h^{(c)} \cdot \d_\mu \chi \d^\mu \phi
- \tilde{\lambda}  \, h^{(c)} \cdot \chi^2 \right]
\nonumber \\
&& =
 m_G^2\left[ m_G^2  \l \d_\mu \phi + \frac{\lambda}{m_G}
\, h^{(c)} \cdot \d_\mu \chi \r^2 -
\l \frac{\tilde{\lambda}  h^{(c)}}{m_G} \r^2 (\d_\mu \chi)^2
- \tilde{\lambda}  h^{(c)} \cdot \chi^2 \right],
\nonumber
\end{eqnarray}
where $\chi$ is a new field.  The first term in the last expression
corresponds to the modified Fierz--Pauli mode $[\phi + (\lambda/m_G)
h^{(c)} \chi]$, while the second term is the kinetic term for the
Boulware--Deser mode $\chi$.  The latter has negative sign, and hence
the Boulware--Deser mode is a ghost (depending on the sign of
$\tilde{\lambda} h^{(c)}$ it may be a tachyon-ghost at sufficiently
low momenta). The local value of its mass squared is of order
\begin{equation}
m_{BD}^2 \simeq \frac{m_G^2}{ h^{(c)}}.
\label{nov16-29+}
\end{equation}
This explains why the high momentum limit and Minkowski limit do not
commute, as we have seen explicitly when studying the theory about
cosmological background.

The mass (\ref{nov16-29+}) diverges as the background approaches the
Minkowski limit.  The latter property might raise a hope that the
Boulware--Deser instability is not so dangerous. If the mass of the
ghost turns out to be larger than the UV scale $\Lambda_{UV}$, one
cannot trust the above analysis, since at energies exceeding
$\Lambda_{UV}$ one has to deal with the unknown UV-completion of the
theory. This observation, however, does not save the Fierz--Pauli
theory~\cite{Creminelli:2005qk}.  Indeed, away from an astrophysical
source of mass $M$ one has
\[
h^{(c)} \simeq \frac{M}{M_{Pl}^2 r},
\]
so that
\begin{equation}
   m_{BD}^2 \simeq \frac{r}{r_*^3},
\label{24-5*}
\end{equation}
where $r_*$ is the radius given by (\ref{oct24-add1}).  Hence, the
Boulware--Deser instability definitely occurs in the interval $(r_*
\Lambda_{UV})^2 r_* > r > r_*$, in which $m_{BD} <\Lambda_{UV}$ and at
the same time the linear approximation is valid. This interval is not
empty, unless $\Lambda_{UV}^{-1} \gtrsim r_*$.  Recall that the value
of $r_*$ for the Sun is of order of 10~pc. Hence, the Fierz--Pauli
theory with $\Lambda_{UV}^{-1} \gtrsim r_*$ cannot be trusted in the
solar system; had the graviton Lorentz-invariant mass, one would
either encounter rapid instabilities or have to deal with an unknown
UV-completion instead of the effective Fierz--Pauli theory.

To conclude, Lorentz-invariant massive gravities in four dimensions
are full of pathologies. One way towards getting around these
pathologies is to give up Lorentz invariance.

\section{Lorentz-violating theories: generalities}
\label{sec:lorentz-invar-viol}

\subsection{Lorentz-violating mass terms}
\label{sub:LVmass-gen}

In this and following Sections we are going to study the class of
theories with Lorentz-violating mass terms. We assume that Minkowski
space-time is a solution of the corresponding field equations, and
that the Euclidean symmetry of 3-dimensional space is not explicitly
broken in perturbation theory about this background. Then the
quadratic action for perturbations about Minkowski background is
\begin{equation}
   S^{(2)} = S_{EH}^{(2)} + S_{m},
\label{genac}
\end{equation}
where $S_{EH}^{(2)}$ is the quadratic part of the Einstein--Hilbert term,
explicitly given by (\ref{EHquadratic-LI}), and $S_{m}$ is the
graviton mass term.  The Lagrangian of the latter is
\begin{equation}
L_{m} = \frac{1}{4} [m_0^2 h_{00} h_{00} + 2m_1^2 h_{0i}h_{0i} - m_2^2
h_{ij} h_{ij} + m_3^2 h_{ii} h_{jj} - 2 m_4^2 h_{00} h_{ii}].
\label{Mh}
\end{equation}
Here, as before, $h_{\mu \nu}$ are perturbations about Minkowski
metric.  The Fierz--Pauli Lagrangian is obtained when all masses in
eq.~(\ref{Mh}), except for $m_0$, are taken to be equal,
\[
  \mbox{FP}: \;\;\; m_0^2 = 0 \;, \;\;
  m_1^2 =m_2^2 = m_3^2 =m_4^2 = m_G^2 .
\]
The latter property explains the conventions used in eq.~(\ref{Mh}).
In what follows we denote by $m$ the overall scale of the masses 
$m_0, \dots, m_4$.

Let us again make use of (3+1)-decomposition (\ref{nov14-1}).  This
formalism is particularly appropriate here, as it fully takes into
account 3-dimensional Euclidean invariance, the only symmetry which is
not explicitly broken by general mass term. The Lagrangian in the
tensor sector is the sum of the kinetic term (\ref{T-kinetic}) and the
mass term
\[
L_{m \, , \;T} = - \frac{m_2^2}{4} h_{ij}^{TT} \,  h_{ij}^{TT}.
\]
Hence there are two propagating  tensor modes with relativistic
dispersion relation
\[
\omega^2 = {\bf p}^2 + m_G^2,
\]
where
\begin{equation}
       m_G = m_2
\label{nov25-4}
\end{equation}
is the mass of tensor gravitons. The requirement that these modes are
not tachyonic gives
\[
   m_2^2 \geq 0.
\]
We will assume in what follows that this is the case.

In the vector sector, the quadratic Lagrangian is the sum of the
Einstein--Hilbert part (\ref{V-kinetic}) and the mass term
\[
L_{m \, , \;V} = \frac{m_1^2}{2} S_i S_i - \frac{m_2^2}{2} \d_i F_j
\, \d_i F_j.
\]
A novelty here, with respect to the Fierz--Pauli case, occurs at the
special value $ m_1 = 0$.  In this case, the field $S_i$ is the
Lagrange multiplier, leading to the constraint $F=0$. Hence, there is
no propagating modes in the vector sector, unlike in the Fierz--Pauli
theory,
\begin{equation}
  m_1 = 0 \; : ~~~~~\mbox{no propagating vector modes}.
\label{nov25-3}
\end{equation}
For $m_1 \neq 0$, the analysis of the vector modes parallels that
given in Section \ref{sub:LI-1}. For $m_1^2 > 0$ the vector sector
contains two normal propagating modes.  The canonically normalized
propagating field is now
\begin{equation}
{\cal F}_i ({\bf p})
= M_{Pl} m_1 \sqrt{\frac{{\bf p}^2}{{\bf p}^2 + m_1^2}}
F_i ({\bf p})
\nonumber
\end{equation}
with the dispersion relation
\[
    \omega^2 = \frac{m_2^2}{m_1^2} ({\bf p}^2 + m_1^2).
\]
In the case $m_1^2 < 0$ and $m_2 \neq 0$, the modes are ghosts or
tachyons at high spatial momenta, so we impose the restriction
\[
   m_1^2 \geq 0.
\]

Let us now turn to the scalar sector.  The full quadratic Lagrangian
is
\begin{eqnarray}
L_S^{(2)} = &&
 2 \left[\d_k \psi \d_k \psi - 3 \d_0 \psi \d_0 \psi
 + 2 (\d_k \varphi \d_k \psi + \d_k B \d_0 \d_k \psi + \d_0 \Delta E
\d_0 \d_k \psi) \right]
\nonumber \\
&&+ \left[ \frac{m_0^2}{2} \varphi^2 + \frac{m_1^2}{4} (\d_i B)^2
+ \frac{3 (3 m_3^2 - m_2^2)}{2} \psi^2 \right.
\nonumber \\
&&
\left.
-
(3 m_3^2 - m_2^2) \psi \Delta E + \half (m_3^2 - m_2^2) (\Delta E)^2
+ m_4 \varphi (3\psi - \Delta E) \right].
\label{nov25-5}
\end{eqnarray}
For general masses, there are two propagating modes, one of which is a
ghost. Indeed, for $m_1 \neq 0$, the field $B$ can be integrated out,
giving the following contribution to the Lagrangian, cf. (\ref{LU}),
\begin{equation}
L_B = \frac{8}{m_1^2} \dot{\psi} \Delta \dot{\psi} .
\label{nov24-1}
\end{equation}
The field $\varphi$ can also be integrated out, and the corresponding
contribution to the Lagrangian of the dynamical fields $\psi$ and $E$
does not contain time derivatives. Hence, the terms with time
derivatives in the resulting Lagrangian for $\psi$ and $E$ again have
the structure (\ref{kin-general}), implying that there is a
ghost. Generally, the ghost exists at all spatial momenta and
frequencies, so the observations we will make in
Section~\ref{sub:harmless} do not help. One has to get rid of the
ghost mode.

\subsection{Eliminating the second scalar mode}
\label{sec:elim-second-scal}

While in general the theory is not healthy, at special values of
masses the ghost mode does not exist. This is the case, in particular,
if either $\varphi$ or $B$ or both remain the Lagrange multiplier(s).
The point is that the corresponding constraint kills the second mode
in the scalar sector, while the remaining mode, if any, may well be
normal.  The two choices of the mass pattern that do the job are $m_0
=0$ and $m_1 = 0$. Let us discuss them in turn.

\subsubsection{$m_0 = 0$}
\label{subsubm0}

In the case $m_0 = 0$, the field $\varphi$ is the Lagrange multiplier,
leading to the constraint
\begin{equation}
2 \Delta \psi = m_4^2 (3\psi - \Delta E).
\label{nov25-2}
\end{equation}
Assuming that $m_1 \neq 0$ and $m_4 \neq 0$, one integrates out the
field $B$ with the result (\ref{nov24-1}) and expresses $\Delta E$ in
terms of $\psi$ using the constraint (\ref{nov25-2}). Then $\psi$ is
the only remaining dynamical field.  The terms in its Lagrangian which
are relevant at high momenta and frequencies, $\omega^2, {\bf p}^2 \gg
m^2$, are
\[
L_\psi = 4 \left[
2 \l \frac{1}{m_4^2} - \frac{1}{m_1^2}\r
\d_0 \d_i \psi \, \d_0 \d_i \psi -
\frac{m_2^2 - m_3^2}{m_4^4} (\Delta \psi)^2 \right] + \dots \; ,
\]
where the omitted terms have at most two derivatives. This Lagrangian
is healthy {\bf at} $\omega^2, {\bf p}^2 \gg m^2$
provided that
\[
   m_1^2 > m_4^2 >0 \; , \;\;\;\; m_2^2 > m_3^2.
\]
We will see below, however, that this case 
is problematic. 

Within the class of theories with $m_0 =0$ there are subclasses in
which more conditions on masses are imposed.  As an example, already
from the above analysis it follows that the case $m_4 =0$, the case
$m_4 = m_1$ and the case $m_2=m_3$ are all special. Detailed study of
these ``boundaries'' is given in Ref.~\cite{Dubovsky:2004sg}.

\subsubsection{$m_1 = 0$}
\label{subsubm1}

For $m_1 =0$ the field $B$ is the Lagrange multiplier.  The
corresponding constraint is $\dot{\psi} =0 $, implying $\psi = 0$ for
propagating modes. Inserting $\psi = 0$ back into the action, one
finds that there remain no terms with time derivatives, so there are
no propagating modes in the scalar sector.  The vector sector has the
same property, see (\ref{nov25-3}).  Thus, the only propagating modes
in the theory with $m_1=0$ are tensor gravitons with mass
(\ref{nov25-4}). We will discuss this theory in detail in 
Section~\ref{sec:minim-model-mass}.

\subsubsection{$m_2 = m_3$, $m_4=0$}

By inspection of the Lagrangian (\ref{nov25-5}) one uncovers one more
special case, $m_2 = m_3$ and at the same time $m_4=0$. It is now the
field $E$, rather than the non-dynamical fields $\varphi$ and $B$,
that plays special role.  The field $\Delta E$ enters the Lagrangian
linearly, and the corresponding field equation is
\[
 2 \ddot{\psi} + (3 m_3^2 - m_2^2) \psi = 0.
\]
Thus, there are no high frequency modes of $\psi$ irrespectively of
spatial momenta. If one is interested in high frequency modes only,
one sets $\psi =0$, and after that obtains the Lagrangian without time
derivatives. Hence, there are no propagating modes of high frequencies
in this case.

So, we see that there are special cases in which Lorentz-violating
massive gravity does not contain ghosts in linearized theory about
Minkowski background. Further analysis of numerous issues raised in
Section~\ref{sec:fierz-pauli-model}, as well as other points to worry
about, is conveniently performed in the St\"uckelberg formalism.
 
\subsection{Symmetries vs fine tuning}
\label{sub:symmetries-gen}

As we have seen in Section \ref{sec:boulware-deser-mode}, getting rid
of the second scalar mode in Minkowski background is by itself
insufficient to make the theory healthy.  In the Lorentz-invariant
theory, the absence of the second mode in Minkowski background is due
to the fine-tuning relation $\alpha = - \beta$ imposed on the mass
term in the Lagrangian (\ref{eq:LI-mass-term}). This fine-tuning is,
however, destroyed in curved backgrounds, and the second,
Boulware--Deser mode reappears. Likewise, similar fine-tuning
relations are problematic in Lorentz-violating theories.  Let us see
this explicitly in the theory with $m_0 = 0$ and no other relations
between the masses. It is convenient to make use of the St\"uckelberg
formalism, and proceed in analogy to Section
\ref{susbsub:stuck-BD}. At the quadratic order, the St\"uckelberg part
of the metric (\ref{nov16-1*}) that contains the derivatives of $\xi$
is
\begin{equation}
g_{\mu \nu} \equiv \eta_{\mu \nu} + h_{\mu \nu}
=\eta_{\mu \nu} + \d_\mu \xi_\nu + \d_\nu \xi_\mu
+ \d_\mu \xi^\lambda \d_\nu \xi_\lambda \; .
\label{dec7-1*}
\end{equation}
Let us concentrate on the terms involving the field $\xi_0$. In
Minkowski background, these come from the second and fifth terms in
the Lagrangian (\ref{Mh}) and read
\[
L_m = \half m_1^2 (\d_0 \xi_i + \d_i \xi_0)  (\d_0 \xi_i + \d_i \xi_0)
- m_4^2 \d_0 \xi_0 \d_i \xi_i + \dots\; ,
\]
where omitted terms contain $\xi_i$ only.  Upon integrating the second
term by parts, one observes that $\xi_0$ is not a dynamical field, so
that there is at most one propagating degree of freedom in the scalar
sector, the longitudinal part of $\xi_i$. This is in accord with the
discussion in Section \ref{subsubm0}.

Once the background is slightly different from Minkowski,
$g^{(c)}_{\mu \nu} = \eta_{\mu \nu} + h_{\mu \nu}^{(c)}$, the latter
property is lost. Indeed, due to the quadratic term in
(\ref{dec7-1*}), the mass terms themselves include the combination
\[
- \half m_4^2 h_{00} h_{ii}^{(c)} = 
- \half m_4^2 (\d_0 \xi_0)^2  h_{ii}^{(c)} + \dots
\]
The field $\xi_0$ becomes dynamical, the second mode reappears, and
in some backgrounds
(appropriate sign of $h_{ii}^{(c)}$) this mode is a ghost.

There is an elegant way out of this fine-tuning problem,
however~\cite{Dubovsky:2004sg}.  Relations between the masses, instead
of being results of fine tuning, may be consequences of unbroken gauge
symmetries, which are parts of the gauge symmetry of General
Relativity.  These residual gauge symmetries may then be expected to
protect the theory from becoming pathological when one extends it to
curved backgrounds and/or generalizes it to include possible UV
effects (the latter will be discussed later on). In several cases this
approach does lead to healthy infrared modified gravities.

One can think of various residual gauge
symmetries~\cite{Dubovsky:2004sg}.  In this review we discuss only a
few of them, which either are known to give rise to interesting
theories or serve as examples of the failure of this approach. The
first unbroken symmetry we are going to elaborate on is
\begin{equation}
(i)~~: ~~~~~~~ x^i \to x^i + \zeta^i (x^i,t) \; .
\label{eq:ghost-cond-symmetry}
\end{equation} 
This symmetry implies that all masses but $m_0$ vanish; this is
the symmetry of the ghost condensate theory~\cite{Arkani-Hamed:2003uy}.

The second symmetry to be discussed is
\begin{equation}
(ii)~: ~~~~~~~ t \to t + \zeta^0 (x^i, t) \; .
\label{dec8-1}
\end{equation}
This symmetry leads to the constraint $m_0=m_1=m_4 =0$.  We will see
that the corresponding theory has problems with the stability against
UV effects, see Section \ref{sub:UVunstable}.

The third symmetry is
\begin{equation}
(iii): ~~~~~~~ x^i \to x^i + \zeta^i (t) \; .  
\label{dec7-iii}
\end{equation}
This symmetry is sufficient to ensure that $m_1 = 0$, while other
masses are unconstrained.  We have found in Section \ref{subsubm1}
that in this case the linearized theory in Minkowski background is
free of pathologies. We will see that the corresponding
theory~\cite{Dubovsky:2004sg,Dubovsky:2004ud,Dubovsky:2005dw} is
healthy both in nearly Minkowski and in general cosmological
backgrounds. It is UV-stable as well.  In fact, as we discuss in
Section~\ref{sec:minim-model-mass}, this theory is quite interesting
from the phenomenological viewpoint.

\subsection{Lorentz-violating scalars}
\label{sub:LVscalars}

A convenient way to analyze the behavior of an infrared modified
gravity of the type we discuss in this review, and also to promote the
perturbation theory about Minkowski background to a full low energy
effective theory, is to start off with a generally covariant theory
with additional scalar fields $\phi^\alpha$, $\alpha =0,1,2,3$, which
we will call Goldstone fields. Breaking of Lorentz-invariance occurs
when these fields obtain background values which depend on space-time
coordinates. In this approach, Lorentz-invariance is broken {\it
spontaneously}, as the original action of the theory is
Lorentz-invariant, while the background is not.

As an example, in Minkowski space-time, the background
fields are
\begin{eqnarray}
\bar{\phi}^0 &=& a \Lambda^2 t \; , 
\nonumber\\
\bar{\phi}^i &=& b \Lambda^2 x^i,
\label{dec7a-1*}
\end{eqnarray}
where $\Lambda $ is a parameter with dimension of mass, and $a$ and
$b$ are coefficients of order one.  In our convention, the fields
$\phi^\alpha$ have dimension of mass.  The background fields
(\ref{dec7a-1*}) are solutions to the equations of motion if the
Lagrangian contains their derivatives only.  The latter property
automatically implies that the Lagrangian is invariant under shift
symmetry $\phi^\alpha (x) \to \phi^\alpha (x) + \lambda^\alpha$ with
constant $\lambda^\alpha$.  This means that the translational symmetry
of $(3+1)$-dimensional space-time is unbroken by the background
(\ref{dec7a-1*}), since one can undo a translation by shifting the
fields $\phi^\alpha$.  Likewise, to preserve spatial rotation symmetry
one requires that the Lagrangian is invariant under $SO(3)$ rotations
of the fields, $\phi^i \to \Lambda^i_j \phi^j$. Thus, one is lead to
consider theories whose actions, at the one-derivative level, have the
general form
\begin{equation}
S = S_{EH} + S_\phi,
\label{dec7-aa}
\end{equation}
where $S_{EH}$ is the Einstein--Hilbert action, while
\begin{equation}
S_\phi = \int~d^4x~\sqrt{-g} \Lambda^4
F(X, V^i, Y^{ij}, Q)
\label{dec7-a2}
\end{equation}
and 
\begin{eqnarray}
   X&=& \frac{1}{\Lambda^4} g^{\mu \nu} \d_\mu \phi^0 \d_\nu \phi^0,
\nonumber \\
  V^i &=& \frac{1}{\Lambda^4} g^{\mu \nu} \d_\mu \phi^0 \d_\nu \phi^i,
\nonumber \\
  Y^{ij}&=& \frac{1}{\Lambda^4} g^{\mu \nu} \d_\mu \phi^i \d_\nu\phi^j,
\label{dec8-2}
 \\
Q &=& \frac{1}{\Lambda^8} \frac{1}{\sqrt{-g}}
\epsilon^{\mu \nu \lambda \rho} \epsilon_{ijk}
\d_\mu \phi^0  \d_\nu \phi^i \d_\lambda \phi^j \d_\rho \phi^k.
\nonumber
\end{eqnarray}
Internal indices $i,j,k$ are to be contracted in the action
(\ref{dec7-a2}) with either $\delta_{ij}$ or $\epsilon_{ijk}$.
Hereafter we will {\it not} use the convention (\ref{dec9-6}) when
writing the Lagrangian for the Goldstone fields; this will simplify
power counting. The combination $Q$ is in fact not independent (apart
from possible subtleties related to the presence of the
$\epsilon$-symbol): its square can be expressed in terms of $X$, $V^i$
and $Y^{ij}$. So, in what follows we only consider functions $F$
depending on the former three combinations.

The energy-momentum tensor of the configuration (\ref{dec7a-1*})
vanishes in Minkowski space-time, and hence Minkowski space-time is a
legitimate background, provided that $a$ and $b$ are such that
\begin{eqnarray}
    - \half F + a^2 \frac{\d F}{dX}
&=& 0 \; ,
\nonumber \\
   \half F \delta_{ij} + b^2 \frac{\d F}{\d Y^{ij}},
&=& 0
\nonumber \\
\frac{\d F}{\d V^i} &=& 0
\label{dec8-5}
\end{eqnarray}
at $ X= a^2$, $Y^{ij} = -b^2 \delta^{ij}$ and $V^i = 0$. In what
follows we often set $a=b=1$ by field redefinition.

The theory with the action (\ref{dec7-a2}) is to be considered as an
effective field theory valid at low energies only\footnote{What is
exactly meant by low energies becomes clear after effects of
higher-order terms are understood: these are energies and/or momenta
at which higher-order terms blow up. Since the actions we discuss are
Lorentz-invariant, the way the UV cutoff $\Lambda$ enters the action
is dictated by Lorentz-invariance.  On the other hand, the value of
energy at which the low energy theory ceases to work may be different
from the value of spatial momentum, due to the spontaneous
Lorentz-violation by background scalar fields.}. The UV cutoff in
this theory $\Lambda_{UV}$ is to be somewhat below $\Lambda$,
cf. Section \ref{sec:strong-coupling}.  Indeed, expanding the fields
about the background (\ref{dec7a-1*}),
\[
\phi^\alpha = \bar{\phi}^\alpha + \pi^\alpha,
\]
one obtains the following structure of the Lagrangian for the
perturbations,
\begin{equation}
L_\pi = (\d \pi)^2 + \frac{1}{\Lambda} (\d \pi)^3 + \dots
\label{dec7-3a*}
\end{equation}
implying that $\Lambda_{UV} \lesssim \Lambda$. In this regard, an
important issue is the UV stability of the
theory~\cite{Dubovsky:2004sg}.  In low energy effective theories,
there is no reason to think that the low energy Lagrangian contains
terms with first derivatives only.  So, one has to worry about the
effects of higher-derivative terms like $\Lambda^{-2} g^{\mu \nu}
g^{\lambda \rho} \d_\mu \d_\nu \phi^\alpha \d_\lambda \d_\rho
\phi^\alpha$.  Naively, these terms are suppressed below the cutoff
scale, i.e., at ${\bf p}^2, \omega^2 \ll \Lambda^2$.  However, if the
kinetic terms in (\ref{dec7-3a*}) do not have generic structure, the
higher-derivative terms may become important.  We will encounter
examples of this sort in what follows.

Turning on gravity, still in Minkowski space-time and in background
(\ref{dec7a-1*}), one observes that the gauge transformation $x^\mu
\to x^\mu + \zeta^\mu (x)$ corresponds to the following transformation
of the fields $\pi^\alpha$,
\[
\pi^\alpha (x) \to \pi^\alpha (x) + \Lambda^2 \zeta^\alpha (x).
\]
Hence, 
\begin{equation}
\xi^\alpha = \Lambda^{-2} \pi^\alpha
\label{dec9-9}
\end{equation}
are the St\"uckelberg fields of the previous Sections. In the unitary
gauge, $\pi^\alpha = 0$, one has $X= 1- h_{00}$, $V^{ij} = -1 -
h_{ij}$, etc., so the quadratic in $h_{\mu \nu}$ part of the action
contains the mass term (\ref{Mh}), the scale of graviton masses being
\begin{equation}
  m = \frac{\Lambda^2}{M_{Pl}} \; ,
\label{dec9-8}
\end{equation}
which is in accord with (\ref{oct12-1-3++}). Hence, the class of
theories (\ref{dec7-aa}) indeed has all expected properties of
Lorentz-violating massive gravity. A convenient feature of this
construction is that the behavior of the theory at ${\bf p}^2,
\omega^2 \gg m_G^2$ in or near Minkowski background can be analyzed by
studying the Goldstone sector only.  Also, the theory away from
Minkowski background is well defined.

Needless to say, for general Lagrange function $F$ the theory is
pathological. As we discussed in Section \ref{sub:symmetries-gen}, it
may not be pathological if a part of the gauge symmetry of General
Relativity remains unbroken. In that case the Goldstone action
(\ref{dec7-a2}) does not have the generic form.  As an example, the
residual gauge invariance $t \to t + \zeta^0 (x^i, t)$, see
(\ref{dec8-1}), in the Goldstone language implies that the Lagrange
function $F$ is invariant under the change of variables
\begin{equation}
\phi^0 \to \phi^0 + \Xi^0 (\phi^i, \phi^0)
\label{dec9-1}
\end{equation}
with arbitrary function $\Xi^0 (\phi^i, \phi^0)$. Indeed, only in this
case the background (\ref{dec7a-1*}) is invariant under the gauge
transformation $t \to t + \zeta^0 (x^i, t)$ supplemented by a field
redefinition. It is in this way that the Lagrangian, and hence
graviton mass terms, get constrained by the requirement of residual
gauge invariance in the Goldstone framework.

\subsection{An example of UV unstable theory}
\label{sub:UVunstable}

To illustrate the problem with UV stability that one may encounter in
otherwise healthy theory, let us consider the model with the residual
gauge symmetry (\ref{dec8-1}), implying the constraint on the
Lorentz-violating graviton masses $m_0=m_1=m_4 =0$.  In the Goldstone
language, this symmetry translates into the field transformation
(\ref{dec9-1}).  This can be a symmetry of the Goldstone action
(\ref{dec7-a2}) only if the field $\phi^0$ is absent altogether.
Hence, the Goldstone sector of the theory has three fields $\phi^i$
and at the one-derivative level the action is
\[
S_\phi = \int~d^4x~\sqrt{-g} \Lambda^4
F(Y^{ij}).
\]
where $Y^{ij}(\phi^i)$ is given by (\ref{dec8-2}).  As pointed out in
Section \ref{sub:symmetries-gen}, the general Lagrangian for the
theory of Goldstone fields, viewed as low energy effective theory,
contains higher order terms, e.g.
\begin{equation}
\Delta F = \frac{1}{\Lambda^4} g^{\mu \nu} g^{\lambda \rho}
\d_\mu \d_\nu \phi^i \d_\lambda \d_\rho \phi^i.
\label{dec9-2}
\end{equation}
We will see that in the model discussed here, these terms are
important and, in fact, give rise to pathologies in the spectrum.

Let us consider this theory in Minkowski background, discarding the
higher-order terms for the time being.  Lorentz-invariance is broken
by the background
\begin{equation}
   \bar{\phi}^i = \Lambda^2 x^i 
\label{dec8-4}
\end{equation}
which obeys the field equation for the Goldstone fields.  Expanding
the fields near this background, $\phi^i = \bar{\phi}^i + \pi^i$, one
obtains the quadratic Lagrangian for the St\"uckelberg fields
$\pi^i$. This Lagrangian involves the first and second derivatives of
the Lagrange function $F$ evaluated at $Y^{ij} = Y^{ij} (\bar{\phi}^i)
= - \delta^{ij}$, which we parameterize as
\begin{eqnarray}
\frac{\d F}{\d Y^{ij}} (\bar{\phi})&=& F_1 \delta_{ij},
\nonumber \\
\frac{\d^2 F}{\d Y^{ij} \d Y^{kl}} (\bar{\phi})
&=& F_{21} \delta_{ij} \delta_{kl}
+ F_{22} \left( \delta_{ik} \delta_{jl} + \delta_{il} \delta_{jk} \right).
\nonumber
\end{eqnarray}
The quadratic Lagrangian is
\begin{equation}
L_\pi = F_1 \d_\mu \pi^i \d^\mu \pi^i
+ 2 F_{21} \d_i \pi^i \d_j \pi^j
+ 2 F_{22} (\d_i \pi^j \d_i \pi^j + \d_i \pi^j
\d_j \pi^i) \; .
\label{dec8-7}
\end{equation}
At first sight, this Lagrangian describes three scalar fields with
healthy kinetic term. This, however, is inconsistent with the fact
that for $m_1 = 0$ there are no propagating modes in the vector
sector, see (\ref{sub:LVmass-gen}). The resolution of this discrepancy
has to do with the requirement that the energy-momentum tensor of the
background field configuration (\ref{dec8-4}) vanishes, so that 
Minkowski metric is a solution of the complete set of field equations.
The corresponding conditions are read off from (\ref{dec8-5}), which
in the absence of the combinations $X$ and $V^i$ yield
\[
   F = 0 \; , \;\;\;\;\; \frac{\d F}{\d Y^{ij}} = 0
\;\;\;\;\; \mbox{at} \;\;\; \phi^i = \bar{\phi^i} \;.
\]
Hence, $F_1 = 0$ in (\ref{dec8-7}), so the one-derivative action
actually corresponds to a theory with no propagating modes: at this
level, all St\"uckelberg fields enter the action without time
derivatives, so none of them is a dynamical field.

Once the higher order terms are added, the situation changes.  The
terms like (\ref{dec9-2}) contain time derivatives, and there is no
symmetry that would forbid them.  In terms of the fields $\pi^i$,
these contributions have the following structure,
\begin{equation}
   \Delta L_\pi = 
\frac{1}{\Lambda^2} \left[(\d_0^2 \pi^i)^2 - (\d_0 \d_i \pi^j)^2 + 
\dots \right].
\label{dec9-4}
\end{equation}
These contributions dominate at high {\it frequencies}, precisely
because the Lagrangian (\ref{dec8-7}) does not contain time
derivatives, i.e., precisely because the fields $\pi^i$ are not
dynamical at the one-derivative level. With the higher order terms
included, the fields $\pi^i$ become propagating, and their dispersion
relation is
\[
   \omega^4 = \mbox{\const} \cdot {\bf p}^2 \Lambda^2 \; .
\]
This means that at least one of the modes for each $\pi^i$ is
tachyonic, and the corresponding ``frequency'' is high even at
moderate spatial momenta (being, nevertheless, smaller than the
cut off scale $\Lambda_{UV}$). Because of that, the model is
unacceptable.

Hence, the fields that are non-dynamical in Minkowski background and
at the level of one-derivative Lagrangian, need to be treated as
suspects.  They may become propagating in curved backgrounds and/or
due to higher order terms in the Lagrangian. We will refer to the
former possibility as the Boulware--Deser instability, while the
second is called UV sensitivity~\cite{Dubovsky:2004sg}.

To conclude the discussion of the model studied here, let us point out
that in the language of metric perturbations, its UV sensitivity is
sensitivity against derivative terms in the Lagrangian for $h_{\mu
\nu}$. As an example, the first, most unwelcome contribution in
(\ref{dec9-4}), in terms of metric perturbations corresponds to the
term
\begin{eqnarray}
\Delta S_h &=& \int~d^4x~\Lambda^2 
\l \half \d_i h_{00} - \d_0 h_{0i} \r^2
\nonumber \\ 
&=& M_{Pl}^2  \int~d^4x~m_G^2 \cdot \frac{1}{\Lambda^2}
\left( \half
 \d_i h_{00} - \d_0 h_{0i} \right)^2,
\nonumber
\end{eqnarray}
where we recall the relation (\ref{dec9-9}) that implies the
correspondence $h \simeq \Lambda^{-2} \d \pi$, and use (\ref{dec9-8}).
This term is invariant under residual gauge transformations
(\ref{dec8-1}), is suppressed by the anticipated UV scale
(\ref{oct12-1-3++}) as compared to the graviton mass terms, so there
is no reason for it not to be present. In this language, what we have
done was to find out that the theory with two non-vanishing graviton
masses $m_2$ and $m_3$ has tachyons in the spectrum, once generic
one-derivative terms in $h_{\mu \nu}$ consistent with the symmetry
(\ref{dec8-1}) are added.

\subsection{Not-so-dangerous instabilities}
\label{sub:harmless}

To end up this Section, let us add digress to more phenomenological
discussion of instabilities in Lorentz-violating theories.  In these
theories one can allow for tachyons and/or ghosts provided they exist
at low frequencies (particle energies) only. In viable theories, the
frequency cutoff $\Lambda_{tc}$ for tachyons can be somewhat higher
than the present Hubble scale $H_0$, while for ghosts the cutoff
$\Lambda_{gh}$ can be many orders of magnitude higher than $H_0$. Let
us discuss this issue in some detail, assuming that ghosts and
tachyons interact with ordinary matter only gravitationally.

Consider tachyons first, and suppose, as an example, that the
dispersion relation is
\begin{equation}
    \omega^2 = - {\bf p}^2
\label{nov23-1}
\end{equation}
for $|{\bf p}| \ll \Lambda_{tc}$, while for $|{\bf p}| > \Lambda_{tc}$
the frequency is normal, $\omega^2 >0$ (an example of such a
dispersion law is $\omega^2 = - {\bf p}^2 + \Lambda_{tc}^{-2} {\bf
p}^4$). Then in expanding Universe, $\omega$ scales at $|{\bf p}| \ll
\Lambda_{tc}$ as
\[
     |\omega (t)| = \frac{\Omega}{a(t)},
\]
where $\Omega$ is constant conformal frequency.  There is a
characteristic moment of time $t_\Lambda$ in the history of the
Universe, at which
\[
    H(t_\Lambda) = \Lambda_{tc}.
\]
Before that time, would-be tachyonic modes with $\omega (t) \lesssim
\Lambda_{tc}$ are over-damped and do not develop, so exponential growth
of any mode is possible only after $t_\Lambda$.  The largest growth
factor corresponds to modes that become tachyonic just at the time
$t_\Lambda$, i.e., modes with
\[
  \omega (t_\Lambda) \equiv \frac{\Omega}{a(t_\Lambda)} \simeq \Lambda_{tc}.
\]
Indeed,
modes of higher conformal frequency still oscillate at $t =
t_\Lambda$, while modes of lower conformal frequency still do not develop
at $t \sim t_\Lambda$. By now, the largest growth factor for
the field amplitude is
\[
\mbox{exp} \left( \int_{t_\Lambda}^{t_{0}} ~ 
\frac{a(t_\Lambda)}{a(t)} \Lambda_{tc}~dt \right) ,
\]
where $t_0$ denotes the present time.  The inhomogeneities in the
tachyon field produce gravitational potentials comparable to those of
ordinary matter with energy density perturbations
\[
\delta \rho \simeq \Lambda_{tc}^4 \cdot
\mbox{exp} \left( 2 \int_{t_\Lambda}^{t_{0}} ~ 
\frac{a(t_\Lambda)}{a(t)} \Lambda_{tc}~dt \right),
\]
where we estimated the pre-exponential factor on dimensional grounds
and neglected redshift of energy when writing this factor.  The bound
on $\Lambda_{tc}$ comes from the requirement that this inhomogeneous
energy density does not exceed observationally allowed value, say,
$10^{-4} \rho_c$ (the exact number is unimportant here). Approximating
the cosmological expansion by $a \propto t^{2/3}$ (matter domination),
we find
\begin{eqnarray}
\delta \rho &\approx& \Lambda_{tc}^4 \cdot \mbox{exp} \l
3 t_0^{1/3} t_\Lambda^{2/3} \Lambda_{tc}\r
\nonumber \\
&=& 
\Lambda_{tc}^4 \cdot \mbox{exp} \left[4 \l \frac{\Lambda_{tc}}{H_0} \r^{1/3}
\right].
\nonumber
\end{eqnarray}
Demanding that $\delta \rho \lesssim 10^{-4} \rho_c \sim 10^{-4}
M_{Pl}^2 H_0^2$ we find
\[
\frac{\Lambda_{tc}}{H_0} \lesssim
\frac{1}{64} \left[ \ln \l 10^{-4} \frac{M_{Pl}^2}{H_0^2} \r
\right]^3 \sim 3\cdot 10^5.
\]
We conclude that the frequency cutoff for tachyons with dispersion
relation (\ref{nov23-1}) must be of order $\Lambda_{tc} \sim 10^5 H_0$
or lower.

The bound on $\Lambda_{tc}$ rather strongly depends on the form of the
dispersion relation for tachyons.  In any case, it is somewhat higher,
but not very much higher, than $H_0$.

\begin{figure}
\begin{picture}(400,150)(0,0)
\put(120,10){
\epsfig{file=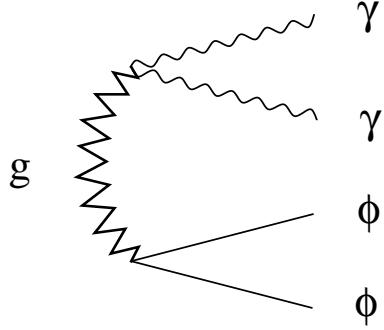,%
width=170pt,height=140pt,%
}}
\end{picture}
\caption{The decay of the vacuum into two ghosts and two photons via
  creation of a virtual graviton pair. }
\label{fig:vacuum-instability}
\end{figure}
Let us now turn to ghosts. The instability in this case is due to pair
creation of ghosts plus usual particles from vacuum, the
process allowed by energy-momentum conservation due to negative energy
of ghost particles.  The strongest bound~\cite{Cline:2003gs} on the
frequency cutoff $\Lambda_{gh}$ comes from the process
\begin{equation}
\mbox{vacuum} \to \phi+ \phi + \gamma + \gamma,
\label{nov23-2}
\end{equation}
where $\phi$ and $\gamma$ denote ghost and photon, respectively.  We
are assuming that ghosts experience gravitational interactions only.
Then this process is described by the diagram of
Fig.~\ref{fig:vacuum-instability} and its rate per unit volume is
estimated on dimensional grounds as\footnote{We assume here that
3-momentum cutoff is also of order $\Lambda_{gh}$.}
\[
\Gamma \simeq \frac{\Lambda_{gh}^8}{M_{Pl}^4}.
\]
Note that all particles in (\ref{nov23-2}) are on-shell, so
$\Lambda_{gh}$ is the cutoff on energy, not the usual UV cutoff on
momentum transfer. In Lorentz-invariant theories one has $\Lambda_{gh}
= \infty$, and the rate is infinite.  This corresponds to infinite
volume of the Lorentz group.  In other words, in Lorentz-invariant
theories, the process (\ref{nov23-2}) with certain momenta of outgoing
particles has its boosted counterparts, so the phase space is
infinite. This is not the case in Lorentz-violating theories. In the
latter, photons created in the process (\ref{nov23-2}) have energies
$E_\gamma \lesssim \Lambda_{gh}$, and their number density in the
present Universe, and hence the flux near the Earth, is of order
\[
F \simeq \Gamma t_0.
\]
The flux per energy interval is
\[
   \frac{dF}{dE_\gamma} (E_\gamma \sim \Lambda_{gh})
\simeq \frac{\Lambda_{gh}^7}{M_{Pl}^4} t_0.
\]
This flux has to be smaller than the EGRET differential flux,
\[
 \frac{dF}{dE_\gamma} = 7 \cdot 10^{-9} \l \frac{E_\gamma}{450~\mbox{MeV}}
\r^{-2.1} \mbox{cm}^{-2} \mbox{s}^{-1} \mbox{sr}^{-1} \mbox{MeV}^{-1}.
\]
This requirement gives~\cite{Cline:2003gs}
\[
\Lambda_{gh} \lesssim 3~\mbox{MeV}.
\]
Hence, the frequency cutoff in Lorentz-violating theories with ghosts
may be relatively high.

\section{Ghost condensate: modification of gravity without graviton
  mass }
\label{sec:ghost-cond-modif}

Consider now an example of the UV-stable theory, namely, the model of
the ``ghost condensate'' \cite{Arkani-Hamed:2003uy}. The issues
discussed above --- the absence of the extra scalar mode near
Minkowski space-time, protection by a residual symmetry against its
reappearance in curved backgrounds and due to higher-derivative
corrections --- play a key role in the construction of this
model. Although the graviton remains massless in the ghost condensate
model, the simplicity of this model makes it a good introduction into
more complicated models of massive gravity.

As we discussed in Section~\ref{sec:elim-second-scal}, a convenient
way to modify the gravitational interaction in the infrared is to
introduce additional scalar fields. In the simplest case this is just
a single scalar field $\phi$ (cf. Section~\ref{sub:LVscalars}) with
the action reminiscent of eq.~(\ref{dec7-a2}),
\begin{equation}
S_\phi = \Lambda^4 \int d^4x \sqrt{-g} F(X)
\label{eq:Sphi-GhCond}
\end{equation}
with $X$ given by the first of eqs.~(\ref{dec8-2}), $X =
g^{\mu\nu}\partial_\mu\phi \partial_\nu\phi/\Lambda^4$. The field
equation derived from this action is
\begin{equation}
\frac{1}{\sqrt{-g}} \d_\mu \left[
F_X (X) \sqrt{-g} g^{\mu \nu} \d_\nu \phi \right] = 0 \;,
\label{dec28-1}
\end{equation}
where $F_X = \d F/\d X$. In Minkowski space-time, and with gravity
switched off, this equation has solution linearly growing in time
(``ghost condensate'')
\begin{equation}
\phi_v =\alpha \Lambda^2 t \; ,
\label{eq:phi_vac}
\end{equation} 
where $\alpha$ is an arbitrary constant.  This background obviously
breaks the Lorentz symmetry. The time translations are also broken,
but the diagonal combination of the time translations and shifts of
$\phi$ by a constant remain a symmetry, so there is a conserved
energy. In the ``unitary'' gauge $\phi=\phi_v$ the variable $X$
reduces to $X=\alpha^2 g^{00}$ and the action (\ref{eq:Sphi-GhCond})
becomes the function of $g^{00}$. This action is invariant under the
space-time dependent transformations of spatial coordinates
\begin{equation}
x^i\to \tilde x^{i}=\tilde x^{i}(x^i,t),
\label{eq:GC-symmetry}
\end{equation} as discussed in
the end of Section~\ref{sub:symmetries-gen}, see
eq.~(\ref{eq:ghost-cond-symmetry}). This symmetry plays an important
role in the construction of the ghost condensate model.

Once the back reaction of ghost condensate on gravitational background
is switched on, the parameter $\alpha$ is no longer arbitrary. In
order that Minkowski space be a solution to the Einstein equations for
$\phi=\phi_v$, the energy-momentum tensor of $\phi_v$ has to vanish,
\[
T_{\mu\nu} = \left[ 2\partial_\mu\phi  \partial_\nu\phi 
F_X - \eta_{\mu\nu} F\right]_{\phi=\phi_v}
=0.
\]
This equation leads to the following two conditions
(cf. (\ref{dec8-5})),
\begin{eqnarray}
\nonumber
2\alpha^2 F_X(\alpha^2)-F(\alpha^2)&=&0,\\
F(\alpha^2)&=&0.
\label{eq:vacuum-eqs-GhC}
\end{eqnarray}
The second of these two conditions is the usual tuning of the
cosmological constant to zero. When this condition is satisfied, the
first of eqs.~(\ref{eq:vacuum-eqs-GhC}) implies that $F_X
(\alpha^2)=0$. We will assume in what follows that that extrema of
$F(X)$ occur at $X \neq 0$. Thus, $\alpha$ does not vanish, and one
can redefine the field in such a way that the conditions
(\ref{eq:vacuum-eqs-GhC}) are satisfied for $\alpha=1$.

In expanding Universe,  ghost condensate is automatically
driven to the point $F_X = 0$. This follows from the
field equation (\ref{dec28-1}). Indeed, this equation can be viewed as
the covariant conservation equation for the current
\[
  J^\mu = F_X (X) g^{\mu \nu} \d_\nu \phi \; .
\]
In the cosmological setting, the field $\phi$ is consistently taken to
depend on time only, so that the only non-vanishing component of this
current is the density $J^0$.  Its covariant conservation implies
that, like other densities, it decays in time,
\[
   J^0 \propto \frac{1}{a^3} \; ,
\]
which means that $F_X (X)$ becomes negligibly small at late
times.

In the unitary gauge and at the quadratic level in metric
perturbations $h_{\mu\nu}$, the action (\ref{eq:Sphi-GhCond}) in
Minkowski background becomes
\begin{equation}
S^{(2)}_\phi = \Lambda^4 \frac{F_{XX}}{2} \int d^4x h_{00}^2
= \half M_{Pl}^2 m_0^2   \int d^4x h_{00}^2       \;,
\label{eq:Sphi-quadratic}
\end{equation}
where $F_{XX}=[d^2F/dX^2](\alpha=1)$ is a constant. With the
Einstein-Hilbert part of the action added, this gives eq.~(\ref{Mh})
with all masses equal to zero except for $m_0$. Thus, at the level of
the two-derivative action there are no propagating degrees of freedom,
cf. Section~\ref{sec:elim-second-scal}.

The same can be seen in the St\"uckelberg language by replacing
$h_{00} \to 2 (M_{Pl} m_0)^{-1}\partial_0\pi$ in
eq.~(\ref{eq:Sphi-quadratic}). The resulting action for the
St\"uckelberg field $\pi$,
\begin{equation}
2 \int d^4x (\partial_0\pi)^2 ,
\label{eq:GC-goldstone-action}
\end{equation}
does not have the gradient term and describes a mode with the
dispersion relation
\begin{equation}
\omega^2=0.
\label{eq:GC-disprelation0}
\end{equation}
If the action of the ghost condensate model contained the contribution
(\ref{eq:Sphi-GhCond}) only, its effect would be simply (partial)
gauge fixing of General Relativity, so in the sector with the initial
condition $X=1$ the theory would describe the Einstein gravity in a
particular gauge. This situation is specific to the ghost condensate
model; we will see in the next Section that in the generic case
modifications of gravity arise already from the first term in the
derivative expansion of the action.

The degeneracy of the action (\ref{eq:GC-goldstone-action}) (the
absence of the spatial gradient terms) signals that the
non-propagating St\"uckelberg mode can become propagating once
higher-derivative corrections are added. These corrections are routine
in the effective low energy theories, but here they play a crucial
role.  The symmetry (\ref{eq:ghost-cond-symmetry}) restricts the
general form of these corrections. In the Goldstone language, the next
order contribution to the action contains higher derivatives acting on
the Goldstone field $\phi$, such as $\partial^2\phi$ and
$\partial_\mu\partial_\nu\phi$, suppressed by powers of
$\Lambda$. When expanded to the quadratic order in $\pi$, these terms
modify the action (\ref{eq:GC-goldstone-action}) to
\[
2\int d^4x 
\left\{ (\partial_0\pi)^2 
+ \frac{c_0}{ \Lambda^2} (\partial_0^2\pi)^2 
+ \frac{c_1}{ \Lambda^2} \partial_0^2\pi \partial_i^2\pi 
+ \frac{c_2}{ \Lambda^2} (\partial_i^2\pi)^2 +\ldots\right\},
\]
where $c_i$ are numerical coefficients roughly of the oder of unity. 
The dispersion relation
for the St\"uckelberg mode becomes
\begin{equation}
\omega^2 = \frac{c_0}{ \Lambda^2} \omega^4 + \frac{c_1}{ \Lambda^2}
\omega^2 {\bf p}^2 + \frac{c_2}{ \Lambda^2}  {\bf p}^4.
\label{eq:disp-relation}
\end{equation}
The are two solutions to this equation. The first one is 
\[
\omega^2 = \Lambda^2/c_0 + {\cal O} ({\bf p}^2).
\]
This solution is irrelevant, since it falls outside of the region of
validity of the low-energy effective theory, the latter being
$\omega\ll\Lambda$.  The second solution represents the modification
of the dispersion relation $\omega^2=0$ which now reads
\begin{equation}
\omega^2 = \frac{c_2}{ \Lambda^2} {\bf p}^4 +
{\cal O}\left(\frac{{\bf p}^6}{\Lambda^4}\right). 
\label{dec28-2}
\end{equation}
This solution describes a slowly propagating mode
\cite{Arkani-Hamed:2003uy} which is non-tachyonic provided that
$c_2>0$.  According to (\ref{eq:GC-goldstone-action}), this mode is
not a ghost for $F_{XX} (\alpha=1) > 1$.  Hence, the theory is healthy
at high spatial momenta.

The mode (\ref{dec28-2}) modifies the gravitational interaction (in
particular, the Newtonian potential) at distances larger than
$r_c=1/m_0$. On the other hand, the time scale at which these
modifications build up is parametrically larger, $t_c=\Lambda/m_0^2$
\cite{Arkani-Hamed:2003uy}. The reason is again that the modification
of gravity only occurs in the next-to-leading order in the derivative
expansion. When the mass $m_0$ tends to zero, the scales $r_c$ and
$t_c$ tend to infinity, and the modifications are smoothly switched
off. In this sense the van~Dam--Veltman--Zakharov phenomenon is absent
in the ghost condensate model. Note, though, that according to
(\ref{eq:Sphi-quadratic}), the mass $m_0$ is related to the UV scale
as
\[
  m_0^2 M_{PL}^2 = F_{XX} \Lambda^4 \; .
\]
Hence, the limit of vanishing mass corresponds to the limit
$\Lambda \to 0$ (at fixed $M_{Pl}$), so the region of validity
of the low energy effective theory shrinks to zero in this limit.

The ghost condensate model does not exhibit the Boulware--Deser
instability either. In contrast to the example considered in
Section~\ref{sub:UVunstable}, the only scalar field $\pi$ present in the
theory has, in flat background, the dispersion relation
(\ref{eq:disp-relation}), which is a consequence of the residual
symmetry (\ref{eq:GC-symmetry}). In slightly curved background, this
dispersion relation may acquire additional terms with small
coefficients controlled by the background curvature. The appearance of
new contributions to the dispersion relation --- for instance, a term
proportional to ${\bf p}^2$ with a negative coefficient --- may cause
tachyonic instability at low spatial momenta.  This is precisely what
happens in some cosmological backgrounds~\cite{Creminelli:2006xe},
this instability being, however, not particularly dangerous.  On the
other hand, the terms induced by slightly curved background cannot
change the sign of the leading $\omega^2$ term, so the propagating
mode does not become a ghost.  The situation, therefore, is different
from the case of the Boulware--Deser mode of
Section~\ref{sec:boulware-deser-mode} where one of the scalar modes is
necessarily a ghost in curved background.

The ghost condensate model and its modifications have unusual
properties. Some of these properties are potentially interesting from
the viewpoint of phenomenology and cosmology, others serve as examples
of novel phenomena that may emerge once Lorentz-invariance is
broken. Let us briefly describe some of them.

Due to Lorentz-violation and mixing of the slowly propagating field
$\pi$ with metric perturbations, gravitational fields of moving
sources are different from gravitational fields of sources that are at
rest with respect to ghost condensate. In particular, there is a
memory effect: moving bodies leave ``star tracks'' in ghost
condensate~\cite{Dubovsky:2004qe,Peloso:2004ut}.

Ghost condensate itself may be viewed as matter with rather unusual
properties. In particular, lumps of this matter can in principle
anti-gravitate \cite{Arkani-Hamed:2003uy}. More generic is the
property that the presence of ghost condensate in space leads to an
instability of the Jeans type with the time scale which is
parametrically large as compared to ordinary fluids of the same energy
density \cite{Arkani-Hamed:2003uy}. The latter property is related
again to the presence of the slowly propagating mode $\pi$.

Non-linear dynamics of ghost condensate is also quite rich.  Evolving
ghost condensate tends to form caustics~\cite{ArkaniHamed:2005gu},
much in common with caustics in some other scalar
theories~\cite{Felder:2002sv}.  Away from the caustics, the dynamics
of ghost condensate is the same as the dynamics of fluid with equation
of state $p \propto \rho^2$.  Another possible effect is the
non-perturbative instability of the background (\ref{eq:phi_vac})
leading to the formation of microscopic ``holes'' of negative
energy~\cite{Krotov:2004if}.

Lorentz-violation makes physics of black holes considerably different
from that in General Relativity.  The least dramatic effect is the
accretion of the ghost condensate onto black
holes~\cite{Frolov:2004vm,Mukohyama:2005rw}.  More exotic are the
possibilities that black hole systems may violate the second law of
thermodynamics~\cite{Dubovsky:2006vk},  signals may escape from
black holes~\cite{Babichev:2006vx}, and  black holes may have
hair~\cite{Dubovsky:2007zi}.

Cosmologically interesting class of models is obtained by adding a
potential term to the action, so that instead of
(\ref{eq:Sphi-GhCond}) one writes
\[
S_\phi = \Lambda^4 \int d^4x \sqrt{-g} \left[ F(X) - V(\phi)\right].
\]
Then both the kinetic term $F(X)$ and the potential term $V(\phi)$
contribute to the energy-momentum tensor. The field $\phi$ keeps
growing, albeit not quite according to (\ref{eq:phi_vac}). This may be
used for constructing models of inflation with ghost condensate
serving as inflaton~\cite{ArkaniHamed:2003uz} and models for dark
energy driving the present accelerated expansion of the
Universe~\cite{Piazza:2004df,Krause:2004bu}.  Interestingly, the field
$\phi$ grows even if potential increases as $\phi$ increases; in that
case the field $\phi$ rolls {\it up} the potential. This gives rise to
phantom behavior~\cite{Senatore:2004rj,Creminelli:2006xe} in which
energy density grows in time, and the equation of state is $p = w
\rho$ with $w < -1$ (and generically $w$ depends on time). This is one
of a few examples of phantom matter without UV pathologies: in most
other cases phantom equation of state is obtained in theories with
unacceptable tachyons and/or ghosts in UV (see, however,
Refs.~\cite{Rubakov:2006pn,Libanov:2007mq}).  If phantom behavior
occurs at inflationary stage of the cosmological evolution, the
consequence is the blue-tilted spectrum of primordial tensor
perturbations (as opposed to the red-tilted spectrum predicted by
theories where inflaton is an ordinary scalar field). The dark energy
driving the present accelerated expansion may also have phantom
equation of state, the feature potentially detectable by future
observations of SNe1a (see, e.g., Ref.~\cite{Mukohyama:2006be}).
Perhaps the most striking possibility is that phantom may give rise to
bouncing cosmology: in General Relativity the relation $p < -\rho$
implies that
\[
\dot{H}  > 0 \; ,
\]
where $H$ is the Hubble parameter, so a transition from contracting to
expanding Universe (from $H<0$ to $H>0$) becomes possible.  Indeed,
solutions of this sort have been
found~\cite{Creminelli:2006xe,Buchbinder:2007ad} and
explored~\cite{Creminelli:2007aq,Buchbinder:2007tw} in ghost
condensate models with suitable potentials $V(\phi)$.  The bounce in
these models occurs in a controllable and self-consistent way.

\section{A minimal model of massive graviton}
\label{sec:minim-model-mass}

An interesting
theory~\cite{Dubovsky:2004sg,Dubovsky:2004ud,Dubovsky:2005dw}, without
obvious pathologies and with massive gravitons, is obtained by
considering the case of the residual gauge symmetry
(\ref{dec7-iii}). This symmetry, $x^i \to x^i + \zeta^i (t)$,
translates into the following symmetry of the Goldstone Lagrangian,
\begin{equation}
  \phi^i  \to \phi^i  + \Xi^i (\phi^0 ) \; ,
\label{dec11-2}
\end{equation}
with three arbitrary functions $\Xi^i$.  At the one-derivative level,
there are two combinations of the Goldstone fields that respect this
symmetry,
\begin{eqnarray}
   X &=& \frac{1}{\Lambda^4} \; g^{\mu \nu} \d_\mu \phi^0 \d_\nu \phi^0 \; ,
\nonumber \\
W^{ij} &=&  \frac{1}{\Lambda^4} \; 
\l g^{\mu \nu} \d_\mu \phi^i \d_\nu \phi^j -
g^{\mu \nu} \d_\mu \phi^0 \d_\nu \phi^i \cdot
 g^{\lambda \rho} \d_\lambda \phi^0 \d_\rho \phi^j/X \r 
\nonumber\\
&=& Y^{ij} - \frac{V^i V^j}{X} \; ,
\nonumber 
\end{eqnarray}
where $Y^{ij}$ and $V^i$ are defined in (\ref{dec8-2}).  Hence, at
this level the Goldstone action is
\begin{equation}
 S_\phi = \int~d^4x~\sqrt{-g} F(X, W^{ij})
\label{dec10-2}
\end{equation}
where indices $i,j$ are contracted using $\delta_{ij}$.

\subsection{Linearized theory}
\label{sec:linearized-theory}

Let us first discuss this theory at the linearized level about
Minkowski background. The background Goldstone fields are given by
(\ref{dec7a-1*}). By field redefinitions we set $a=b=1$ and write the
background fields simply as
\begin{eqnarray} 	
\bar{\phi}^0 &=& \Lambda^2 t \; , 
\label{dec11-1} \\
\bar{\phi}^i &=& \Lambda^2 x^i \; .
\label{dec11-5}
\end{eqnarray}
From (\ref{dec8-5}) we obtain that the energy-momentum of this
configuration vanishes, provided that
\begin{eqnarray}
   - \half F + \frac{\d F}{\d X} = 0 \; ,
\label{dec10-1}\\
    \half F \delta_{ij} +  \frac{\d F}{\d W^{ij}} = 0 \;
\nonumber
\end{eqnarray}
at $X = 1$, $W^{ij} = - \delta^{ij}$. Switching metric perturbations
on, and using the unitary gauge $\phi^\alpha = \bar{\phi}^\alpha$, one
finds that the theory about Minkowski background is gravity with the
Lorentz-violating mass terms (\ref{Mh}) with the only constraint
\[
   m_1 = 0 \; .
\]
Other mass parameters are independent of each other, and are expressed
through $F$ and its first and second derivatives at $X = 1$, $W^{ij} =
- \delta^{ij}$.  Hence, at the level of one-derivative Goldstone
action neither vector nor scalar sector contains propagating modes, as
we discussed in Section \ref{subsubm1}, while tensor gravitons (two
degrees of freedom) have mass $m_G = m_2$.  In this sense the model
can be viewed as the minimal model of massive graviton.

It is instructive to switch off gravity and consider the Goldstone
sector of this theory, in Minkowski background but away from the point
(\ref{dec10-1}).  The quadratic Lagrangian for the perturbations
$\pi^\alpha =\phi^\alpha - \bar{\phi}^\alpha$ is obtained from
(\ref{dec10-2}) and has the following generic form,
\begin{equation}
L_{\pi} = \frac{a}{2} (\dot{\pi}^0)^2 -  \frac{b}{2} (\d_i \pi^0  )^2
+ c \dot{\pi}^0 \d_i \pi^i
 + \frac{d_1}{2} (\d_i \pi^i)^2 + \frac{d_2}{2} (\d_i \pi^j)^2 \; ,
\label{dec11-3}
\end{equation}
where $ b = 2 (\d F/ \d X) (\bar{\phi})$ while other constants contain
second derivatives of $F$ at $\phi = \bar{\phi}$. One observes that
the fields $\pi^i$ are non-dynamical. Their equations of motion in the
vector sector give $\pi^{T \, i} = 0$, where $\pi^{T \, i}$ is the
transverse part, $\d_i \pi^{T \, i} = 0$. Hence, there is no
non-trivial modes in the vector sector even for general linearly
rising background. The equation of motion for the longitudinal part of
$\pi^i$ gives
\[
   \pi^i = \mbox{const} \cdot \frac{\d_i}{\Delta} \dot{\pi}^0 \; .
\]
One plugs this expression into the equation of motion for $\pi^0$ and
obtains
\[
    \tilde{a} \ddot{\pi}^0 - b \Delta \pi^0 = 0 \; ,
\]
where $\tilde{a}$ is a combination of the constants $a$, $c$, $d_1$
and $d_2$.  For general linearly rising Goldstone background, the
dispersion relation is $\omega^2 = \mbox{const} \cdot {\bf p}^2$. One
can see that with suitable choice of parameters, this mode is neither
tachyon nor ghost~\cite{Dubovsky:2004sg,Dubovsky:2005dw}.

At the point
\begin{equation}
\frac{\d F}{\d X} =0
\label{dec11-4}
\end{equation}
the dispersion relation is $\omega^2 =0$ at the level of
one-derivative action. In fact, this special point is basically
coincident with the Minkowski point (\ref{dec10-1}), since we neglect
gravity here and therefore cannot discriminate between different
values of $F$ at $\phi = \bar{\phi}$. Overall, the situation in the
scalar and vector sectors is very similar to what one finds in the
ghost condensate theory.

The absence of propagating modes associated with the fields $\pi^i$
(rather than $\pi^0$) is by no means an accident. Given the background
(\ref{dec11-1}), the symmetry (\ref{dec11-2}) implies that the theory
is invariant under infinitesimal transformations
\[
\pi^i \to \pi^i + \Xi^i (t) .
\]
This means that at the one derivative level, the Lagrangian does not
contain time derivatives of the fields $\pi^i$, so these fields are
not dynamical.  This is of course explicit in (\ref{dec11-3}). So, the
dispersion relation ${\bf p}^2 =0$, characteristic of non-propagating
modes, is protected in this model by the symmetry (\ref{dec11-2}).

The latter observation is useful for the discussion of the UV
sensitivity issue in this model. Under the assumption that higher
derivative terms respect the symmetry (\ref{dec11-2}), these terms
cannot contain $\d_0^2 \phi^i$ (in the reference frame where the
background $\bar{\phi}^0$ has the form (\ref{dec11-1})), and in terms
of perturbations $\pi^\alpha$ they are quadratic combinations of
\[
\Lambda^{-1} \d_0 \d_j \pi^i \; , \;\;\;
\Lambda^{-1} \d_j \d_k \pi^i  \; , \;\;\;
\Lambda^{-1} \d_0^2 \pi^0 \; , \;\;\;
\Lambda^{-1} \d_0 \d_j \pi^0 \; , \;\;\;
\Lambda^{-1} \d_j \d_k \pi^0  \; .
\]
Once these terms are added to the Lagrangian (\ref{dec11-3}), the
fields $\pi^i$ formally become dynamical, but it is straightforward to
see that the equation for the corresponding dispersion relation has
the form
\[
  {\bf p}^2 \left[ \omega^2 - \mbox{const} \cdot
\Lambda^2 + O({\bf p}^2)\right] = 0
\]
Hence, the would-be new  propagating modes have the dispersion relation
\[
\omega^2 = \mbox{const} \cdot \Lambda^2 +  O({\bf p}^2) \; .
\]
Since the frequencies are of order of the UV cutoff, these modes are
actually absent in the low energy theory. In this sense the theory is
UV stable: upon switching on higher-derivative terms, there remains
the mode with the dispersion relation ${\bf p}^2=0$ only.

At this point it is worth discussing the physical interpretation of
the modes with the dispersion relation ${\bf p}^2=0$. They can be
thought of as degrees of freedom with infinite propagation velocity
(unlike the ghost condensate mode which has zero velocity at the
one-derivative level and acquires a small velocity due to
higher-derivative terms). Physically, they describe sound waves
propagating through the rigid coordinate frame selected in space by
the functions $\phi^i$. The rigidity of this frame is ensured by the
symmetry (\ref{dec7-iii}) and $SO(3)$ symmetry of the Goldstone action
that allow to move and rotate this frame only as a whole.  Infinitely
fast propagating modes do not imply the violation of causality in the
absence of Lorentz invariance, but allow for instantaneous transfer of
information. The latter leads to a number of unusual effects related
to black hole physics \cite{Dubovsky:2006vk,Dubovsky:2007zi}.  A
detailed discussion of the properties of these modes in a toy QED
model can be found in Refs.~\cite{Dvali:2005nt,Gabadadze:2004iv}.

The higher order terms are important also for the remaining dynamical
field $\pi^0$, if the background obeys (\ref{dec11-4}).  In that case
the one-derivative dispersion relation $\omega^2=0$ gets transformed
into
\begin{equation}
  \omega^2 = \mbox{const} \cdot \frac{{\bf p}^4}{\Lambda^2} \; .
\label{dec11-6}
\end{equation}
So, the spectrum of the low energy effective theory is the same as in
the ghost condensate case, except that tensor gravitons are massive in
the model discussed here.

The symmetry (\ref{dec11-2}) protects the theory from the
Boulware--Deser instability as well. In nearly Minkowski space-time,
and for the background nearly the same as in (\ref{dec11-1}),
(\ref{dec11-5}), one can choose a reference frame in which the
background $\bar{\phi}^0$ has precisely the form (\ref{dec11-1}).  In
that frame, the above analysis goes through: the fields $\pi^i$ are
non-dynamical in the low energy effective theory, at least for
$\omega^2, {\bf p}^2 \gg m_G^2$, and there remains one dynamical mode
associated with the field $\pi^0$. Its dispersion relation coincides
with (\ref{dec11-6}) modulo corrections proportional to the deviation
of the background from Minkowski space.

\subsection{Phenomenology}
\label{sec:line-theory-near}

By analogy to conventional field theory one might expect that non-zero
graviton mass leads to the exponential suppression of the
gravitational potential at distances greater than the inverse graviton
mass. The latter would then be constrained by observations to be very
small.  This {\it is not} the case in the model described by the
action (\ref{dec10-2}), the reason being the violation of
Lorentz-invariance. We will see below that the gravitational potential
remains unchanged at the linear level, at least in some region of
parameter space. In this region the behavior of the model is similar
to General Relativity in many respects and may be phenomenologically
acceptable. At the same time, there may exist a number of interesting
and potentially detectable effects, the non-zero graviton mass being
one of them.

\subsubsection{Newton's law}
\label{sec:newtons-law}

Newton's law emerges from General Relativity in the linear
approximation.  In order to derive its analog in the model described
by the action (\ref{dec10-2}), it is instructive to go back to the
unitary gauge where the perturbations of the Goldstone fields are
absent, and the only perturbations are those of metric. This will make
the comparison to  General Relativity  straightforward. As in
Section~\ref{sub:LI-1}, it is convenient to decompose the metric
perturbations according to eq.~(\ref{eq:h-parameterization}). The
quadratic part of the action is then given by
\begin{equation}
L^{(2)} = L_{EH}^{(2)} + L_m + L_s,
\label{quadratic_action}
\end{equation}
where $L_{EH}^{(2)}$, $L_m$ and $L_s$ come from the Einstein-Hilbert, mass
and source terms, respectively. 
The Einstein--Hilbert term is given by eq.~(\ref{jan25-11}),
while the mass and source terms are
\begin{gather}
L_m = M_{Pl}^2\Bigl\{ -\frac{1}{ 4} m_2^2 (h_{ij}^{TT})^2 
-  \frac{1}{ 2} m_2^2 (\d_i F_j)^2 
+ m_0^2\varphi^2 + \l m_3^2-m_2^2\r\l\Delta E\r^2-\nonumber\\
- 2(3m_3^2-m_2^2)\psi\Delta E + 3\l 3m_3^2-m_2^2\r\psi^2+
2m_4^2\Delta E - 6m_4^2\varphi\psi \Bigr\} ,
\label{massterm2}
\end{gather}
\begin{equation}
L_s=-T_{00}\l\varphi+\d_0B-\d_0^2E\r-T_{ii}\psi + (S_i + \d_0F_i)T_{0i} 
+ \frac{1}{ 2} h_{ij} T_{ij}.
\label{eq:source-term}
\end{equation}
Notations for the masses are the same as in (\ref{Mh}); the masses
$m_i^2$ are combinations of the first and second derivatives of the
function $F$, the parameter $\Lambda$ and the Planck mass. As
discussed above, their overall scale is $m \sim \Lambda^2/M_{Pl}$.
The source term contains an external energy-momentum tensor
$T_{\mu\nu}$ which we assume to be conserved.  All combinations
coupled to the components of $T_{\mu\nu}$ are gauge invariant. The one
multiplying $T_{00}$,
\[
\Phi\equiv\varphi+\d_0B-\d_0^2E,
\] 
plays the role of the Newtonian potential in the non-relativistic
limit of  General Relativity. 

In the tensor sector, only the transverse traceless perturbations
$h^{TT}_{ij}$ are present (two degrees of freedom).  Their field
equation is that of a massive field with the mass $m_G=m_2$.  Note
that massive tensor field does not necessarily have five polarizations
in a Lorentz-breaking theory. The examples of this phenomenon have
already been discussed in the preceeding Sections.

In the vector sector the field equations read
\begin{eqnarray}
\label{dSi1}
&&- \Delta (S_i + \d_0 F_i)=- T_{0i},\\
\label{dSi2}
&&\d_0\Delta ( S_i + \d_0F_i) + m_2^2 \Delta F_i = \d_0 T_{0i}.
\end{eqnarray}
Taking the time derivative of Eq.~(\ref{dSi1}) and adding it to 
Eq.~(\ref{dSi2}) gives
\[
F_i =0,
\]
provided that $m_2^2\neq 0$.  Thus, the vector sector of the model
behaves in the same way as in the Einstein theory in the gauge
$F_i=0$. There are no propagating vector perturbations and interaction
of sources is not modified in the vector sector unless one takes into
account non-linear effects or higher derivative terms.

The interaction potential between static sources ( Newton's potential)
is determined by the scalar sector of the model.  The field equations
for scalar perturbations are
\begin{eqnarray}
\label{psieq}
&&2\Delta \psi+m_0^2\varphi+m_4^2\Delta E
-3m_4^2\psi=\frac{T_{00}}{ 2M_{Pl}^2},\\
\label{phieq}
&&2\Delta \Phi-2\Delta \psi+6\d_0^2\psi-
\l 3m_3^2-m_2^2\r\Delta E+3\l 3m_3^2-m_2^2\r\psi-3m_4^2\varphi=
\frac{T_{ii}}{ 2M_{Pl}^2},\\
\label{Eeq}
&&-2\Delta \d_0^2\psi+\left( m_3^2-m_2^2\right)\Delta^2 E-
\left( 3m_3^2-m_2^2\right)\Delta\psi+
m_4^2\Delta\varphi=-\frac{\d_0^2T_{00}}{ 2M_{Pl}^2},\\
\label{Beq}
&&2\Delta \d_0\psi=\frac{\d_0T_{00}}{ 2M_{Pl}^2}.
\end{eqnarray}
Eq.~(\ref{Beq}) implies 
\begin{equation}
\psi=\frac{1}{ \Delta} \frac{T_{00}}{ 4M_{Pl}^2}+\psi_0(x^i),
\label{psi}
\end{equation}
where $\psi_0(x^i)$ is an arbitrary time-independent function.
From Eqs.~(\ref{psieq}) and (\ref{Eeq}) one finds
\begin{eqnarray}
\label{phi}
&&\varphi=\frac{2m_2^2m_4^2}{ \mathfrak{M}}\psi + 
\frac{2(m_3^2-m_2^2)}{  \mathfrak{M}} 
\Delta \psi_0, \\
\label{E}
&&\Delta E=  \l 3 - \frac{2m_0^2m_2^2}{  \mathfrak{M}}\r \psi 
- \frac{2m_4^2}{  \mathfrak{M}} \Delta \psi_0,
\end{eqnarray}
where 
\[
 \mathfrak{M} = m_4^4-m_0^2(m_3^2-m_2^2).
\]
Finally, substituting Eqs.~(\ref{psi}), (\ref{phi}) and (\ref{E}) into
eq.~(\ref{phieq}) one  finds the gauge-invariant potential $\Phi$,
\begin{gather}
\label{eq:Phi}
\Phi=\frac{1}{\Delta} \frac{T_{00}+ T_{ii}}{ 4M_{Pl}^2} 
- 3\frac{\d_0^2}{ \Delta^2} \frac{T_{00}}{ 4 M_{Pl}^2} 
+  \l 3 - \frac{ 2m_0^2 m_2^2}{ \mathfrak{M}} \r 
\frac{m_2^2}{ \Delta} 
\l \frac{1}{ \Delta} \frac{T_{00}}{ 4 M_{Pl}^2} + \psi_0 \r
+ \l 1 -\frac{2m_2^2m_4^2}{ \mathfrak{M} }\r \psi_0,
\end{gather}
The first two terms in the right hand side of eq.~(\ref{eq:Phi})
are the standard contributions in the Einstein theory, the first one
becoming the Newtonian potential in the nonrelativistic limit. Thus,
except for the $\psi_0$-dependent terms, the gauge-invariant potentials
$\Phi$ and $\psi$ differ from their analogs in the
Einstein theory $\Phi_E$ and $\psi_E$ by the mass-dependent third term
in the right hand side of eq.~(\ref{eq:Phi}),
\begin{eqnarray}
\nonumber
&&\psi=\psi_E,\\
\label{PPhi}
&&\Phi=\Phi_E + \l 3 - \frac{2m_0^2m_2^2}{ \mathfrak{M}} \r
\frac{m_2^2}{ \Delta^2}
\frac{T_{00}}{ 4M_{Pl}^2}.
\end{eqnarray}
The second term in eq.(\ref{PPhi}) vanishes if all masses uniformly
tend to zero, so both of the potentials $\psi$ and $\Phi$ become the
same as in General Relativity in the massless limit.  This means the
absence of the vDVZ discontinuity in the model.

For a static source, Eq.~(\ref{PPhi}) leads to the modification
of the Newtonian potential of a point mass $M$ which in coordinate
space takes the form
\[
\Phi = G_N M \left(-\frac{1}{ r} + \mu^2 r\right),
\]
where 
\begin{equation}
\mu^2 = -\frac{1}{ 2} m_2^2\l 3 - \frac{2 m_0^2 m_2^2}{  \mathfrak{M}}\r.
\label{mu}
\end{equation}
Since the potential is growing, the perturbation theory breaks down at
distances $r\gtrsim 1/(G_NM\mu^2)$.  This would be unacceptable for
relatively large graviton masses.  However, the modification of the
potential is absent in the case $3 \mathfrak{M}=2m_0^2m_2^2$ (and $
\mathfrak{M}\neq 0$).  We will see in what follows that this condition
can be ensured by a particular dilatation symmetry\footnote{Like other
symmetries which we have discussed in Sect.~\ref{sub:symmetries-gen},
this dilatation symmetry, eq.~(\ref{dilatation}), may be viewed as an
unbroken part of the diffeomorphism invariance.}  which is
automatically enforced at the cosmological attractor, i.e., at late
times of the cosmological evolution.

The freedom of choosing the time-independent function $\psi_0(x)$
which enters the above gravitational potentials corresponds to the
presence of the scalar mode with the dispersion relation $\omega^2=0$.
As discussed in Section~\ref{sec:linearized-theory}, this mode is an
analog of the ghost condensate mode and becomes dynamical with the
account of higher-derivative terms in the action, acquiring the
dispersion relation $\omega^2\propto {\bf p}^4$.  The value of
$\psi_0$ is determined by the initial conditions. In the linear
regime, a non-zero value of $\psi_0$ would mean the presence of the
incoming ``ghost condensate wave''. So, for the purpose of finding the
potential between sources, the physical choice is $\psi_0(x^i)=0$. We
note, however, that this choice is not so evident in the cosmological
context.

\subsubsection{Cosmological solutions}
\label{sec:cosm-solut}

At the time of writing this review, only spatially flat cosmological
solutions are known in the model (\ref{dec10-2}). The flat
cosmological ansatz is\footnote{ In principle, one may write a more
general time-dependent ansatz for the scalar fields, namely,
$\phi^i=\Lambda^2 C(t) x^i$, where $C(t)$ is an arbitrary function of
time. For models respecting the symmetry (\ref{dilatation}), which we
mainly consider in what follows, this ansatz leads to the same
cosmological evolution as the ansatz (\ref{simple_ansatz}), see
Ref.~\cite{Dubovsky:2005dw} for details.}
\begin{gather}
\nonumber
ds^2= dt^2-a^2(t)dx_i^2,\\\label{simple_ansatz}
\phi^0=\phi(t), \quad
\phi^i=\Lambda^2x^i.
\end{gather}
For this ansatz $W^{ij}=-a^{-2}\delta^{ij}$, so the function $F$ in
eq.~(\ref{dec10-2}) depends only on $X$ and $a$, $F = F(X,a)$. The
Einstein equations reduce to the Friedman equation,
\begin{gather}
\label{00eq}
\left(\frac{\dot{a}}{ a}\right)^2=
\frac{1}{ 6M_{Pl}^2}\Bigl\{ \rho_m +  2\Lambda^4X F_X 
- \Lambda^4F\Bigr\}
\equiv \frac{1}{ 6M_{Pl}^2}\Bigl\{ \rho_m +  \rho_1 + \rho_2\Bigr\}\;,
\end{gather}
where $\rho_m$ is the energy density not including
Goldstone fields. The field
equation for $\phi^0$ is
\begin{equation}
\label{eq:phi0}
\d_t\left( {a^3 \sqrt{X} F_X}\right)=0\;.
\end{equation}
The field equations for $\phi^i$ are satisfied automatically. In
principle, it is straightforward to solve this system of equations for
any given function $F(X,a)$. Upon integration, eq.~(\ref{eq:phi0})
gives an algebraic equation which determines $X$ as a function of the
scale factor $a$.  This makes eq.~(\ref{00eq}) a closed equation for
the scale factor $a(t)$.

From the point of view of cosmological applications, of particular
interest are solutions where the scale factor $a(t)$ tends to infinity
at late times. Since the graviton masses are linear combinations of
the function $F(X,a)$ and its derivatives, one may wonder whether they
remain finite or tend to zero in this limit, and whether the
effective-theory description remains valid. Indeed, eq.~(\ref{eq:phi0})
implies that at late times either $X$ or $F_X$ tend to zero, which
suggests that the graviton masses might tend to zero as well. 

Consider a particular class of functions $F$ such that $X(a)$ as found
from eq.~(\ref{eq:phi0}) asymptotes to some power of $a$ at large
$a$. This is not a very restrictive assumption --- for instance, it is
satisfied by any algebraic function $F(X,a)$.  Then there exists 
a real constant $\gamma$ such
that the combination $X^\gamma/a^2$ tends to a
non-zero value as $a \to \infty$. Eq.~(\ref{eq:phi0}) implies
that $XF_X=\mbox{const}\cdot \sqrt{X}/a^3$; this determines the
dependence of the energy component $\rho_1$ on the scale factor,
\begin{equation}
\rho_1 = 
{\rm const} 
\frac{1}{ a^{3-{1/\gamma}}}.
\label{rho1}
\end{equation}
This relation generalizes the behavior found in the ghost condensate
model where the energy density of the ghost condensate scales like
$1/a^3$ \cite{Arkani-Hamed:2003uy} (the latter behavior is recovered
from eq.~(\ref{rho1}) at $\gamma\to\infty$).

For $\gamma>1/3$ the energy density $\rho_1$ behaves like the dark
energy component with the the negative pressure.  Its equation of
state varies between that of cold dark matter, $w=0$ (for
$\gamma=+\infty$), and that of the cosmological constant, $w=-1$ (for
$\gamma=1/3$).  For $0<\gamma<1/3$ the term $\rho_1$ grows with
$a$. This corresponds to the energy density component with a highly
negative equation of state, $w<-1$. Without fine tuning, this
contribution cannot be canceled out by the term $\rho_2$, so that the
Hubble rate diverges as $a\to\infty$ leading to the breakdown of the
low-energy effective theory and rapid instabilities
\cite{Dubovsky:2005xd}.  In what follows we assume that $\gamma$ does
not belong to this range. For $\gamma<0$ the energy density $\rho_1$
corresponds to fluid with positive pressure.

In order to see that the graviton masses remain finite and the
effective field theory description is valid in the limit $a\to\infty$,
it is convenient to replace $X$ by a new variable $Z=X^\gamma/a^2$.
The function $F(X,a)$ becomes the function of $Z$ and $a$, $\tilde
F(Z,a)= F(Z^{1/\gamma}a^{2/\gamma},a)$. Note that it satisfies the
relation $\gamma Z \tilde F_Z = XF_X$, where $\tilde F_Z = \d \tilde
F/\d Z$.  In these notations Eq.~(\ref{eq:phi0}) reads
\begin{equation}
\gamma a^{3-\frac{1}{ \gamma}} Z^{1-\frac{1}{ 2\gamma}} \tilde F_Z(Z,a) = A, 
\label{F_Z}
\end{equation}
where $A$ is an integration constant. This equation determines $Z$ as
a function of $a$. By construction, this dependence is such that
$Z(a\to\infty) = Z_0$, where $Z_0$ is some constant. 

If one assumes further that the function $\tilde F(Z,a)$ is regular at
$a\to\infty$, then at late times one has
\begin{equation}
F(X,a) = \tilde F(Z,a) \to F_0(Z). 
\label{FofZ}
\end{equation}
In terms of the original variables this means that in the limit
$a\to\infty$ the function $F(X,W^{ij})$ depends only on the
combination $X^\gamma W^{ij}$. This corresponds to the following
dilatation symmetry of the Goldstone action,
\begin{eqnarray}
\nonumber
\phi_0 &\to& \lambda \phi_0,\\
\phi_i &\to& \lambda^{-\gamma} \phi_i, 
\label{dilatation}
\end{eqnarray}
which is equivalent, in the unitary gauge, to the following unbroken
part of diffeomorphism invariance, $t\to\lambda t$,
$x^i\to\lambda^{-\gamma}x^i$. In this case one has
\[
\rho_2 = - \Lambda^4 F_0(Z_0),
\]
which behaves like a cosmological constant (assuming $F_0(Z_0)\neq
0$).  Likewise, at $a\to\infty$ the graviton masses become functions
of $Z_0$ and in general remain finite.

The models obeying eq.~(\ref{FofZ}) have an interesting feature which
is a consequence of the symmetry (\ref{dilatation}). It is
straightforward to check that Eq.~(\ref{dilatation}) implies the
following relations among graviton masses in Minkowski space,
\begin{equation}
m_0^2= - 3\gamma m_4^2, \qquad \gamma(m_2^2 - 3 m_3^2) = m_4^2. 
\label{relation-for-masses}
\end{equation}
These relations ensure that the parameter $\mu^2$ defined by
Eq.~(\ref{mu}) is zero, i.e., the correction to the Newtonian
potential (the last term in eq.~(\ref{PPhi})) vanishes. Thus, apart
from the effects of the higher derivative terms, at late times the
only modification of gravity at the linear level is the non-zero mass
of the two polarizations of the graviton. 

A particularly simple case occurs when the function $F$ depends only
on the combination 
\begin{equation}
Z^{ij}= X^\gamma W^{ij} \; .
\label{vr-feb5-2}
\end{equation}
If $\gamma > 1/3$ or $\gamma<0$, the evolution drives the system to
the point $\tilde F_Z=0$, in full similarity with the ghost condensate
model. In the case $0<\gamma < 1/3$ and regular $\tilde F$, $Z^{ij}$
diverges at large $a$. This breaks the validity of the low energy
effective theory.

\subsubsection{Massive gravitons} 
\label{sec:phen-lorentz-break}

Let us consider in more detail the properties of massive gravitons,
namely, the experimental constraints on the graviton mass and the
possibility of the graviton creation in the early Universe.  For
simplicity, in this Section we limit ourselves to the model with the
action 
\begin{equation}
\label{Zgammaaction}
S_G=\Lambda^4\int \sqrt{-g}d^4xF(Z^{ij}) \; ,
\end{equation}
where $Z^{ij}$ are given by eq.~(\ref{vr-feb5-2}).  In this model
there are no corrections to the Newton's potential at the linear
level, so the tests of (linearized) gravity based on the Solar system
and Cavendish-type experiments \cite{Esposito-Farese:2000ij} are
automatically satisfied. The constraints on the graviton mass come
from the emission and/or propagation of gravitational waves.

Observations of the slowdown of the orbital motion in binary pulsar
systems~\cite{RevModPhys.66.711} are considered as an indirect proof
of the existence of gravitational waves. The agreement of these
observations with General Relativity implies that the mass of the
graviton cannot be larger than the characteristic frequency of the
emitted gravitational waves. This frequency is set by the period of
the orbital motion which is of order 10 hours, implying the following
limit on the graviton mass,
\begin{equation}
\label{pulsarlimit}
\frac{m_G}{ 2\pi}\equiv \nu_G \lesssim  3\cdot 10^{-5}~{\mbox{Hz}}\approx
\l 10^{15}~{\mbox{cm}}\r^{-1} \sim (70 {\rm AU})^{-1}.
\end{equation}
Thus, the maximum allowed graviton mass is comparable to the inverse
size of the Solar system, which is a very large mass (short distance)
in cosmological standards. Gravitons of such a mass can serve as dark
matter candidates provided they can be produced in in the early
Universe in sufficient numbers. Indeed, if the graviton mass is large
enough, $ (mv)^{-1}\lesssim 1~\mbox{kpc}\sim 3\cdot 10^{21}~\mbox{cm}
$, where $v\sim 10^{-3}$ is typical velocity in the halo, massive
gravitons may cluster in galaxies and account for the dark matter in
galactic halos.

It is straightforward to estimate the cosmological abundance of relic
massive gravitons.  The massive gravitons are described by the
transverse traceless perturbation of the metric, $h_{ij}^{TT}$.  The
quadratic action for $h_{ij}$ (we omit the superscript in what
follows) in the expanding Universe has the following form,
\begin{equation}
M_{Pl}^2\int d^3kd\eta a^2(\eta)
\left[ (h^{\prime}_{ij})^2- \left(\d_kh_{ij}\right)^2
-m_G^2a^2(\eta)h^2_{ij}\right], 
\label{eq:massive-grav-action}
\end{equation}
where $\eta$ is conformal time and prime denotes $\d /\d \eta$. 
Eq.~(\ref{eq:massive-grav-action})
has the form of the action of a minimally coupled massive scalar
field. Therefore, similarly to scalar bosons, massive gravitons are
produced efficiently during inflation
(cf. Ref.~\cite{Rubakov:1982df}).

To be concrete, consider a scenario where the Hubble parameter $H_i$
is constant during inflation.  This scenario may be realized, for
instance, in hybrid models of inflation~\cite{Linde:1993cn}.  First,
one has to check that the phenomenologically relevant values of
parameters correspond to the regime below the cutoff scale of the
effective theory, i.e. $H_i\lesssim\Lambda$. This implies for the
energy scale of inflation $E_i\sim\sqrt{H_iM_{Pl}}$ that
\begin{equation}
\label{energy}
E_i<m_G^{1/4}M_{Pl}^{3/4}\approx 10^7~{\mbox{GeV}}
\l m_G\cdot 10^{15}~{\mbox{cm}}\r^{1/4}.
\end{equation}
This value is high enough to be consistent with everything else in
cosmology (in particular, to allow for successful baryogenesis) even
for graviton masses of the order of the current Hubble scale.

Consider now the production of massive gravitons.  Assuming the above
scenario of inflation, the perturbation spectrum for the massive
gravitons is that for the minimally coupled massive scalar field in de
Sitter space~\cite{Bunch:1978yq},
\begin{equation}
\label{spectrum}
\langle h_{ij}^2\rangle\simeq \frac{1}{ 4\pi^2} \l \frac{H_i}{M_{Pl}}\r^2\int
 \frac{dk}{ k}
\l \frac{k}{ H_i}\r^\frac{2m_G^2}{ 3H^2}\;.
\end{equation}
Importantly, for long enough inflation,
the present physical momenta of most of the gravitons are smaller than the
present Hubble scale.

Metric fluctuations remain frozen until the Hubble parameter becomes
smaller than the graviton mass, and afterwards they start to oscillate
with the amplitude decreasing as $a^{-3/2}$.  The energy density in
massive gravitons at the beginning of oscillations is of order
\begin{equation}
\rho_*\sim M_{Pl}^2m_G^2\langle h_{ij}^2\rangle\simeq
\frac{3 H_i^4}{ 8\pi^2}\;,
\nonumber
\end{equation}
where we neglected a pre-factor which is roughly of order 1.  
Today the fraction of the energy density in the massive
gravitons is
\begin{equation}
\label{Omega}
\Omega_g=\frac{\rho_*}{ z_*^3\rho_c}=
\frac{\rho_*}{ z_e^3\rho_c}\l \frac{H_e}{ H_*}\r^{3/2},
\end{equation}
where $z_*$ is the redshift at the start of oscillations, $H_*\sim
m_G$ is the Hubble parameter at that time, $H_e\approx 0.4\cdot
10^{-12}$~s$^{-1}$ is the Hubble parameter at the matter/radiation
equality, and $z_e\approx 3200$ is the corresponding redshift.
Combining all factors together we get
\begin{equation}
\label{finalOmega}
\Omega_g\sim 3\cdot 10^3 (m_G\cdot 10^{15}{\mbox{cm}})^{1/2}\l 
\frac{H_i}{\Lambda}\r^4 \;.
\end{equation}
This estimate assumes that the number of e-foldings during inflation
is large, $\ln N_e>H^2/m_G^2$, which is quite natural in the
model of inflation considered here. 

According to eq.~(\ref{finalOmega}), massive gravitons are produced
efficiently enough to comprise all of the cold dark matter, provided
the value of the Hubble parameter during inflation is about one order
of magnitude below the scale $\Lambda$.  Interestingly, one obtains
$\Omega_g\sim 1$ when the initial energy density in the metric
perturbations is close to the cutoff scale,
$\rho_*^{1/4}\sim\Lambda$. This suggests that other mechanisms of
production unrelated to inflation may naturally lead to the same
result, $\Omega_g\sim 1$.

If massive gravitons have been produced in substantial amounts during
the evolution of the Universe, they can be observed by the
gravitational wave detectors. At distances shorter than the
wavelength, the effect of a transverse traceless gravitational wave on
test massive particles in Newtonian approximation is described by the
acceleration $\ddot{h}_{ij}x^j/2$ (see, e.g.,
Ref.~\cite{Thorne:1987af} for a review). The same is true for massive
gravitational waves, the only difference being that the wavelengths
are longer in the non-relativistic case, so the Newtonian description
works for larger range of distances. Thus, the non-relativistic waves
act on the detector in the same way as massless waves of the same
frequency.

To estimate the amplitude of the gravitational waves we assume that they
comprise all of the dark matter in the halo of our Galaxy.  The energy
density in non-relativistic gravitational waves is of order
$M_{Pl}^2m_G^2h_{ij}^2$.  Equating this to the local halo density
$\rho_0 \sim 0.3 \,{\rm GeV/cm}^3$ 
one gets
\begin{equation}
\label{hestimate}
h_{ij} \sim 10^{-10}\l \frac{3\cdot 10^{-5}\mbox{Hz}}{\nu_G}\r.
\end{equation}
At frequencies $ 10^{-6}\div 10^{-5}$~Hz this value is many orders of
magnitude above the expected sensitivity of the LISA
detector~\cite{Bender:2003uv}. Thus, LISA may observe massive
gravitational waves even if their abundance is much lower than that
required to play the role of the dark matter.  Note that in the nearby
frequency range $10^{-9}\div 10^{-7}$~Hz there is a restrictive
bound~\cite{Lommen:2002je} at the level $\Omega_g<10^{-9}$ on the
stochastic background of gravitational waves coming from timing of
millisecond pulsars~\cite{Sazhin1978SvA}. So, it is possible that the
model can be tested by the re-analysis of the already existing data on
pulsar timing. This re-analysis would have to take into account that,
unlike the usual gravitational waves, relic massive gravitons produce
a monochromatic line at the frequency equal to the graviton mass. Such
a narrow line with the relative width $\Delta\nu/\nu\sim 10^{-3}$ is a
distinctive signature of the model.

Another possible signature is the time delay of a gravitational wave
signal as compared to electromagnetic radiation. In terms of the
frequency of the wave $f$ and the distance $D$ to the source, the time
delay is given by
\[
\Delta t = \frac{D}{ 2} \left( \frac{m_G}{ 2\pi f} \right)^2
\]
(assuming $f\gg m_G$).  Consider as an example gravitational waves
emitted during merger of two massive black holes --- one of the
promising processes from the standpoint of gravitational wave
detection.  The frequency of these waves is of order of the
gravitational radius of the resulting black hole,
\[
f\sim R_S^{-1}  = \frac{M_{\rm Pl}^2}{ 2M},
\]
where $M$ is the black hole  mass. Thus, for 
$m_G \sim 10^{-15} \mbox{cm}^{-1} $ the time delay is
\[
\Delta t \sim \frac{D}{ 2} \left(\frac{M m_G}{ \pi M_{\rm Pl}^2 }
\right)^2
\sim 5\times 10^{-6} \left(\frac{D}{ {\rm Mpc}} \right) 
\left( \frac{M}{M_\odot}\right)^2 \mbox{s}.
\]
This is probably too small to be detected for solar mass black holes,
but may be detectable for heavier ones.

\subsubsection{Refined cosmological tests: growth of perturbation}
\label{sec:refin-cosm-tests}

Given that some models of massive gravity pass the most obvious
experimental tests, the question arises whether they may provide a
viable alternative to General Relativity in describing more subtle
issues. One of these issues is the theory of structure formation. In
the standard cosmology based on General Relativity, the formation of
the observed structure in the Universe is explained by the growth of
primordial perturbations, mostly during the matter-dominated stage
(see,. e.g., \cite{Dodelson:2003ft,Mukhanov:2005sc} and references
therein). The conventional theory is in good agreement with
observations provided the dark matter component has the right
properties \cite{Tegmark:2003ud,Spergel:2006hy,Seljak:2006bg}. It is
not obvious that General Relativity can be modified without spoiling
this agreement. We demonstrate in this Section that massive gravity
model described by the action (\ref{Zgammaaction}) is an example of
such a modification, i.e., this model successfully passes structure
formation test even though the graviton mass is very large in
cosmological standards. This again illustrates the fact that in
Lorentz-violating theory, the mass of transverse traceless graviton
has very little to do with the properties of 3-dimensionally scalar
modes.

Perturbations relevant for structure formation are 3-dimensional
scalars.  In massive gravity the scalar sector contains additional
scalar fields which may alter the growth rate and make the model
incompatible with observations. Without gauge fixing, the scalar
sector contains metric perturbations $\varphi$ (not to be confused
with the Goldstone fields $\phi^0$, $\phi^i$), $B$, $\psi$ and $E$
defined according to eqs.~(\ref{nov15-3}) and (\ref{nov14-1}),
perturbations of the Goldstone fields $\pi_0$ and $\pi_L$ (the
longitudinal part of $\pi_i$), and perturbations of ordinary matter.
In total there are 9 scalar perturbations of which one can form 7
gauge-invariant combinations whose dynamics is responsible for the
structure formation. The complete set of equations that govern the
behavior of these perturbations can be found in
Ref.\cite{Bebronne:2007qh}.

The system of equations for perturbations can be reduced to two
equations for the gauge-invariant gravitational potentials $\Phi$ and
$\Psi$.  In General Relativity they satisfy the relation
$\Phi-\Psi=0$. In massive gravity this relation changes to
\begin{eqnarray}
\Phi - \Psi = \vartheta \left( x^i \right) \, a^{1/\gamma - 1}, 
\label{eq:036}
\end{eqnarray}
where $\vartheta(x^i)$ is an arbitrary function of spatial coordinates
which arises as an integration constant. The origin of this constant
is the presence of the mode with the dispersion relation
$\omega^2=0$. We have already encountered the appearance of such a
constant in Sect.~\ref{sec:newtons-law}.

The second equation is the closed equation for $\Psi$, 
\begin{eqnarray}
 && \frac{\d^2 \Psi}{\partial a^2} + \frac{1}{a} \left( 4 + 3 c_s^2 
+ \frac{H^\prime}{H^2} \right) \frac{\d \Psi}{\partial a} 
+ \frac{1}{a^2} \left[ \left( 1 + 3 c_s^2 \right) 
+ 2 \frac{H^\prime}{H^2} 
- \frac{c_s^2 \Delta}{H^2} \right] \Psi \nonumber \\
&& ~~~~~~~~ =  \left[ \frac{\gamma c_s^2 \Delta}{H^2} 
- \left( 3 c_s^2 + \frac{1}{\gamma} 
+ 2 \frac{H^\prime}{H^2} \right) \right] \vartheta 
\, a^{1/\gamma - 3}.
\label{eq:037} 
\end{eqnarray}
In terms of the solutions to this equation, the density contrast is
expressed as follows, 
\begin{eqnarray}
\delta_\rho = \frac{2 M_{Pl}^{2}}{\rho_m} \left(
\gamma \Delta - 3 H^2 \right) 
a^{1/\gamma-3} \vartheta \label{eq:drho/rho=}
- \frac{2 M_{Pl}^{2}}{a^2 \rho_m} \left[ 3
H^2 \left( 1 + a \frac{\partial}{\partial a} \right) -
\Delta \right] \Psi, 
\label{eq:drho/rho-gen}
\end{eqnarray}
where $\vartheta$ is the same time-independent function of the spatial
coordinates as in eq.~(\ref{eq:036}).

The standard cosmological perturbations are recovered by setting the
graviton masses to zero, $m_i^2=0$. In this case one has
$\Phi-\Psi=0$, i.e., $\vartheta(x^i)=0$. Then the equations for
perturbations become identical to those in the Einstein theory. Note
that the function $\vartheta$ is determined by the initial
conditions. Setting $\vartheta=0$ would eliminate the
$\vartheta$-dependent terms in eqs.~(\ref{eq:037}) and
(\ref{eq:drho/rho=}) and bring these equations to the conventional
form even in the case $m_G^2\neq 0$. Hence, there always exist initial
conditions such that the model (\ref{Zgammaaction}) exhibits the
standard rate of perturbation growth and, therefore, is compatible
with observations. Furthermore, at some values of the parameter
$\gamma$ the part of the perturbations that is proportional to
$\vartheta(x^i)$ grows slower than the conventional part and hence it
is subdominant, so the agreement with observations is achieved for any
function $\vartheta(x^i)$ provided it is not too large.

In the case of matter perturbations in matter-dominated Universe
eq.~(\ref{eq:037}) reduces to the following equation,
\begin{eqnarray*}
\frac{\partial^2 \Psi}{\partial a^2} + \frac{7}{2 a} 
\frac{\partial \Psi}{\partial a} 
+ \left( \frac{1}{\gamma} - 1 \right) a^{1/\gamma - 3} \vartheta= 0,
\label{eq:matter-domination}
\end{eqnarray*}
which differs from the standard case by the presence of the
inhomogeneous term proportional to $\vartheta$. The solution to this
equation reads
\begin{eqnarray*} \label{043}
\Psi = - \frac{2 \gamma}{2 + 3 \gamma} a^{1/\gamma - 1} \vartheta(x^i) 
+ a^{-5/2}  c_1(x^i)
+ c_2(x^i),
\end{eqnarray*}
where $c_i(x^i)$ are the integration constants. Substituting this
solution into eq.~(\ref{eq:drho/rho=}) one finds the density contrast 
\begin{eqnarray}
\nonumber
\delta_\rho
&=& \left( \frac{2 M_{Pl}^{2} a}{ \rho_0} 
\Delta + 3 \right) \frac{c_1(x^i)}{a^{5/2}}
+ 2 \left( \frac{a M_{Pl}^{2}}{ \rho_0} \Delta -1\right) c_2(x^i)
\\ && \label{eq:drho/rho-matter} + \frac{6 \gamma}{ 2 + 3 \gamma} 
a^{1/\gamma - 1} \left( \frac{a \gamma M_{Pl}^{2}}{ \rho_0}
\Delta - 1 \right) \vartheta (x^i),
\end{eqnarray}
where $\rho_0$ is the energy density of matter at present. The first
two terms in this equation are precisely the ones which appear in the
standard Einstein theory, the second term describing the linear growth
of the perturbations, $\delta_\rho \propto a$.  The difference with
the conventional case is in the third term on the right hand side of
eq.~(\ref{eq:drho/rho-matter}). The perturbations corresponding to
this term grow proportionally to $a^{1/\gamma}$. For $\gamma > 1$ or
$\gamma < 0$ these ``anomalous'' perturbations grow slower than the
standard ones.

At the epoch of radiation domination the situation is similar. For
relativistic fluid one has $c_s^2=w = 1 / 3$, so that
eq.~(\ref{eq:037}) becomes
\begin{eqnarray}
\frac{\partial^2 \Psi}{\partial a^2} + \frac{4}{a}
\frac{\partial \Psi}{\partial a} - \frac{M_{Pl}^2
\Delta}{\rho_r} \Psi 
+ \left( \frac{1}{\gamma} - 1 - \frac{a^2 \gamma M_{Pl}^2
\Delta}{\rho_r} \right) a^{1/\gamma-3}\vartheta = 0 \; ,
\label{eq:psi_radiation}
\end{eqnarray}
where $\rho_r$ is the energy density of radiation at present.  For
generic value of $\gamma$ the solution to this equation is
cumbersome. For simplicity let us concentrate on the modes which are
shorter than the horizon size, $k^2 \gg H^2$. The density contrast
calculated according to eq.~(\ref{eq:drho/rho-gen}) has the standard
oscillating piece and the extra part proportional to $\vartheta$,
\begin{eqnarray}
\nonumber
\delta_\rho &\sim& c_1(x^i) \sin y + c_2(x^i) \cos y 
+ 2 \gamma \left( \dfrac{\rho_r}{ k^2 M_{Pl}^2}
\right)^{(1/\gamma-1)/2} 
\\ \label{051} 
& & \times
\left[ - y^{1+1/\gamma}+ \int_0^y \textrm{d}x \, 
x^{1+1/\gamma} \sin (y-x) \right] 
 \vartheta ,\
\end{eqnarray}
where $y = \eta k / \sqrt{3}$ is proportional to the scale factor,
while $c_i(x^i)$ are two integration constants. As one can see from
this expression, for $-1\leq \gamma<0$ the $\vartheta$-dependent
contribution to the density contrast decays in time, so that only the
standard contribution remains. Thus, in this range of $\gamma$ the
perturbations behave just as predicted by General Relativity in both
matter and radiation-dominated epochs.

Another case of interest is $\gamma=1$. This case is special because
at $\gamma=1$ the $a$-dependence of the last term in
eq.~(\ref{eq:psi_radiation}) disappears. In fact, one can show that in
this case the dependence on $\vartheta$ cancels out in the density
contrast, so that only the standard part of perturbations remains.

At other values of $\gamma$, the $\vartheta$-dependent contributions
to perturbations grow in the radia\-tion dominated Universe. Whether
the model of this sort is compatible with observations depends on the
unknown function (``integration constant'') $\vartheta(x^i)$. It is
worth {\bf pointing out} that this function may become slowly varying
in time when higher-derivative corrections to the action
(\ref{Zgammaaction}) are taken into account. It remains to be
understood whether these corrections can drive $\vartheta(x^i)$ to
zero during inflation, in which case the dependence on the initial
value of $\vartheta(x^i)$ will be eliminated and the model will be
compatible with observations at any value of the parameter $\gamma$.

\subsubsection{Non-linear solutions: black holes}
\label{sec:non-line-solut}

The approach based on the Goldstone fields with the action
(\ref{dec7-a2}) (as compared to, say, the Fierz-Pauli model) is fully
non-linear. We have already taken advantage of this fact in
Section~\ref{sec:cosm-solut} where we derived cosmological solutions
in massive gravity. Another interesting question related to non-linear
gravitational dynamics is the existence and properties of black
holes. Rapid progress in observational techniques will allow for
quantitative study of astrophysical black holes in near future,
including mapping the metric near the black hole horizon
\cite{Ryan:1995wh,2000ApJ...528L..13F,Collins:2004ex,Narayan:2005ie,Broderick:2005at,Glampedakis:2005cf}.
It is therefore important to understand what kind of deviations from
General Relativity are possible at least in principle.

Black hole properties are universal in General Relativity, in the
sense that the black hole metric is uniquely characterized by the
black hole mass and angular momentum. This property is related to the
causal structure of the black hole space-time and is a consequence of
the ``no-hair'' theorems
\cite{Bekenstein:1971hc,Bekenstein:1972ky,Teitelboim:1972qx,Price:1971fb}.
Thus, properties of black holes are extremely ``resistant'' to
modifications. As an example, they remain unchanged in scalar-tensor
theories
\cite{Bekenstein:1971hc,Bekenstein:1972ky,1971ApJ...166L..35T,Psaltis:2007cw}.
For this reason, constructing an alternative model of a black hole is a
challenging problem.

We will see in this Section that the black hole properties in massive
gravity do differ from those in General Relativity. In other words, in
massive gravity black holes do have ``hair''. The origin of these hair
lies in the instantaneous interaction present in Lorentz-violating
massive gravity. Their existence is thus related to the mode with the
dispersion relation ${\bf p}^2=0$, which is in turn a consequence of
the symmetry (\ref{dec7-iii}) as has been pointed out in
Section~\ref{sec:linearized-theory}.  For simplicity, we again limit
our discussion to the particular class of models with the action
(\ref{Zgammaaction}). Since the presence of the instantaneous
interaction is a generic property of Lorentz-violating massive
gravities (in particular, models with the action (\ref{dec10-2})), we
expect our conclusions to apply to a wider class of models than
considered in this Section.

The most straightforward approach to the problem would be to try to
find the black hole solutions explicitly and to see if they differ
from General Relativity black holes. However, this appears to be a
prohibitively difficult task. One may simplify the problem by asking a
slightly different question: does massive gravity possess a black hole
solution with exactly the same metric as in General Relativity? To
answer this question one would have to find the configuration of the
Goldstone fields such that for the given black hole metric all
equations of motion (the Einstein equations and the equations of
motion of the scalar fields) are satisfied. If this is possible, then
the solution with the given metric exists. Alternatively, if this is
not possible, the black hole solutions are modified in massive
gravity.

In order that the black hole metric be a solution to the Einstein
equations, the energy-momentum tensor of the Goldstone fields $\phi^0$
and $\phi^i$ must vanish in the exterior of the black hole,
\begin{eqnarray}
0=T_{\mu\nu} =
&-& g_{\mu\nu} F 
+2 \frac{\delta F}{ \delta W^{ij} } 
\left\{ \left(\gamma \frac{W^{ij}}{ X} 
+ \frac{V^i V^j}{ X^2} \right) \partial_\mu\phi^0 \partial_\nu\phi^0
\right.
\nonumber \\
&+& \left. X^\gamma \partial_\mu\phi^i \partial_\nu\phi^j
- \frac{V^i}{ X} \left(\partial_\mu\phi^0 \partial_\nu\phi^j
- \partial_\nu\phi^0 \partial_\mu\phi^j\right) \right\}, 
\label{eq:Tmunu-curveed}
\end{eqnarray}
where $g_{\mu\nu}$ is the black hole metric. It is clear from
eq.~(\ref{eq:Tmunu-curveed}) that, except perhaps for some very
special functions $F$, the energy-momentum tensor does not vanish as
it would require that 10 equations are satisfied with 4 unknowns.

Recall now that our model is constructed in such a way that the
energy-momentum tensor of the Goldstone fields vanishes in Minkowski
space. This is achieved by choosing the vacuum solutions for the
Goldstone fields, eqs.~(\ref{dec11-1}) and (\ref{dec11-5}), in such a
way that eqs.~(\ref{dec10-1}) are satisfied. For the model
(\ref{Zgammaaction}) the latter equations imply
\[
F=0, \qquad 
\frac{\delta F}{ \delta W^{ij}}=0
\]
in the Minkowski vacuum where 
\begin{equation}
W^{ij}=-\delta^{ij}.
\label{eq:Z=delta}
\end{equation}
Thus, we can make $T_{\mu\nu}$ vanish if we find a configuration of
the Goldstone fields such that eqs.~(\ref{eq:Z=delta}) are satisfied
{\it in the background metric of the black hole}.

There are fewer equations in (\ref{eq:Z=delta}) as compared to
(\ref{eq:Tmunu-curveed}), but they are still too many~--- the system
(\ref{eq:Z=delta}) contains 6 differential equations for only 4
unknown functions $\phi^0$ and $\phi^i$. Consequently, if there are no
degeneracies, these equations are impossible to satisfy and we expect
that the Goldstone fields cannot be adjusted in such a way that their
energy-momentum tensor is zero.

An equivalent form of eq.~(\ref{eq:Z=delta}) can be obtained by going
into the unitary gauge. In this gauge eq.~(\ref{eq:Z=delta}) becomes
\begin{equation}
(g^{00})^\gamma g_{ij}^{-1} = \delta^{ij}. 
\label{eq:Z=delta-unitarygauge}
\end{equation}
In geometrical terms, solving eq.~(\ref{eq:Z=delta-unitarygauge}) is
equivalent to finding, for a given metric, the coordinate frame in
which the constant time slices are conformally flat. This
reformulation of eq.~(\ref{eq:Z=delta}) is particularly convenient.

In the case of the Schwarzschild black hole there actually {\it
exists} a solution to eqs.~(\ref{eq:Z=delta}). Equivalently, there
exists a coordinate frame in which the spatial part of metric is
conformally flat, the so-called Gullstrand-Painleve frame. In this
frame the black hole metric has the form
\[
ds^2 = d\tau^2 - \left( dx^i-\frac{R_s^{1/2}}{ r^{3/2}} x^i d\tau
\right)^2, 
\]
where $R_s$ is the Schwarzschild radius of the black hole and
$r=\sqrt{x_i^2}$, while the scalar field configuration that solves
eqs.~(\ref{eq:Z=delta}) is simply
\begin{equation}
\phi^0= \Lambda^2 \tau, \qquad \phi^i = \Lambda^2 x^i. 
\label{eq:phi-GPframe}
\end{equation}
Transforming back to the Schwarzschild coordinates one finds
\[
\phi^0 = \Lambda^2 \left[t+ 2 \sqrt{r R_s} + 
R_s \ln \left( \frac{\sqrt{r} - \sqrt{R_s}}{\sqrt{r} + \sqrt{R_s}} \right)
\right],
\]
while $\phi^i$ are still given by eqs.~(\ref{eq:phi-GPframe}). Thus,
the Schwarzschild black holes are solutions of massive gravity as
well.

The situation is different in the case of a rotating black hole: the
above miracle does not happen and, as expected,
eqs.~(\ref{eq:Z=delta}) or (\ref{eq:Z=delta-unitarygauge}) do not have
solutions. In fact, conformally flat spatial slicings are an important
ingredient in the numerical simulations of the black hole mergers, so
their existence for various solutions of the Einstein equations has
been extensively studied~\cite{Garat:2000pn,ValienteKroon:2004gj}. In
particular, it was proven that the conformally flat slicing of the
Kerr metric is impossible due to the existence of the non-trivial
invariant of the quadrupole origin \cite{ValienteKroon:2004gj}
\begin{equation}
\label{obstruction}
{\Upsilon}=-112\pi J^2.
\end{equation}
Moreover, the results of Ref. \cite{ValienteKroon:2004gj} imply that
not only the Kerr metric, but an arbitrary axisymmetric vacuum
solution of the Einstein equations with non-zero angular momentum has
a non-vanishing value of $\Upsilon$ and, consequently, does not allow
conformally flat spatial slicings. Therefore, there are no
configurations of the Goldstone fields such that their energy-momentum
tensor is zero in the background of the Kerr or any other metric with
non-zero angular momentum. Consequently, rotating black holes in
massive gravity have to be different from the Einstein theory.

The fact that the rotating black holes are modified in the presence of
the Goldstone fields as compared to their General Relativity
counterparts is in accord with the expectation that in massive gravity
black holes may have ``hair''. The existence of  hair can be
demonstrated explicitly in a toy model of Lorentz-violating
electrodynamics with the action \cite{Dubovsky:2007zi}
\begin{equation}
S = S_{EH} + \int d^4x \sqrt{-g} 
\left\{ F(X)-\frac{1}{ 4} F_{\mu\nu}^2 + m^2 G^{\mu\nu}A_\mu A_\nu \right\}.
\label{eq:massiveQED-action}
\end{equation}
Here $S_{EH}$ is the Einstein-Hilbert action, $X$ is given by
eq.~(\ref{dec8-2}) while $G^{\mu\nu}$ is the ``effective metric''
\[
G^{\mu\nu} = g^{\mu\nu} - 
\frac{\partial^\mu \phi^0 \partial^\nu\phi^0}{ X}. 
\] 
This model is analogous to massive gravity in that it possesses the
instantaneous interactions~\cite{Dvali:2005nt} which are responsible
for the presence of black hole hair. Moreover, one can
show~\cite{Dubovsky:2007zi} that the standard charged rotating black
holes are not solutions in this model, in full similarity with
rotating black holes not being solutions in massive gravity.

To demonstrate the existence of electromagnetic hair in the model
(\ref{eq:massiveQED-action}) one has to show that there exist
non-trivial static finite-energy solutions for the linearized
perturbations of the electromagnetic field in the background of the
Schwarzschild black hole. In the sector with the angular momentum
$l=1$, the vector field can be parameterized by 4 real functions of
the radial variable $\rho$ (see Ref.~\cite{Dubovsky:2007zi} for
explicit expressions). The equation for the perturbations of the
vector field translates into a coupled system of ordinary differential
equations for the radial functions. One has to show that there exists
a solution to this system that is regular both at infinity and at the
black hole horizon $\rho\to-\infty$.

The existence of a regular solution can be demonstrated by counting of
decreasing and growing modes in the asymptotic regions. Here we only
present the results; details can be found in
Ref.~\cite{Dubovsky:2007zi}. One can show that one of the four radial
functions decouples and the corresponding equation does not have
regular solutions. The equations for the remaining three radial
functions can be rewritten in terms of a single fourth-order
equation. Thus, an arbitrary solution is parameterized by four real
parameters, one of which is an overall normalization. At infinity,
there are two decreasing and two growing solutions. Requiring the
general solution to decrease at infinity fixes two of these three
parameters. At the horizon one finds three regular and one singular
solutions. The remaining free parameter can, therefore, be used to
eliminate the singular part and obtain the solution regular everywhere
--- the dipole hair.

Overall, the following picture emerges.  Black holes in massive
gravity have no reason to be universal. In particular, the metric of a
rotating black hole can (and must, according to the direct analysis)
be different from that in the Einstein theory. The differences between
different possible metrics -- black hole hair -- depend on the
collapse history. It has been argued \cite{Dubovsky:2007zi} that these
differences, as well as the deviations from the standard metric, are
of order one only at distances much larger than the inverse graviton
mass $m^{-1}$, and are likely to be suppressed by the factor $\sim
(ml)^2$ at distances $l\ll m^{-1}$, unless the parameters of the model
are tuned. Given the existing constraint on the graviton mass
(\ref{pulsarlimit}), in the simplest models the effects of the black
hole hair are observable only for the largest black holes with masses
of $\mbox{(a few)}\times 10^{9}$~M$_\odot$.

\section{Conclusions}

To summarize, Lorentz-invariant massive gravity in 4 dimensions has
severe self-consistency problems. It has either ghosts in the
perturbation spectrum about Minkowski space, or unacceptably low UV
energy scale at which strong coupling sets in, plus Boulware--Deser
ghost mode away from Minkowski background. Because of
Lorentz-invariance, the pathological ghost modes exist at arbitrarily
high spatial momenta, so vacuum in this theory is catastrophically
unstable. Presently, no way of curing these problems is known, and it
appears rather unlikely that this theory can be made healthy and
phenomenologically acceptable.

Infrared modified gravities may be less problematic in theories with
extra spatial dimensions and brane-worlds. Among the most widely
discussed models of this sort is the DGP model whose normal (as
opposed to self-accelerated) branch does not have ghosts in the
spectrum and may or may not have acceptably high UV strong coupling
scale.

In this review we followed another route and discussed
Lorentz-violating theories. Among those, we concentrated on a subclass
of theories which, in the unitary gauge, have only metric as dynamical
field, and which have Minkowski space as a solution to the field
equations. Under these conditions, the lowest order terms in the
action in Minkowski background are mass terms for metric
perturbations.  Hence, the emphasis in this review was on
Lorentz-violating massive gravities.  There is a plethora of other
possibilities, some of which are reviewed, e.g., in 
Refs.~\cite{Magueijo:2003gj,Jacobson:2008aj}.

Once the spectrum of a theory is
not Lorentz-invariant, ghosts, and to lesser extent tachyons
become phenomenologically acceptable, provided they exist only
at sufficiently low spatial momenta and energies and only weakly
(e.g., gravitationally) interact with matter. Furthermore,
some Lorentz-violating massive gravities do not have
obvious pathologies at all, and are phenomenologically acceptable
even for relatively high energy scale of Lorentz-violation.
Unlike the Fierz--Pauli theory that has vDVZ discontinuity,
at the classical level
these theories are smooth, perturbative deformations
of General Relativity, while at the quantum level
their UV strong coupling scale is not dangerously low. 
The most appealing among these theories
are the ones that leave unbroken some part of the diffeomorphism
invariance of General Relativity, the feature that ensures the
stability of these theories against deformation of the background
and/or generation of higher order terms in the action.

A general problem we should mention in this regard is the
UV completion of these theories. Unlike General Relativity which
is believed to be an effective low energy theory descending from string 
theory, massive gravities do not have obvious string theory
completions.
We believe this issue is worth investigating in the future.

Massive gravities of the sort we discuss in this review are
conveniently analyzed by making use of the St\"uckelberg--Goldstone
formalism. This formalism involves scalar fields whose background
values roll along either time-like or space-like directions, or both.
The advantage is that the full general covariance is restored, so at
energies and momenta exceeding the graviton mass scale the new modes,
over and beyond gravitons of General Relativity, are perturbations of
these scalar fields, which effectively decouple from gravity (except
for the Fierz--Pauli case).  In this way the spectrum of the theory is
studied rather straightforwardly. Furthermore, one may view the
Goldstone action as a non-linear generalization of the graviton mass
terms, and proceed to study non-linear properties of the resulting
theory, such as cosmology and black holes.

Rolling scalar fields are interesting in many respects, even though
their perturbations may be gauged away so that in the unitary gauge
the theory involves  metric only.  In the cosmological context,
rolling scalar fields are capable of giving rise to the late time
accelerated expansion of the Universe, with non-trivial equation of
state of the effective dark energy.  It is worth noting that other
theories with IR modified gravity are often unable to do that. As an
example, in theories with vector field condensates, the latter may
tend to constant values at late times.  Then there is a general
argument showing that the late time evolution of the Universe is
basically the same as in General Relativity\footnote{The condensates
may still evolve at the present cosmological epoch, leading to dark
energy with non-trivial equation of state. This possibility has been
explored, e.g., in
Refs. \cite{Rubakov:2006pn,Ferreira:2006ga,Libanov:2007mq}.}, possibly
with the cosmological constant~\cite{Eling:2004dk,Libanov:2005vu}.
The argument goes as follows.
Spatially flat, homogeneous and isotropic metric has the general form
\[
ds^2 = N^2 (t) dt^2 - a^2 (t) \delta_{ij} dx^i dx^j \; .
\]
This form is symmetric under time reparameterizations and space
dilations,
\be
t\to t' (t) \; , \;\;\;\;\;
 x^i\to \lambda  x^i
\label{jan30vr}
\ee with arbitrary function $t^\prime(t)$ and arbitrary constant
$\lambda$.  With matter fields settled down at their vacuum values (in
locally Minkowski frame), the only dynamical variables are $N(t)$ and
$a(t)$, and the action for these variables should respect the
symmetries (\ref{jan30vr}).  The only action that is local in time,
consistent with these symmetries and has no more than two time
derivatives is
\[
S (N,a)=\tilde{M}_{Pl}^2 \int
\limits_{}^{}\!dt \frac{1}{N}\left(\frac{\dot{a}}{a} \right)^2
- \tilde{\Lambda} \int \! N dt
\]
where $\tilde{M}_{Pl}$ and $\tilde{\Lambda}$ need not coincide with
the genuine Planck mass and cosmological constant.  This action has
precisely the same form as the action of General Relativity with the
cosmological constant, specified to homogeneous and isotropic space.
No matter what condensates are there in the Universe, its evolution
proceeds according to the Friedman equation, possibly with modified
Newton's and cosmological constants\footnote{In fact, the possibility
that the ``cosmological'' Newton's constant and ``Newton's law''
Newton's constant may be different, is of phenomenological
interest\cite{Carroll:2004ai}.}, provided that the condensates are
independent of space-time point (in locally Minkowski frame) and
consistent with homogeneity and isotropy of space.

Despite apparent generality, this argument does not apply to the
rolling scalar fields just because their background values do depend
on space-time point.  We gave explicit example of non-trivial
late-time cosmological evolution in Section~\ref{sec:cosm-solut}. More
possibilities emerge if one adds a scalar potential for the rolling
field(s), as we discussed in the end of
Section~\ref{sec:ghost-cond-modif}.

Lorentz-violating massive gravities have a number of other interesting
features. Massive gravitons are candidates for dark matter particles;
in that case the dark matter detection is a job for future (and maybe
even present) gravitational wave searches. Unlike in Lorentz-invariant
theories, black holes are expected to have reach properties. On
phenomenological side, this opens up a possibility to search for
Lorentz-violation by measuring the metrics of black holes in the
vicinity of their horizons.  From theoretical viewpoint,
Lorentz-violating massive gravities may be employed to gain better
insight into both classical and quantum aspects of black hole
physics. The studies of these fascinating issues have started only
recently, and one expects rapid progress in this direction.

\vspace{0.5cm}

The authors are indebted to M.~Bebronne, S.~Dubovsky, S.~Sibiryakov
and M.~Libanov for helpful discussions and reading of the manuscript.
The work of V.R is supported in part by Russian Foundation for Basic
Research, grant 08-02-00473. The work of P.T. is supported in part by
the Belspo:IAP-VI/11 and IISN grants.

%
%
%
%

\appendix

\section{Appendix }
\label{appA}

In this Appendix we give details of the treatment of small 
perturbations about cosmological background
in Lorentz-invariant massive gravity.
 We use the setup and notations of Section \ref{subsub-cosmo}.
To give explicit examples, we use the theory with the particular form
of the mass term, which is given by (\ref{FPmasscurved}).
With the cosmological constant term included,
the background equations in conformal time are, in general
\begin{eqnarray}
\H^2 &=&  H_0^2 a^2 + \epsilon_0
\nonumber \\
\label{BD-back-Fr}
2 \H^\prime + \H^2 &=& 3 H_0^2 a^2
+ \epsilon_s
\label{BD-backgr}
\end{eqnarray}
where $\epsilon_0 (a, n)$ and $\epsilon_s (a,n)$ come from the mass term. 
As pointed out in Section
 \ref{subsub-cosmo},
consistency of these equations implies an equation relating $n (\eta)$
and $a(\eta)$,
which generically has the form
$  n^\prime  = f(n,a) a^\prime $, but this is irrelevant for the discussion
here, as $a$ and $n$ can take arbitrary values at a given moment of time.
We will be interested in nearly Minkowski backgrounds,
for which $|a-1| \ll 1$, $|n-1| \ll 1$. 
Since the mass term is quadratic in metric perturbations 
about Minkowski background, the source functions
$\epsilon_0$ and $\epsilon_s$
vanish for $a=1$, $n=1$ and near these values one has
\begin{equation}
    \epsilon_0 \; , \; \epsilon_s = O(a-1) + O(n-1) 
\nonumber\\
\end{equation}
As an example, in the case
(\ref{FPmasscurved}) the source functions are
\begin{equation}
\epsilon_0 = - \frac{1}{2} m_G^2
(a^2 -1) n \; , \;\;\;\;
\epsilon_s =
- \frac{m_G^2}{2n}[2 (a^2 -1) + (a^2 n^2 -1)]
\label{nov15-9}
\end{equation}

We begin with the range of momenta ({\it i}) in (\ref{nov15-7}). 
In this range, the parameters $\H^2$, $|a-1|$ and $|n-1|$
are the smallest parameters in the problem; we take them formally
to be
of one and the same order.
We make use of the fact that the Minkowski values for the parameters
 entering (\ref{dSaction-mass}) are the ones in the Fierz--Pauli
theory, so that
\bea
m_\psi^2 &=&  3 m_G^2 + O(\H^2) \; , \; \;\;
m_B^2 = - \half m_G^2 + O(\H^2) \; , \; \;\;
\nonumber \\
\mu_1 &=& 3 m_G^2 + O(\H^2) \; , \;\;\;
\mu_2 = - m_G^2  + O(\H^2) \; , \; \;\;
\mu_3 =  - 2 m_G^2  + O(\H^2)
\label{nov15-8}
\eea
while the rest of the parameters are $ O(\H^2)$, meaning that
they vanish in the Minkowski limit.
To find the number of dynamical modes and obtain their dispersion
relations, we
write down the system of
linear equations for perturbations,
and calculate its determinant to order $\H^2$. The equations are
\bea
\varphi && : \;\;\;
m_\varphi^2 \varphi + 2 \H \Delta B + (-2 \Delta \psi + 6 \H \psi^\prime
+ \mu_1 \psi) + (-2 \H \Delta E^\prime + \mu_2 \Delta E) = 0
\label{Phieq}
\\
B/\Delta && : \;\;\; 2\H \varphi + m_B^2 B - 2\psi^\prime = 0
\label{Ueq}
\\
\psi && : \;\;\; (-2\Delta \varphi - 6 \H \varphi^\prime
- 6q \varphi + \mu_1 \varphi) +
(2 \Delta B^\prime + 4 \H \Delta B)
\nonumber \\
&& \;\;\;\;\;\;\;\;\;
+ (6 \psi^{\prime \prime} - 2 \Delta \psi + 4 \H \psi^\prime
+ 2q \psi + m_\psi^2 \psi) + (-2 \Delta E^{\prime \prime}
-4 \H \Delta E^\prime + \mu_3 \Delta E) = 0
\nonumber \\
E/\Delta && : \;\;\;
(2 \H \varphi^\prime + 2q \varphi + \mu_2 \varphi)
+ (-2 \psi^{\prime \prime} - 4 \H \psi^\prime + \mu_3 \psi)
+ m_E^2 \Delta E =0
\nonumber
\label{fulleq-FRW}
\eea
where $q=\H^\prime + 2 \H^2 = O(\H^2)$.
We now calculate the determinant of this system to figure out
the number of modes and their dispersion relations.
We find, after going to Fourier space and 
using (\ref{nov15-8}),
\bea
\frac{1}{\Delta} \mbox{Det} = &&m_G^6 (3 \omega^2 - 3{\bf p}^2 - 3m_G^2)
\nonumber \\
&& + (12 m_G^2 \H^2 + 2 m_G^2 m_\varphi^2) \cdot \omega^4 + 
O(m_G^2 \H^2 \omega^2 {\bf p}^2) + O(m_G^2 \H^2 {\bf p}^4)
\label{Det} 
\eea
Here we assume that  ${\bf p}^2 \gg m_G^2$, $\omega^2 \gg m_G^2$,
and keep only those new (with respect to Minkowski space)
terms which are proportional to the highest
power of $\omega$. 
Note that there have been cancellations: in particular,
the terms of order $\H^2 {\bf p}^2 \omega^4$ cancelled out.
These cancellations are  remnant of the gauge invariance:
the terms of order  $\H^2 {\bf p}^2 \omega^4$ would be independent of the
graviton mass, so they would remain in de~Sitter space for
massless gravitons, which would be inconsistent with gauge invariance.

Since the determinant is of the fourth order in $\omega$,
there are two modes. One of them has the dispersion relation
of the (longitudinal component of) massive graviton. This result is valid
in the range of momenta ({\it i}) only; indeed, the terms neglected in
(\ref{Det}) are large at high momenta.
We will discuss the high momentum limit later on.
In the range of momenta under discussion here,
the second, Boulware--Deser 
mode has $\omega^2 \gg {\bf p}^2$, and its frequency
is given by
\begin{equation}
\omega^2 = - \frac{3 m_G^4}{12 \H^2 + 2 m_\varphi^2}
\label{BDfreq}
\end{equation}
The  discussion here
is valid for arbitrary mass term, not necessarily
 (\ref{FPmasscurved}). 
For any mass term having the Fierz--Pauli form in Minkowski
background, one has $m_\varphi^2 \sim \H^2$, 
so that there is no smooth limit
to Minkowski space for the frequency of the
Boulware--Deser mode. 
This mode can be both  tachyonic and non-tachyonic;
in the example (\ref{FPmasscurved}) this depends on the sign
of $(a-1)$. Indeed, in this example $m_\varphi^2 = -6 \H^2
- 3\epsilon_0$, so the frequency is given by
\[
\omega^2 
=  \frac{ m_G^4}{2\epsilon_0} 
\]
and its sign  is inverse to the sign of $(a-1)$, see (\ref{nov15-9}).

To see whether the Boulware--Deser mode is a ghost, we take advantage 
of the fact that its frequency (\ref{BDfreq}) does not depend on
spatial momenta, provided they belong to the region
 ({\it i}). Thus, to obtain the action for this mode,
we can omit terms with the Laplacian in eqs.~(\ref{Phieq}) and
(\ref{Ueq}), except for the terms containing $\Delta E$
(here we treat $\Delta E$ as a field, on equal footing with $\psi$).
We thus obtain
\begin{eqnarray}
\varphi &=& \frac{1}{m_\varphi^2} [2\H (\Delta E - 3 \psi)^\prime
+ m_G^2 (\Delta E - 3 \psi)] 
\nonumber \\
B &=& \frac{1}{m_B^2}(2 \psi^\prime - 2 \H \varphi)
\nonumber
\end{eqnarray}
where we used leading-order expressions for $\mu_1$ and $\mu_2$,
namely $\mu_1 = 3 m_G^2$, $\mu_2 = - m_G^2$.
Plugging these expressions back into the action (\ref{dSaction-mass}),
integrating by parts 
and again omitting terms with the spatial Laplacian
and terms suppressed by the ratio $\H^2 /m_G^2$,
we find the action for the dynamical fields $\psi$ and $\Delta E$,
\begin{equation}
S^{(2)}_{EH + \Lambda + m_G} = 2 M_{Pl} \int~d^3x d\eta~a^2 \left[
- \frac{6\H^2 + m_\varphi^2}{3 m_\varphi^2 } \l \Delta E^\prime - 
3 \psi^\prime \r^2 - \frac{m_G^4 }{2 m_\varphi^2} \l \Delta E - 3 \psi \r^2
 + \frac{1}{3} (\Delta E^\prime)^2
\right]
\nonumber
\end{equation}
This expression is again valid for any mass term, not necessarily
 (\ref{FPmasscurved}).
Here we neglected the terms $m_\psi^2 \psi^2$ and
$\mu_3 \psi \Delta E$, as their contributions to the action
 for the Boulware--Deser mode are suppressed by 
$\H^2 /m_G^2$. It is clear from this action that the
Boulware--Deser field with the dispersion relation (\ref{BDfreq})
is $(\Delta E - 3 \psi)$, and that in the case when
this field is not a tachyon, it is a ghost. Indeed,
this field is not a tachyon for $(6\H^2 + m_\varphi^2) < 0$,
which also implies $m_\varphi^2 <0$, so that the kinetic term 
for this field is negative. Note also that the action 
for this field is singular in the Minkowski limit, i.e., in the
limit $\H \to 0$, in which $m_\varphi^2 \to 0$ as well.
 
Let us repeat that this analysis is valid at ${\bf p}^2 \ll
m_G^4/\H^2$ only.  Going to high-momentum limit using the expression
(\ref{Det}) would be incorrect, partially because the terms not
explicitly written in (\ref{Det}) are important at high momenta,
partially because the terms of higher order in $\H$ potentially are
higher order in ${\bf p}^2$.

Let us now proceed to high momentum limit ({\it ii}) in  (\ref{nov15-7}).
The analysis leading to the action (\ref{finalaction-mass}) applies 
to any mass term. This action has the following general form,
\begin{equation}
S^{(2)}_{EH + \Lambda + m_G} = 2 M_{Pl} \int~d^4x~a^2 \left\{
A \psi \Delta \psi
+ C (\psi^\prime)^2  +  B \psi^\prime \Delta E
+ \frac{m_B^2}{2} E^\prime \Delta E^\prime 
+ \frac{m_E^2}{2} (\Delta E)^2 \right\}
\nonumber
\end{equation}
 To proceed further, let us consider the mass term (\ref{FPmasscurved}).
After straightforward calculation one finds 
in that case
\begin{eqnarray}
m_\varphi^2 = - 6 \H^2 - 3 \epsilon_0 \; ,  &&
m_B^2 = - 3 \epsilon_0 - \half m_G^2 a^2 n \; , \;\;\;\;\;\;
\mu_2 = - 3 \epsilon_0 - m_G^2 a^2 n
\nonumber \\
1 - \frac{\H^\prime}{\H^2}  
 &=& \frac{3 \epsilon_0 - \epsilon_s}{2 \H^2} \; , \;\;\;\;\;\;\;
m_E^2 = - \epsilon_s
\nonumber
\end{eqnarray}
Hence
\begin{eqnarray}
A&\equiv&  1 - \frac{\H^\prime}{\H^2}
+ \frac{m_B^2}{2\H^2} = \frac{1}{4\H^2} (-2\epsilon_s - m_G^2 a^2 n)
\nonumber \\
B &\equiv&   \frac{\mu_2 - m_B^2}{\H} =
- \frac{m_G^2 a^2 n}{2\H}
\nonumber \\
C &\equiv& 
3 + \frac{m_\varphi^2}{2\H^2}
= - \frac{3\epsilon_0}{2 \H^2}
\nonumber
\end{eqnarray}
The dispersion relations are obtained by solving the equations of motion 
for $\psi$ and $E$. There are two modes, one with
\begin{equation}
  \omega_1^2 =  {\bf p}^2
\nonumber
\end{equation}
and another with
\begin{equation}
  \omega_2^2 = \frac{\epsilon_s}{3\epsilon_0}
\cdot {\bf p}^2
\nonumber
\end{equation}
These expressions are valid for all $a$ and $n$, not
necessarily close to 1.

One of these modes is a tachyon or ghost. Indeed, positivity of energy
requires
\begin{equation}
 A>0 \; , \;\;\;  C > 0 \; ,   \; \;\; m_B^2 < 0 \; , \;\;\;
m_E^2 > 0 \; .
\nonumber
\end{equation}
Now, 
\[
\omega_1^2 = 2\H^2 \cdot \frac{A}{m_B^2}
\]
so the latter requirements would imply $\omega_1^2 < 0$, that is,
tachyon. Note that in certain backgrounds, the second mode is
superluminal.


\begin{thebibliography}{100}

\bibitem{Weinberg:1988cp}
S.~Weinberg,
\newblock Rev. Mod. Phys. {\bf 61}, 1 (1989).

\bibitem{Sahni:1999gb}
V.~Sahni and A.~A. Starobinsky,
\newblock Int. J. Mod. Phys. {\bf D9}, 373 (2000), [astro-ph/9904398].

\bibitem{Dolgov:1997za}
A.~D. Dolgov,
\newblock astro-ph/9708045.

\bibitem{Chernin:2001sy}
A.~D. Chernin,
\newblock Phys. Usp. {\bf 44}, 1099 (2001).

\bibitem{Padmanabhan:2002ji}
T.~Padmanabhan,
\newblock Phys. Rept. {\bf 380}, 235 (2003), [hep-th/0212290].

\bibitem{Peebles:2002gy}
P.~J.~E. Peebles and B.~Ratra,
\newblock Rev. Mod. Phys. {\bf 75}, 559 (2003), [astro-ph/0207347].

\bibitem{Copeland:2006wr}
E.~J. Copeland, M.~Sami and S.~Tsujikawa,
\newblock Int. J. Mod. Phys. {\bf D15}, 1753 (2006), [hep-th/0603057].

\bibitem{Spergel:2006hy}
WMAP, D.~N. Spergel {\em et~al.},
\newblock Astrophys. J. Suppl. {\bf 170}, 377 (2007), [astro-ph/0603449].

\bibitem{Dolgov:2006xi}
A.~D. Dolgov,
\newblock hep-ph/0606230.

\bibitem{Rubakov:1999aq}
V.~A. Rubakov,
\newblock Phys. Rev. {\bf D61}, 061501 (2000), [hep-ph/9911305].

\bibitem{Steinhardt:2006bf}
P.~J. Steinhardt and N.~Turok,
\newblock Science {\bf 312}, 1180 (2006), [astro-ph/0605173].

\bibitem{Linde:2002gj}
A.~Linde,
\newblock hep-th/0211048.

\bibitem{Rubakov:2001kp}
V.~A. Rubakov,
\newblock Phys. Usp. {\bf 44}, 871 (2001), [hep-ph/0104152].

\bibitem{Charmousis:1999rg}
C.~Charmousis, R.~Gregory and V.~A. Rubakov,
\newblock Phys. Rev. {\bf D62}, 067505 (2000), [hep-th/9912160].

\bibitem{Kogan:1999wc}
I.~I. Kogan, S.~Mouslopoulos, A.~Papazoglou, G.~G. Ross and J.~Santiago,
\newblock Nucl. Phys. {\bf B584}, 313 (2000), [hep-ph/9912552].

\bibitem{Gregory:2000jc}
R.~Gregory, V.~A. Rubakov and S.~M. Sibiryakov,
\newblock Phys. Rev. Lett. {\bf 84}, 5928 (2000), [hep-th/0002072].

\bibitem{Gabadadze:2007dv}
G.~Gabadadze,
\newblock Nucl. Phys. Proc. Suppl. {\bf 171}, 88 (2007), [arXiv:0705.1929
  [hep-th]].

\bibitem{Dvali:2000hr}
G.~R. Dvali, G.~Gabadadze and M.~Porrati,
\newblock Phys. Lett. {\bf B485}, 208 (2000), [hep-th/0005016].

\bibitem{Gabadadze:2003ii}
G.~Gabadadze,
\newblock hep-ph/0308112.

\bibitem{Luty:2003vm}
M.~A. Luty, M.~Porrati and R.~Rattazzi,
\newblock JHEP {\bf 09}, 029 (2003), [hep-th/0303116].

\bibitem{Rubakov:2003zb}
V.~A. Rubakov,
\newblock hep-th/0303125.

\bibitem{Nicolis:2004qq}
A.~Nicolis and R.~Rattazzi,
\newblock JHEP {\bf 06}, 059 (2004), [hep-th/0404159].

\bibitem{Deffayet:2000uy}
C.~Deffayet,
\newblock Phys. Lett. {\bf B502}, 199 (2001), [hep-th/0010186].

\bibitem{Deffayet:2001pu}
C.~Deffayet, G.~R. Dvali and G.~Gabadadze,
\newblock Phys. Rev. {\bf D65}, 044023 (2002), [astro-ph/0105068].

\bibitem{Dvali:2002pe}
G.~Dvali, G.~Gabadadze and M.~Shifman,
\newblock Phys. Rev. {\bf D67}, 044020 (2003), [hep-th/0202174].

\bibitem{Gorbunov:2005zk}
D.~Gorbunov, K.~Koyama and S.~Sibiryakov,
\newblock Phys. Rev. {\bf D73}, 044016 (2006), [hep-th/0512097].

\bibitem{Bekenstein:1984tv}
J.~Bekenstein and M.~Milgrom,
\newblock Astrophys. J. {\bf 286}, 7 (1984).

\bibitem{Bekenstein:1988zy}
J.~D. Bekenstein,
\newblock Phys. Lett. {\bf B202}, 497 (1988).

\bibitem{Bekenstein:2004ne}
J.~D. Bekenstein,
\newblock Phys. Rev. {\bf D70}, 083509 (2004), [astro-ph/0403694].

\bibitem{Bekenstein:2004ca}
J.~D. Bekenstein,
\newblock PoS {\bf JHW2004}, 012 (2005), [astro-ph/0412652].

\bibitem{Bekenstein:2005nv}
J.~D. Bekenstein and R.~H. Sanders,
\newblock astro-ph/0509519.

\bibitem{Logunov:1998ge}
A.~A. Logunov,
\newblock Relativistic theory of gravity, Commack, USA: Nova Sci. Publ. (1998)
  114 p.

\bibitem{Logunov:2001cx}
A.~A. Logunov,
\newblock gr-qc/0210005.

\bibitem{'tHooft:2007bf}
G.~'t~Hooft,
\newblock arXiv:0708.3184 [hep-th].

\bibitem{Boisseau:2000pr}
B.~Boisseau, G.~Esposito-Farese, D.~Polarski and A.~A. Starobinsky,
\newblock Phys. Rev. Lett. {\bf 85}, 2236 (2000), [gr-qc/0001066].

\bibitem{Gannouji:2006jm}
R.~Gannouji, D.~Polarski, A.~Ranquet and A.~A. Starobinsky,
\newblock JCAP {\bf 0609}, 016 (2006), [astro-ph/0606287].

\bibitem{Carloni:2007eu}
S.~Carloni, S.~Capozziello, J.~A. Leach and P.~K.~S. Dunsby,
\newblock gr-qc/0701009.

\bibitem{Jacobson:2008aj}
T.~Jacobson,
\newblock arXiv:0801.1547 [gr-qc].

\bibitem{Gripaios:2004ms}
B.~M. Gripaios,
\newblock JHEP {\bf 10}, 069 (2004), [hep-th/0408127].

\bibitem{Arkani-Hamed:2003uy}
N.~Arkani-Hamed, H.-C. Cheng, M.~A. Luty and S.~Mukohyama,
\newblock JHEP {\bf 05}, 074 (2004), [hep-th/0312099].

\bibitem{Rubakov:2004eb}
V.~A. Rubakov,
\newblock hep-th/0407104.

\bibitem{Dubovsky:2004sg}
S.~L. Dubovsky,
\newblock JHEP {\bf 10}, 076 (2004), [hep-th/0409124].

\bibitem{Dubovsky:2005dw}
S.~L. Dubovsky, P.~G. Tinyakov and I.~I. Tkachev,
\newblock Phys. Rev. {\bf D72}, 084011 (2005), [hep-th/0504067].

\bibitem{Dubovsky:2004ud}
S.~L. Dubovsky, P.~G. Tinyakov and I.~I. Tkachev,
\newblock Phys. Rev. Lett. {\bf 94}, 181102 (2005), [hep-th/0411158].

\bibitem{Fierz:1939ix}
M.~Fierz and W.~Pauli,
\newblock Proc. Roy. Soc. Lond. {\bf A173}, 211 (1939).

\bibitem{Arkani-Hamed:2002sp}
N.~Arkani-Hamed, H.~Georgi and M.~D. Schwartz,
\newblock Ann. Phys. {\bf 305}, 96 (2003), [hep-th/0210184].

\bibitem{Kakushadze:2007hf}
Z.~Kakushadze,
\newblock arXiv:0710.1061 [hep-th].

\bibitem{Porrati:2001db}
M.~Porrati,
\newblock JHEP {\bf 04}, 058 (2002), [hep-th/0112166].

\bibitem{Mukhanov:1990me}
V.~F. Mukhanov, H.~A. Feldman and R.~H. Brandenberger,
\newblock Phys. Rept. {\bf 215}, 203 (1992).

\bibitem{vanDam:1970vg}
H.~van Dam and M.~J.~G. Veltman,
\newblock Nucl. Phys. {\bf B22}, 397 (1970).

\bibitem{Zakharov:1970cc}
V.~I. Zakharov,
\newblock JETP Lett. {\bf 12}, 312 (1970).

\bibitem{Will:2005va}
C.~M. Will,
\newblock gr-qc/0510072.

\bibitem{Ford:1980up}
L.~H. Ford and H.~Van~Dam,
\newblock Nucl. Phys. {\bf B169}, 126 (1980).

\bibitem{Vainshtein:1972sx}
A.~I. Vainshtein,
\newblock Phys. Lett. {\bf B39}, 393 (1972).

\bibitem{Creminelli:2005qk}
P.~Creminelli, A.~Nicolis, M.~Papucci and E.~Trincherini,
\newblock JHEP {\bf 09}, 003 (2005), [hep-th/0505147].

\bibitem{Long:2003dx}
J.~C. Long {\em et~al.},
\newblock Nature {\bf 421}, 922 (2003).

\bibitem{Long:2003ta}
J.~C. Long and J.~C. Price,
\newblock Comptes Rendus Physique {\bf 4}, 337 (2003), [hep-ph/0303057].

\bibitem{Smullin:2005iv}
S.~J. Smullin {\em et~al.},
\newblock Phys. Rev. {\bf D72}, 122001 (2005), [hep-ph/0508204].

\bibitem{Hoyle:2004cw}
C.~D. Hoyle {\em et~al.},
\newblock Phys. Rev. {\bf D70}, 042004 (2004), [hep-ph/0405262].

\bibitem{Gherghetta:2006ha}
T.~Gherghetta,
\newblock hep-ph/0601213.

\bibitem{Hewett:2005uc}
J.~L. Hewett,
\newblock Proc. Les Houches Summer School on Theoretical Physics. Session 84:
  Particle Physics Beyond the Standard Model, 2005.

\bibitem{Rubakov:2005ub}
V.~Rubakov,
\newblock Proc. Les Houches Summer School on Theoretical Physics. Session 84:
  Particle Physics Beyond the Standard Model, 2005.

\bibitem{Aubert:2003je}
A.~Aubert,
\newblock Phys. Rev. {\bf D69}, 087502 (2004), [hep-th/0312246].

\bibitem{Boulware:1973my}
D.~G. Boulware and S.~Deser,
\newblock Phys. Rev. {\bf D6}, 3368 (1972).

\bibitem{Deffayet:2005ys}
C.~Deffayet and J.-W. Rombouts,
\newblock Phys. Rev. {\bf D72}, 044003 (2005), [gr-qc/0505134].

\bibitem{Cline:2003gs}
J.~M. Cline, S.~Jeon and G.~D. Moore,
\newblock Phys. Rev. {\bf D70}, 043543 (2004), [hep-ph/0311312].

\bibitem{Creminelli:2006xe}
P.~Creminelli, M.~A. Luty, A.~Nicolis and L.~Senatore,
\newblock JHEP {\bf 12}, 080 (2006), [hep-th/0606090].

\bibitem{Dubovsky:2004qe}
S.~L. Dubovsky,
\newblock JCAP {\bf 0407}, 009 (2004), [hep-ph/0403308].

\bibitem{Peloso:2004ut}
M.~Peloso and L.~Sorbo,
\newblock Phys. Lett. {\bf B593}, 25 (2004), [hep-th/0404005].

\bibitem{ArkaniHamed:2005gu}
N.~Arkani-Hamed, H.-C. Cheng, M.~A. Luty, S.~Mukohyama and T.~Wiseman,
\newblock JHEP {\bf 01}, 036 (2007), [hep-ph/0507120].

\bibitem{Felder:2002sv}
G.~N. Felder, L.~Kofman and A.~Starobinsky,
\newblock JHEP {\bf 09}, 026 (2002), [hep-th/0208019].

\bibitem{Krotov:2004if}
D.~Krotov, C.~Rebbi, V.~A. Rubakov and V.~Zakharov,
\newblock Phys. Rev. {\bf D71}, 045014 (2005), [hep-ph/0407081].

\bibitem{Frolov:2004vm}
A.~V. Frolov,
\newblock Phys. Rev. {\bf D70}, 061501 (2004), [hep-th/0404216].

\bibitem{Mukohyama:2005rw}
S.~Mukohyama,
\newblock Phys. Rev. {\bf D71}, 104019 (2005), [hep-th/0502189].

\bibitem{Dubovsky:2006vk}
S.~L. Dubovsky and S.~M. Sibiryakov,
\newblock Phys. Lett. {\bf B638}, 509 (2006), [hep-th/0603158].

\bibitem{Babichev:2006vx}
E.~Babichev, V.~F. Mukhanov and A.~Vikman,
\newblock JHEP {\bf 09}, 061 (2006), [hep-th/0604075].

\bibitem{Dubovsky:2007zi}
S.~Dubovsky, P.~Tinyakov and M.~Zaldarriaga,
\newblock JHEP {\bf 11}, 083 (2007), [arXiv:0706.0288 [hep-th]].

\bibitem{ArkaniHamed:2003uz}
N.~Arkani-Hamed, P.~Creminelli, S.~Mukohyama and M.~Zaldarriaga,
\newblock JCAP {\bf 0404}, 001 (2004), [hep-th/0312100].

\bibitem{Piazza:2004df}
F.~Piazza and S.~Tsujikawa,
\newblock JCAP {\bf 0407}, 004 (2004), [hep-th/0405054].

\bibitem{Krause:2004bu}
A.~Krause and S.-P. Ng,
\newblock Int. J. Mod. Phys. {\bf A21}, 1091 (2006), [hep-th/0409241].

\bibitem{Senatore:2004rj}
L.~Senatore,
\newblock Phys. Rev. {\bf D71}, 043512 (2005), [astro-ph/0406187].

\bibitem{Rubakov:2006pn}
V.~A. Rubakov,
\newblock Theor. Math. Phys. {\bf 149}, 1651 (2006), [hep-th/0604153].

\bibitem{Libanov:2007mq}
M.~Libanov, V.~Rubakov, E.~Papantonopoulos, M.~Sami and S.~Tsujikawa,
\newblock JCAP {\bf 0708}, 010 (2007), [arXiv:0704.1848 [hep-th]].

\bibitem{Mukohyama:2006be}
S.~Mukohyama,
\newblock JCAP {\bf 0610}, 011 (2006), [hep-th/0607181].

\bibitem{Buchbinder:2007ad}
E.~I. Buchbinder, J.~Khoury and B.~A. Ovrut,
\newblock Phys. Rev. {\bf D76}, 123503 (2007), [hep-th/0702154].

\bibitem{Creminelli:2007aq}
P.~Creminelli and L.~Senatore,
\newblock JCAP {\bf 0711}, 010 (2007), [hep-th/0702165].

\bibitem{Buchbinder:2007tw}
E.~I. Buchbinder, J.~Khoury and B.~A. Ovrut,
\newblock JHEP {\bf 11}, 076 (2007), [arXiv:0706.3903 [hep-th]].

\bibitem{Dvali:2005nt}
G.~Dvali, M.~Papucci and M.~D. Schwartz,
\newblock Phys. Rev. Lett. {\bf 94}, 191602 (2005), [hep-th/0501157].

\bibitem{Gabadadze:2004iv}
G.~Gabadadze and L.~Grisa,
\newblock Phys. Lett. {\bf B617}, 124 (2005), [hep-th/0412332].

\bibitem{Dubovsky:2005xd}
S.~Dubovsky, T.~Gregoire, A.~Nicolis and R.~Rattazzi,
\newblock JHEP {\bf 03}, 025 (2006), [hep-th/0512260].

\bibitem{Esposito-Farese:2000ij}
G.~Esposito-Farese and D.~Polarski,
\newblock Phys. Rev. {\bf D63}, 063504 (2001), [gr-qc/0009034].

\bibitem{RevModPhys.66.711}
J.~H. Taylor,
\newblock Rev. Mod. Phys. {\bf 66}, 711 (1994).

\bibitem{Rubakov:1982df}
V.~A. Rubakov, M.~V. Sazhin and A.~V. Veryaskin,
\newblock Phys. Lett. {\bf B115}, 189 (1982).

\bibitem{Linde:1993cn}
A.~D. Linde,
\newblock Phys. Rev. {\bf D49}, 748 (1994), [astro-ph/9307002].

\bibitem{Bunch:1978yq}
T.~S. Bunch and P.~C.~W. Davies,
\newblock Proc. Roy. Soc. Lond. {\bf A360}, 117 (1978).

\bibitem{Thorne:1987af}
K.~S. Thorne,
\newblock In *Hawking, S.W. (ed.), Israel, W. (ed.): Three hundred years of
  gravitation*, 330-458. (see Book Index).

\bibitem{Bender:2003uv}
P.~L. Bender,
\newblock Class. Quant. Grav. {\bf 20}, S301 (2003).

\bibitem{Lommen:2002je}
A.~N. Lommen,
\newblock astro-ph/0208572.

\bibitem{Sazhin1978SvA}
M.~V. {Sazhin},
\newblock Soviet Astronomy {\bf 22}, 36 (1978).

\bibitem{Dodelson:2003ft}
S.~Dodelson,
\newblock Amsterdam, Netherlands: Academic Pr. (2003) 440 p.

\bibitem{Mukhanov:2005sc}
V.~Mukhanov,
\newblock Cambridge, UK: Univ. Pr. (2005) 421 p.

\bibitem{Tegmark:2003ud}
SDSS, M.~Tegmark {\em et~al.},
\newblock Phys. Rev. {\bf D69}, 103501 (2004), [astro-ph/0310723].

\bibitem{Seljak:2006bg}
U.~Seljak, A.~Slosar and P.~McDonald,
\newblock JCAP {\bf 0610}, 014 (2006), [astro-ph/0604335].

\bibitem{Bebronne:2007qh}
M.~V. Bebronne and P.~G. Tinyakov,
\newblock Phys. Rev. {\bf D76}, 084011 (2007), [arXiv:0705.1301 [astro-ph]].

\bibitem{Ryan:1995wh}
F.~D. Ryan,
\newblock Phys. Rev. {\bf D52}, 5707 (1995).

\bibitem{2000ApJ...528L..13F}
H.~{Falcke}, F.~{Melia} and E.~{Agol},
\newblock ApJ {\bf 528}, L13 (2000), [arXiv:astro-ph/9912263].

\bibitem{Collins:2004ex}
N.~A. Collins and S.~A. Hughes,
\newblock Phys. Rev. {\bf D69}, 124022 (2004), [gr-qc/0402063].

\bibitem{Narayan:2005ie}
R.~Narayan,
\newblock New J. Phys. {\bf 7}, 199 (2005), [gr-qc/0506078].

\bibitem{Broderick:2005at}
A.~E. Broderick and A.~Loeb,
\newblock Astrophys. J. {\bf 636}, L109 (2006), [astro-ph/0508386].

\bibitem{Glampedakis:2005cf}
K.~Glampedakis and S.~Babak,
\newblock Class. Quant. Grav. {\bf 23}, 4167 (2006), [gr-qc/0510057].

\bibitem{Bekenstein:1971hc}
J.~D. Bekenstein,
\newblock Phys. Rev. {\bf D5}, 1239 (1972).

\bibitem{Bekenstein:1972ky}
J.~D. Bekenstein,
\newblock Phys. Rev. {\bf D5}, 2403 (1972).

\bibitem{Teitelboim:1972qx}
C.~Teitelboim,
\newblock Phys. Rev. {\bf D5}, 2941 (1972).

\bibitem{Price:1971fb}
R.~H. Price,
\newblock Phys. Rev. {\bf D5}, 2419 (1972).

\bibitem{1971ApJ...166L..35T}
K.~S. {Thorne} and J.~J. {Dykla},
\newblock Astrophys. J. {\bf 166}, L35+ (1971).

\bibitem{Psaltis:2007cw}
D.~Psaltis, D.~Perrodin, K.~R. Dienes and I.~Mocioiu,
\newblock arXiv:0710.4564 [astro-ph].

\bibitem{Garat:2000pn}
A.~Garat and R.~H. Price,
\newblock Phys. Rev. {\bf D61}, 124011 (2000), [gr-qc/0002013].

\bibitem{ValienteKroon:2004gj}
J.~A. Valiente~Kroon,
\newblock Class. Quant. Grav. {\bf 21}, 3237 (2004), [gr-qc/0402033].

\bibitem{Magueijo:2003gj}
J.~Magueijo,
\newblock Rept. Prog. Phys. {\bf 66}, 2025 (2003), [astro-ph/0305457].

\bibitem{Ferreira:2006ga}
P.~G. Ferreira, B.~M. Gripaios, R.~Saffari and T.~G. Zlosnik,
\newblock Phys. Rev. {\bf D75}, 044014 (2007), [astro-ph/0610125].

\bibitem{Eling:2004dk}
C.~Eling, T.~Jacobson and D.~Mattingly,
\newblock gr-qc/0410001.

\bibitem{Libanov:2005vu}
M.~V. Libanov and V.~A. Rubakov,
\newblock JHEP {\bf 08}, 001 (2005), [hep-th/0505231].

\bibitem{Carroll:2004ai}
S.~M. Carroll and E.~A. Lim,
\newblock Phys. Rev. {\bf D70}, 123525 (2004), [hep-th/0407149].

\end{thebibliography}

\end{document}